\RequirePackage{snapshot}       
\documentclass[preprints,review,accept,moreauthors,pdftex]{Definitions/mdpi-trunc} 
\usepackage{subcaption}
\firstpage{1} 
\pubvolume{1}
\issuenum{1}
\articlenumber{0}
\pubyear{2021}
\copyrightyear{2020}
\datereceived{} 
\dateaccepted{} 
\datepublished{} 
\hreflink{https://doi.org/} 

\Title{Review of the Advanced LIGO gravitational wave observatories leading to observing run four}

\TitleCitation{Review of the Advanced LIGO gravitational wave observatories leading to observing run four}

\Author{Craig Cahillane $^{1}$\orcidA{}, Georgia Mansell $^{1,2,3}$\orcidB{}}

\AuthorNames{Craig Cahillane and Georgia Mansell}

\AuthorCitation{Cahillane, C.; Mansell, G.}

\address{%
    $^{1}$ \quad LIGO Hanford Observatory, Richland, Washington 99352, USA\\
    $^{2}$ \quad LIGO, Massachusetts Institute of Technology, Cambridge, MA 02139, USA\\
    $^{3}$ \quad Syracuse University, Syracuse, NY 13244, USA
}

\corres{Correspondence: ccahilla@caltech.edu}

\abstract{
    Gravitational waves from binary black hole and neutron star mergers are being regularly detected.
    As of 2021, ninety confident gravitational wave detections have been made by the LIGO and Virgo detectors.
    Work is ongoing to further increase the sensitivity of the detectors for the fourth observing run,
    including installing some of the A+ upgrades designed to lower the fundamental noise that limits the sensitivity to gravitational waves.
    In this review, we will overview how the LIGO detectors work, 
    including their optical configuration and lock acquisition procedure,
    discuss the detectors' fundamental and technical noise limits and compare to the current measured sensitivity,
    and review the A+ upgrades currently being installed in the detectors.
}


\begin{document}

\section{Introduction}
\label{sec:introduction}

On September 14th, 2015, the Advanced LIGO detectors made the first direct detection of gravitational waves (GWs) from a binary black hole merger \cite{GW150914}.
The detectors had just achieved full operation after a five-year hiatus while they were upgraded from initial to Advanced LIGO.
The Advanced LIGO detectors featured new technologies and new optical configuration, designed to improve the signal-to-noise ratio of GW signals across the audio frequency spectrum \cite{AdvLIGOPaper, AdvLIGOFinalDesign, Martynov2016}.
The upgraded detectors drastically improved the sensitivity to intermediate mass black hole merger GW signals, enabling the detection of GW150914 \cite{GW150914DetectorPaper}.

On August 17th, 2017, the Advanced LIGO and Virgo detectors discovered gravitational waves from a binary neutron star merger with a gamma-ray-burst counterpart \cite{GW170817}.
This event triggered telescopes to point in the direction of the merger, to catch electromagnetic radiation from across the energy spectrum \cite{GW170817multimessenger}.

Today, the detectors have progressed significantly toward the goal of achieving design sensitivity \cite{Buikema2020}.
Through the third observing run (O3), ninety confident gravitational wave detections from astrophysical compact binary mergers have been reported, along with many more low-confidence detections \cite{ThirdCatalogPaper}.

Now, in the period between observing runs three and four, major infrastructure improvements known as the A$+$ upgrades are being installed at the LIGO detectors \cite{SkyLocation2020, InstrumentScienceWhitePaper2021}.
These upgrades are focused on lowering the fundamental noise limit of the Advanced LIGO detectors, making higher levels of sensitivity to gravitational waves possible.

Here we will review the design and performance of the Advanced LIGO detectors leading into to observing run four (O4), scheduled to begin in December 2022.
Section \ref{sec:gravitational_waves} will briefly overview the gravitational wave signals we expect.
Section \ref{sec:detectors} will overview the Advanced LIGO optical configuration and lock acquisition process.
Section \ref{sec:sensitivity} will review the fundamental limits of the Advanced LIGO detectors' sensitivity, as well as the current achieved sensitivity.
Section \ref{sec:current_performance} will discuss the current performance of the detectors, introducing the topics of point absorbers on optics and squeezed states of light.
Section \ref{sec:upgrades} will overview the upgrades currently being installed in preparation for O4.
Section \ref{sec:conclusions} will comment on future avenues for increasing detector sensitivity.
Appendices \ref{sec:michelson_interferometer} and \ref{sec:fabry_perot_interferometer} will overview the basics of the Michelson and Fabry-P\'erot interferometric configurations, the fundamental building blocks of the full Advanced LIGO interferometer.


\section{Gravitational waves}
\label{sec:gravitational_waves}

A gravitational wave can be described as a small perturbation $h_{\mu\nu}$ on a flat spacetime metric $\eta_{\mu\nu}$ \cite{MTW, Sathyaprakash2009}:
\begin{align}
    \label{eq:spacetime_metric}
    g_{\mu\nu} = \eta_{\mu\nu} + h_{\mu\nu}.
\end{align}

In the transverse-traceless gauge, a gravitational wave propagating in the $z$ direction can be expressed as
\begin{align}
    \label{eq:gw_metric}
    h_{\mu\nu}(t, x, y, z) = \begin{pmatrix}
        0 & 0        & 0        & 0 \\
        0 & h_+      & h_\times & 0 \\
        0 & h_\times & -h_+     & 0 \\
        0 & 0        & 0        & 0
    \end{pmatrix} \cos(\omega t - k z),
\end{align}
where $h_+$ is the plus-polarization gravitational wave strain,
$h_\times$ is the cross-polarization strain,
and $\omega$ and $k$ are the frequency and wavenumber of the GW.
In Eq.~\ref{eq:gw_metric}, we have defined the usual coordinate system $(t, x, y, z)$ for the Greek indices ranging from 0 to 3.
In this gauge choice the trace of the matrix in Eq.~\ref{eq:gw_metric} is zero,
and the spacetime strain is only in the $x$ and $y$ directions, transverse to the $z$ direction of propagation.

Next we will show that a gravitational wave modulates the spacetime interval $ds$, and show how this can be interpreted as a change in length $\Delta L$ \cite{Saulson1994, Adhikari2014}.
In general, the spacetime interval between any two points is
\begin{align}
    \label{eq:spacetime_interval_general_1}
    ds^2 & = g_{\mu\nu} dx^\mu dx^\nu                                                  \\
    \label{eq:spacetime_interval_general_2}
         & = (\eta_{\mu\nu} + h_{\mu\nu}) dx^\mu dx^\nu                                \\
    \label{eq:spacetime_interval_general_3}
         & = -c^2 dt^2 + (1 + h_+) dx^2 + (1 - h_+) dy^2 + 2 h_\times dx \, dy + dz^2,
\end{align}
where we have set our coordinate vector $dx^\mu = dx^\nu = \left(c dt, dx, dy, dz \right)^T$.

Gravitational wave detectors use laser light to sense spacetime.
Light always has a spacetime interval $ds = 0$.
If we set up a test particle on the $x$-axis a length $L_x$ from the origin, and look at the spacetime interval $ds$ for a light wave traveling between the origin and particle when only a $h_+$ wave is incident,
assuming $h_+ \ll 1$ we get
\begin{align}
    \label{eq:spacetime_interval_test_particle_x_1}
    ds^2 = 0          & = -c^2 dt^2 + (1 + h_+) dx^2                                                                 \\
    \label{eq:spacetime_interval_test_particle_x_2}
    c \int_0^{t_0} dt & = \int_0^{L_x} \sqrt{1 + h_+} dx \approx \int_0^{L_x} \left( 1 + \dfrac{1}{2} h_+ \right) dx \\
    \label{eq:spacetime_interval_test_particle_x_3}
    c t_0             & =  \left( 1 + \dfrac{1}{2} h_+ \right) L_x.
\end{align}
Equation~\ref{eq:spacetime_interval_test_particle_x_3} emphasizes that, in the chosen gauge and coordinates,
the passing gravitational wave $h_+$ modulates the light travel time $t_0$ between the two stationary points $(0,0,0)$ and $(L_x,0,0)$.
Equivalently, the GW strain can be said to modulate the $x$ length: $\Delta L_x = h_+ L_x / 2$.
Along the $y$-axis between the points $(0,0,0)$ and $(0,L_y,0)$, the sign of the $h_+$ GW modulation is flipped as seen from Eq.~\ref{eq:spacetime_interval_general_3}: $\Delta L_y = - h_+ L_y / 2$.

The differential length $\Delta L$ in the $x$- and $y$-axis, assuming $L_x = L_y = L$, yields
\begin{align}
    \label{eq:gw_differential_length_1}
    \Delta L    = \Delta L_x & - \Delta L_y = h_+ L   \\
    \label{eq:gw_differential_length_2}
    h_+                      & = \dfrac{\Delta L}{L}.
\end{align}
Equation~\ref{eq:gw_differential_length_2} is the usual strain-to-length relation used in GW detection based on Michelson interferometers, which feature two orthogonal optical cavities filled with laser light.
This motivates the choice for extremely long interferometer arms: generally, the longer the arms, the larger the differential length change $\Delta L$ will be.
This holds as long as the long-wavelength approximation $\lambda_{GW} \gg L$ is true: if not, Eq.~\ref{eq:gw_differential_length_2} breaks down because the GW oscillates spacetime faster than the light can complete a round-trip \cite{Rakhmanov2008}.

\section{Advanced LIGO detectors}
\label{sec:detectors}

Each Advanced LIGO detector is a long-baseline laser interferometer with two 4~km long orthogonal arms.
The interferometer acts as a transducer, transforming the GW signal into observable laser power fluctuations at the antisymmetric port.

The interferometer is supported by several auxiliary subsystems required to detect gravitational waves.
Auxiliary subsystems include the core optics length controls \cite{Drever1983, Regehr1995, Sigg1998, Fritschel2001, Strain2003, Fricke2012, Izumi2017},
angular controls \cite{Anderson1984, Morrison1994, Mavalvala1998, Sidles2006, Hirose2010, Barsotti2010, Dooley2013, Enomoto2016},
high-powered stabilized laser \cite{Kwee2012, Seifert2006, Kwee2009},
vacuum system \cite{Zucker1996, Dolesi2011, Phelps2013},
optics suspensions \cite{Robertson2002, Aston2012, Carbone2012},
seismic isolation \cite{Daw2004, HEPI2014, Matichard2015, Biscans2018},
and electronics and data acquisition systems \cite{Bork2001, Bartos2010, Rollins2016, Bork2021}.
This review will focus on the optical configuration and operation of the interferometers.

The core of the Advanced LIGO detectors are dual-recycled, Fabry-P\'erot, Michelson interferometers \cite{AdvLIGOPaper, AdvLIGOFinalDesign},
enhanced with an input and an output mode cleaner \cite{Mueller2016, Arai2013}, and
filled with pre-stabilized laser light \cite{Kwee2012}.
The entire LIGO optomechanical control system is based on the Pound-Drever-Hall frequency stabilization technique \cite{Drever1983}.

\begin{figure}[H]
    \includegraphics[width=10.5 cm]{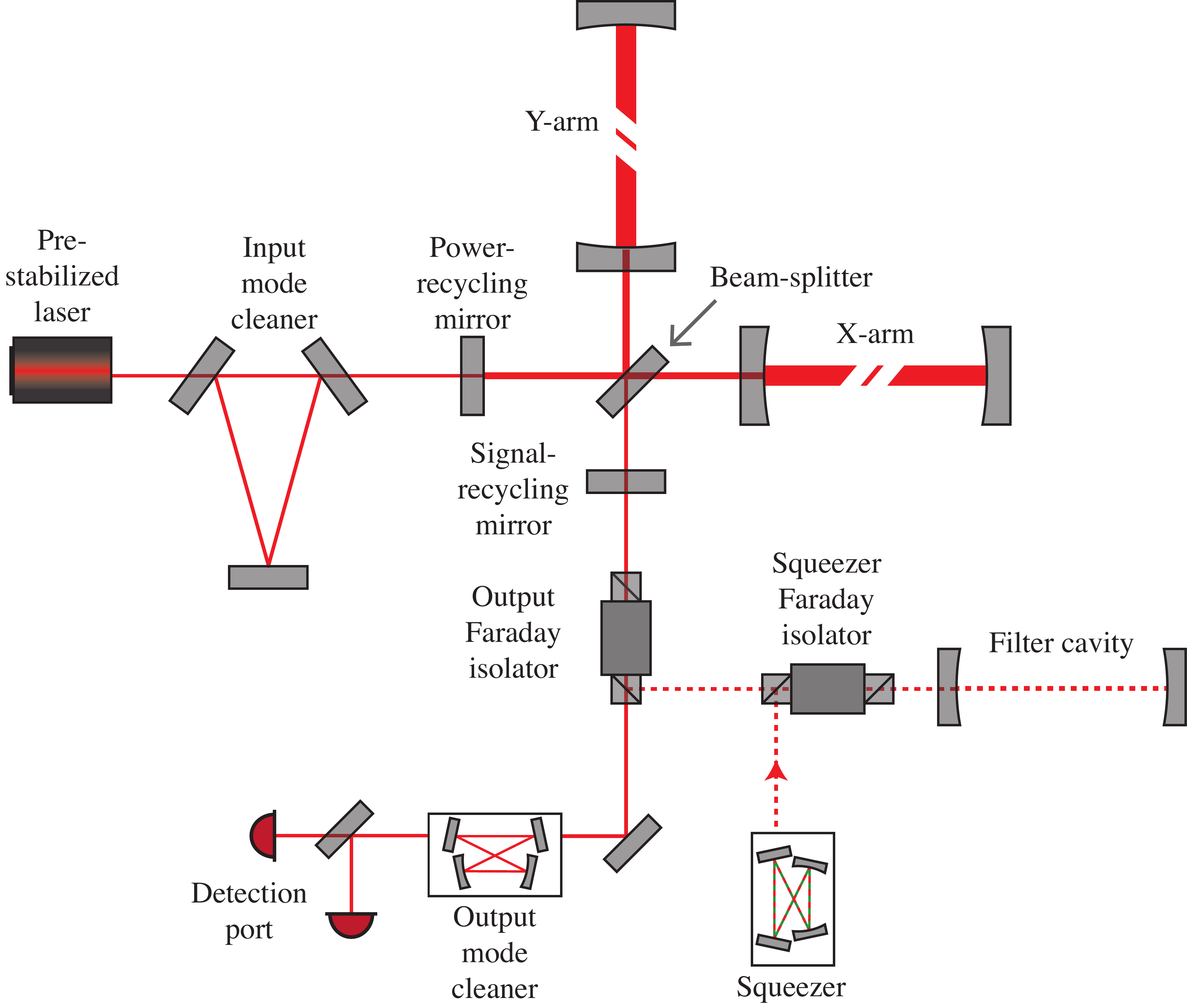}
    \caption{
        Simplified optical layout of the LIGO detectors for the fourth observing run.
        The pre-stabilized laser first traverses the input mode cleaner to further stabilize the laser light before entering the main interferometer.
        The main interferometer consists of the seven core optics in the upper right, including the Fabry-P\'erot arm cavities, 50:50 beamsplitter, and recycling mirrors.
        The GW signal from the main interferometer (solid red) transmits through the signal-recycling mirror, output Faraday isolator, and output mode cleaner to the detection photodetectors at the detection (or antisymmetric) port.
        The dashed red represents the squeezed light input path from the squeezer cavity, reflected off the filter cavity, and input on the back of the signal-recycling mirror.
        The filter cavity is discussed in Section~\ref{subsec:filter_cavity}.
    }
    \label{fig:ligo_o4_layout}
\end{figure}

Figure~\ref{fig:ligo_o4_layout} shows a simplified optical configuration planned for O4.
The optical configuration is the same as O3, except for the addition of the 300~m filter cavity \cite{Evans2013, McCuller2020, filtercavitydesign}.

In this section we will overview the Advanced LIGO optical configuration, as well as briefly overview the lock acquisition process.
Appendix~\ref{sec:michelson_interferometer} overviews the optical components that make up the Advanced LIGO design.

\subsection{Main interferometer}
\label{subsec:main_interferometer}

The main interferometer consists of seven core optics shown in Figure~\ref{fig:ligo_o4_layout}:
the power-recycling mirror,
the signal-recycling mirror,
the beamsplitter,
and the four arm cavity optics known as input test masses (ITMs) and end test masses (ETMs).

The main interferometer relies on \textit{constructive interference} to build up high levels of laser power inside the 4~km Fabry-P\'erot arm cavities (see Appendix~\ref{sec:fabry_perot_interferometer}).
With more laser power built up in the interferometer, more light is modulated by a passing GW, creating a stronger detected response to GWs at the detection port.

In the Michelson interferometer formed by the two arms and the beamsplitter, \textit{destructive interference} occurs at the antisymmetric (detection) port,
where the the beams from the two arms are recombined out-of-phase, so no light appears.
In contrast, at the input port constructive interference occurs, so all the light input is reflected back toward the laser in the Michelson interferometer (see Appendix~\ref{sec:michelson_interferometer}).

Differential phase changes in the light in each arm, such as those caused by gravitational waves, will cause light to exit out the Michelson detection port.
Common phase changes, on the other hand, will have no effect on the light levels at either the input or detection port.
Thus, the Michelson detection port is said to have high \textit{common mode rejection}, as both frequency and intensity fluctuations in the input laser light are largely rejected from the detection port.

\subsubsection{Basic signal}
\label{subsubsec:basic_signal}

The gravitational wave \textit{signal-to-noise ratio} (SNR) at the antisymmetric port is formed by the laser power signal due to GWs $P_\mathrm{as}$,
as well as the laser power noise $\sqrt{ S_P }$, i.e. fluctuations not due to GWs.
The gravitational wave $\mathrm{SNR}$ for the shot noise dominated regime can be approximately written
\begin{align}
    \label{eq:gw_signal_to_noise_approx}
    \mathrm{SNR} \approx \dfrac{P_\mathrm{as}}{\sqrt{ S_P }} \sqrt{t_\mathrm{sig}} \propto \dfrac{ L \sqrt{ P_\mathrm{arm} t_\mathrm{sig}} }{\lambda} h,
\end{align}
where $L$ is the length of the Fabry-P\'erot arm cavities,
$P_\mathrm{arms}$ is the full power buildup in the arms,
$t_\mathrm{sig}$ is the duration the GW signal in the detector bandwidth,
$\lambda$ is the laser wavelength, and
$h$ is the GW strain amplitude.
The full detector response is derived in \cite{Buonanno2001, KLMTV}, and expanded upon in \cite{Izumi2017, WardThesis, HallThesis, CahillaneThesis}.
A more complete understanding of detector signal and noise processing can be found in \cite{NoiseGuide2020, Allen2012}.

Several major considerations in detector design are captured in Eq.~\ref{eq:gw_signal_to_noise_approx}.
First, the simplest way to amplify the signal is to extend the arm length $L$.
The main limit on making detectors longer is cost of the facility, particularly the evacuated beamtube, which currently limits the LIGO detectors to the 4~kilometer scale.
Maximizing the arm power $P_\mathrm{arm}$ increases the detectable laser signal created by GWs, and is limited by input power and losses in the interferometer from absorption and scatter.
Reducing the detector wavelength $\lambda$ would na\"ively increase sensitivity to GWs,
but would require all major detector infrastructure such as the source laser, optical coatings, substrates, and photodetectors to perform at or better than the current noise levels.

\subsubsection{Dual-recycling}
\label{subsubsec:dual_recycling}

\textit{Dual-recycling} refers to the two recycling cavities formed by the mirrors at the input and output of the main interferometer \cite{Meers1988, Strain1991, Heinzel1998, Grote2004}.
The mirror at the input is the \textit{power-recycling mirror}, and is used to reflect light back into the main interferometer,
enabling greater levels of light circulating inside the interferometer \cite{Fritschel1992, KLMTV, AndoThesis}.
The mirror at the output is the \textit{signal-recycling mirror}, and is used to broaden the detector bandwidth \cite{Buonanno2001}.

The Advanced LIGO recycling cavities are designed to be geometrically stable to better control spatial mode of the beam entering and exiting the Michelson \cite{Arain2008},
although point absorbers on the mirrors are suspected of polluting the main spatial mode (see Section~\ref{subsubsec:point_absorbers}).
Control schemes for the interferometer degrees of freedom associated with the recycling cavities have been designed and implemented for length \cite{Regehr1995, Sigg1998, Fritschel2001, Strain2003, Izumi2017}
and angular controls \cite{Fritschel1998, Mavalvala1998, Barsotti2010, Dooley2013}.

\subsubsection{Squeezer}
\label{subsubsec:squeezer}

Heisenberg uncertainty in the form of shot noise and radiation pressure noise (Section~\ref{subsec:quantum_noise}) limits the sensitivity of the interferometer \cite{Caves1980, Caves1981}.
The \textit{squeezer} is a squeezed vacuum source, and refers to the optics producing entangled photons for injection into the antisymmetric port of the interferometer \cite{Tse2019, Oelker2016a, Yu2020, McCuller2021}.
The ensemble of entangled photons produce a quantum \textit{squeezed vacuum} electromagnetic field.
By squeezing the quantum vacuum, quantum shot noise can be lowered across the bandwidth of the detector.
This is known as \textit{frequency-independent} squeezing.

The filter cavity shown in Figure~\ref{fig:ligo_o4_layout} will enable \textit{frequency-dependent} squeezed light injection.
The results of squeezing in O3 are explained in Section~\ref{subsec:quantum_squeezing}.
The filter cavity is explained further in Section~\ref{subsec:filter_cavity}.

\subsubsection{Detector bandwidth and linewidth}
\label{subsubsec:detector_bandwidth_and_linewidth}

Most often, in LIGO the \textit{detector bandwidth} refers to the frequency at which the differential arm (DARM) frequency response begins falling off.
This value is also known as the \textit{DARM coupled-cavity pole} or simply the \textit{DARM pole}.
This frequency is defined primarily by the DARM coupled cavity, which is formed by the arm cavities and the signal-recycling cavity \cite{Izumi2017}.
As mentioned in Section~\ref{subsubsec:dual_recycling}, the signal recycling mirror is locked exactly off-resonance to broaden the detector bandwidth,
in a scheme known as \textit{resonant-sideband extraction} \cite{Buonanno2001, WardThesis}.
During mid-2021 locking, the detector bandwidth at LIGO Hanford was about $450~\mathrm{Hz}$.

Similarly, the \textit{detector linewidth} refers to the full-width half-maximum of the laser frequency noise when the detector is locked.
With a long-baseline, high finesse interferometer like Advanced LIGO, this is identical to twice the frequency at which the common arm (CARM) frequency response begins falling off.
This is known as the \textit{CARM coupled-cavity pole} or the \textit{CARM pole}.
The CARM coupled cavity is formed by the arm cavities and the power-recycling cavity, which in this case is locked on-resonance to enhance the resonating power \cite{Izumi2017}.
This, paired with the $4~\mathrm{km}$ long baseline, makes the linewidth very small, and the laser ultra-stable in the detector bandwidth \cite{Cahillane2021}.
The detector linewidth is estimated to be about $1~\mathrm{Hz}$.

\subsubsection{Calibration}
\label{subsubsec:calibration}

\textit{Calibration} is the process of converting the detector output $P_\mathrm{as}$ into gravitational wave units of strain $h$ \cite{GW150914CalPaper, CalUncPaper, Sun2020, Sun2021, Lindblom2009, CALTimeDependence, Vitale2021}.
The calibration reference is the \textit{photon calibrator}, which uses an auxiliary laser to apply a known force on the optics via radiation pressure \cite{aLIGOPCALPaper, Bhattacharjee2020}.
The O3 calibration response upper limit on systematic error and associated uncertainty is $\sim 11\%$ in magnitude and $\sim 9^\circ$ in phase ($68\%$ confidence interval) in the sensitive frequency band 20-2000 Hz \cite{Sun2020, Sun2021}.
The systematic error alone is estimated at levels of $< 2\%$ in magnitude and $< 4^\circ$ in phase \cite{Sun2021}.

Newtonian calibrators, which employ rapidly spinning masses near the optics, are also under development \cite{Estevez2018, Inoue2018, Schreiner2021}.
During O3, a Newtonian calibrator with a quadrupole and hexapole was installed at Hanford, and successfully induced motion on the X-end test mass (ETMX) \cite{Ross2021}.
Due to problems with precision installation and distance uncertainty analysis, the Newtonian calibrator will not be pursued by LIGO as a precision calibration instrument in O4.

\subsection{Input mode cleaner}
\label{subsec:input_mode_cleaner}

The input mode cleaner is a three-mirror, 33~meter  round trip triangular cavity used to further stabilize the frequency, intensity, and spatial mode content of the input laser before it enters the main interferometer \cite{Mueller2016}.
The RMS laser frequency noise is limited by the linewidth of the interferometer, which is extremely low (1.2~Hz).
The laser frequency is locked to the input mode cleaner length, providing high-gain high-bandwidth feedback ($\sim100$~kHz) to massively suppress frequency noise intrinsic to the NPRO laser \cite{Fritschel2001, Cahillane2021, Kane1985}.
A small sample of the transmission through the input mode cleaner is used to stabilize the intensity of the laser input into the main interferometer.


\subsection{Output mode cleaner}
\label{subsec:output_mode_cleaner}

The output mode cleaner is a four-mirror, 1~meter round trip bowtie cavity used to transmit only the main interferometer GW signal \cite{Arai2013, KorthThesis, HoakThesis, VenugopalanThesis}.

The GW readout scheme is known as \textit{DC readout} \cite{Hild2009, Fricke2012}.
A picometer-scale offset in the differential arm length is deliberately introduced and is controlled to let 20~mW of light leak out to the detection port.
This light used as a local oscillator, beating against the GW signal light, rendering it detectable on a photodetector.

The radio-frequency sidebands used for controlling interferometer degrees of freedom, and higher-order modes from the main interferometer, are both reflected away from the detection port by the output mode cleaner.
Backscatter, i.e. reflection from the output mode cleaner along the main beam path, is rejected by the output Faraday isolator.

\subsection{Lock acquisition}
\label{subsec:lock_acquisition}


The lock acquisition process is a sequence of steps taken to bring the interferometer from a free-swinging uncontrolled state to an observation-ready state \cite{AdvLIGOPaper, Strain2003, MartynovThesis, StaleyThesis}.
The optical cavities shown in Figure~\ref{fig:ligo_o4_layout} must be held on resonance (locked) and in the correct alignment. 
This section will review the lock acquisition process used during O3, which was also described in \cite{Buikema2020}.

Each cavity is locked using the Pound-Drever-Hall (PDH) technique \cite{Drever1983}.
Four sets of radio-frequency (RF) phase-modulated sidebands are added to the input laser using an electro-optic modulator (9~MHz, 24~MHz, 45~MHz, 118~MHz).
The RF sideband frequencies are chosen to be resonant in some cavities and anti-resonant in others.
The RF beat notes are detected on reflection of the interferometer, at the antisymmetric (detection) port, or through a pick-off on transmission of the power-recycling cavity.
RF photodetectors at each port are then used to sense the length and angular degrees of freedom.

The lock acquisition process is coded using the Guardian finite state machine \cite{Rollins2016}.
During O3 the lock acquisition sequence took roughly 25 minutes, but depends strongly on environmental factors including seismic activity and wind speed \cite{Buikema2020}.
The lock acquisition sequence is always undergoing improvements to speed and versatility.

\subsubsection{Pre-stabilized laser and input optics}
\label{subsubsec:psl_and_input_optics}

The first step of the lock process is to ensure a laser stabilized in frequency, intensity, and spatial mode is entering the main interferometer.
Inside a clean room, several important optical components reside on an optical table, making up a full system known as the \textit{pre-stabilized laser}, or PSL \cite{Kwee2012}.
Included in the pre-stabilized laser is
a 2-Watt NPRO 1064~nm laser source,
a high-powered amplifier to increase the input laser power,
a pre-mode cleaner to clean the the laser beam spatial mode,
a reference cavity to stabilize the laser frequency,
and two photodetectors on a pickoff to stabilize the laser intensity.

Next, the pre-stabilized laser beam is input onto the first in-vacuum, suspended cavity, the input mode cleaner \cite{Mueller2016}.
The beam is further cleaned and stabilized by the input mode cleaner (Section~\ref{subsec:input_mode_cleaner}),
and traverses the \textit{input Faraday isolator} which prevents the formation parasitic interferometer with the main interferometer and provides access to the interferometer reflected beam.
Finally, the beam is incident on the first mirror of the main interferometer, the power-recycling mirror.

\subsubsection{Arm length stabilization}
\label{subsubsec:arm_length_stabilization}

\begin{figure}[H]
    \includegraphics[width=10.5 cm]{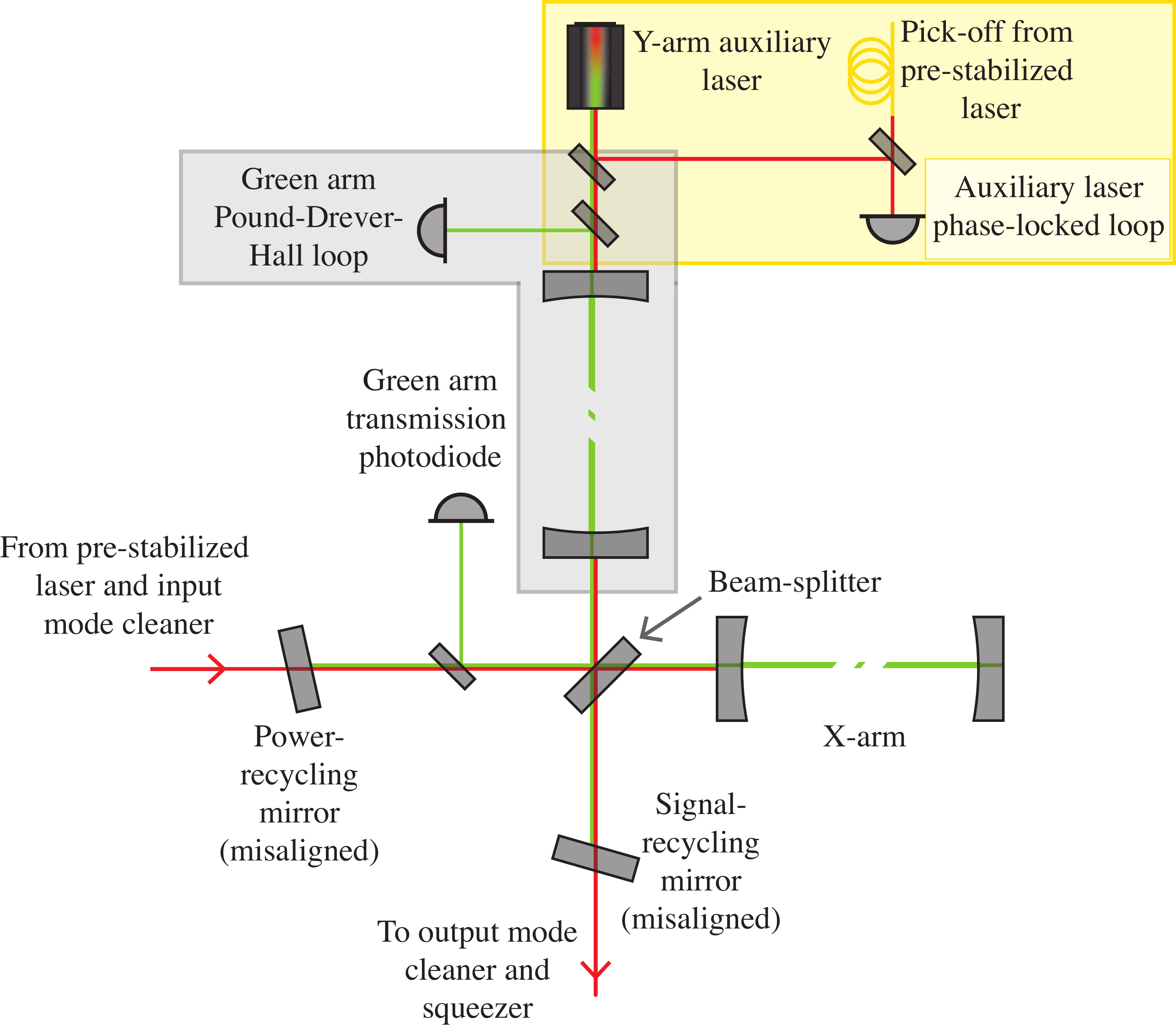}
    \caption{
        Interferometer layout during the first stage of lock acquisition: arm length stabilization.
        The arm length stabilization uses auxiliary lasers in the end stations (only the Y-arm is shown for simplicity), which emit beams at 1064~nm and 532~nm.
        The lasers are phase locked to the pre-stabilized laser which is delivered to the end station by optical fiber (yellow box).
        The green laser is then locked to the arm cavity through a Pound-Drever-Hall loop on reflection (gray box).
    }
    \label{fig:lock_acquisition_als}
\end{figure}

Next, the arms are brought under control using green light, known as the \textit{arm length stabilization} system \cite{Staley2014, Izumi2012, Mullavey2012}.
Green light is used so that the arm lengths can be independently controlled while infrared is used to lock the corner.

The ALS system consists of two auxiliary green laser sources at each end station.
The end station lasers are phase-locked to the main laser, then frequency-doubled to generate 532~nm (green) light which is injected into the arm cavities.
Each ALS laser is then locked to their respective arm lengths.

\subsubsection{Dual-recycled Michelson locking}
\label{subsubsec:dual_recycled_michelson_locking}

\begin{figure}[H]
    \includegraphics[width=10.5 cm]{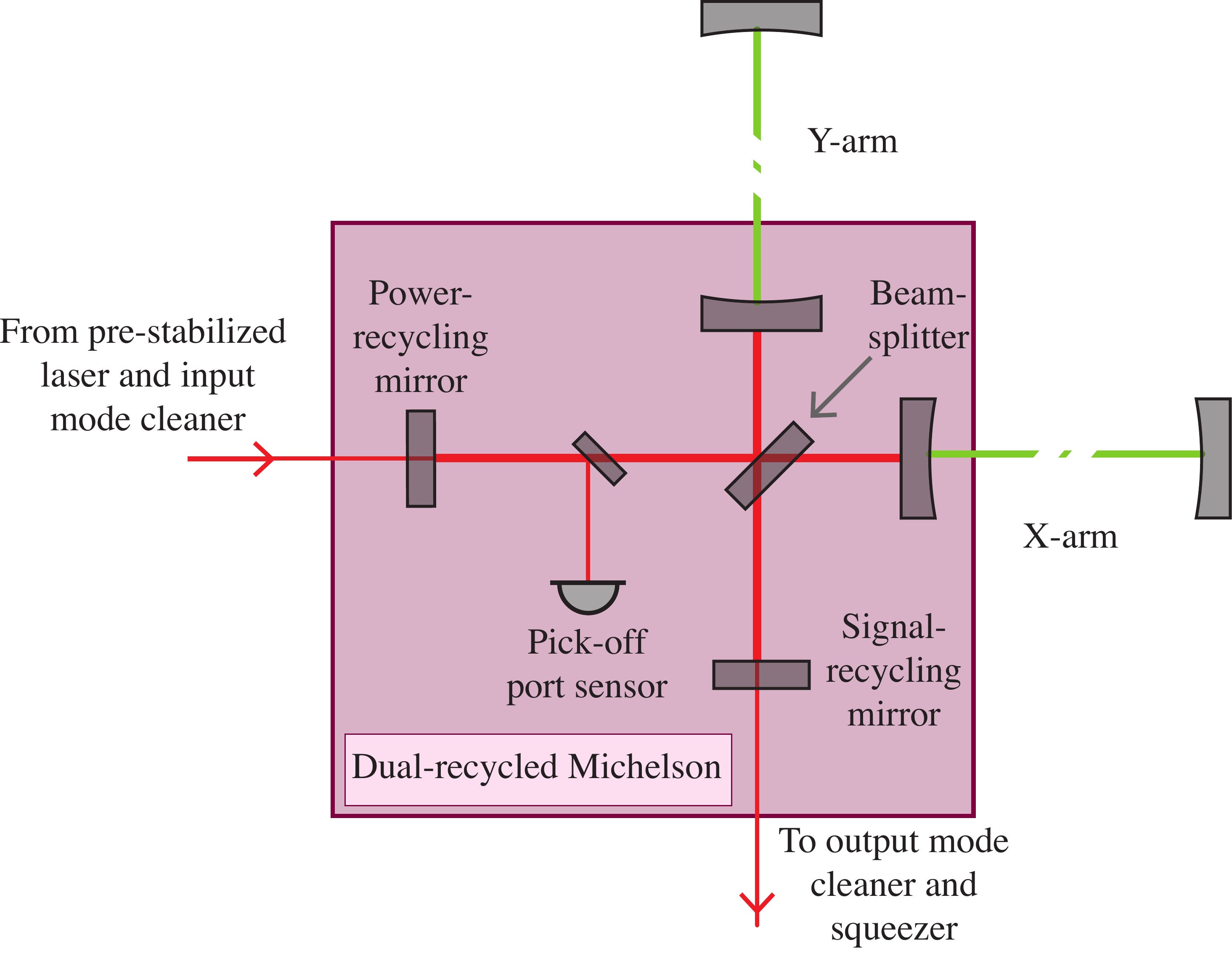}
    \caption{
        Interferometer layout during the second stage of lock acquisition: dual-recycled Michelson locking.
        With the arms locked on green, the arms are held off-resonance for infrared so the corner degrees of freedom can be locked independently of the arms.
        The dual-recycled Michelson is shown in the pink box.
        DRMI has three length degrees of freedom: the Michelson length (MICH), signal-recycling cavity length (SRCL), and power-recycling cavity length (PRCL).
        The photodiode at the pick-off port is used to sense all degrees of freedom, they are decoupled via the different phase modulation sideband frequencies which resonate in different cavities.
    }
    \label{fig:lock_acquisition_drmi}
\end{figure}

The next step in the lock acquisition is to lock the dual-recycled Michelson interferometer (DRMI) with infrared light.
The dual-recycled Michelson interferometer is formed by the five optics in the corner: the power- and signal-recycling mirror, beamsplitter, and input test masses to the arm cavities.
Crucially, the end test masses are \textit{not} included.

During the green locking of the arms, the corner optics are purposely misaligned.
Then, the arms are purposely held off-resonance for infrared using the information from the green lock.
Finally, the three corner degrees of freedom (the Michelson length, power recycling length, and signal recycling length) are brought under control simultaneously using the \textit{3f PDH locking} technique \cite{Arai3f, Staley2014}.
3f locking is used because, as the interferometer arms are brought into resonance for infrared, the usual 1f PDH locking signals flip sign, which would cause the interferometer corner to lose control.

\subsubsection{Full interferometer locking}
\label{subsubsec:full_interferometer}

Having locked the dual-recycled Michelson with infrared light, the arms are brought onto resonance for infrared light.
As the arms move into resonance, $\sim 10~\mathrm{kW}$ of infrared laser power begins resonating in each interferometer arm.

In O3, the noise in the arm length stabilization system is around $\sim 2~\mathrm{Hz}$ \cite{alog43119, alog43214}.
However, the linewidth of the interferometer is $\sim 1~\mathrm{Hz}$.
This renders it impossible to directly transition from ALS to full interferometer PDH locking \cite{MartynovThesis, Staley2014}.

Therefore, the infrared transmission through the arms is used as an error signal to sense the common arm length as it is brought to full power.
Once full power is very nearly reached, the common and differential arm length error signal is transitioned to the PDH error signal.

Once the full interferometer is fully locked on infrared, the corner degrees of freedom are switched from 3f to 1f PDH error signals,
differential arm control is switched to DC readout \cite{Fricke2012},
and all the angular controls are turned and allowed to converged on the best alignment in preparation for high power.

\subsubsection{High power, low noise lock}
\label{subsubsec:high_power_low_noise_lock}

Until this point the laser input power is kept at 2~watts.
Once full lock is achieved, the input power is increased to the highest achievable power.

At this point, the suspension actuators are brought from acquisition mode -- with high range and high noise -- to low-noise mode,
and the control loop bandwidths for the length and angular controls are cut-off to achieve the lowest noise state.

Finally, squeezed light is injected to further reduce quantum shot noise; the squeezer subsystem is further discussed in Sections \ref{subsubsec:squeezer} and \ref{subsec:quantum_squeezing}.
The interferometer is now ready to observe gravitational waves.

During full lock, the circulating power in the interferometer heats the optics until the heat absorbed and emitted reaches a steady-state.
This heating process is known as \textit{thermalization}, and takes about thirty minutes after full lock to reach a steady-state.
Thermalization affects many aspects of the interferometer, most notably the radius of curvature of the main optics, which affects the scattering of laser light out of the laser's fundamental spatial mode \cite{Vajente2014}.
The optical gain of the interferometer is also affected, although this is tracked via calibration lines \cite{CALTimeDependence}.

The \textit{thermal compensation system} is a subsystem dedicated to monitoring and controlling negative changes in the interferometer at full power \cite{Brooks2016}.
The thermal compensation system is comprised of ring heater actuators to adjust the test mass optics radii of curvature,
spatially tunable $CO_2$ laser projectors to heat optical surfaces,
and Hartmann wavefront sensors to monitor optic surface changes.
This system is primarily used to repair spatial distortions in the main beam, enabling higher power buildup inside the interferometer.
It is also used to minimize measured noise couplings to the gravitational wave data channel.

\section{Detector sensitivity to gravitational waves}
\label{sec:sensitivity}

Detector sensitivity refers to the gravitational wave \textit{signal-to-noise ratio}.
The GW signal is imprinted on the laser light resonating in the detector, and sensed by photodetectors at the detection port.

\textit{Noise} is any laser power fluctuations sensed at the detection port that is \textit{not} due to gravitational waves.
The are two main types of noise: \textit{fundamental} and \textit{technical}.

Fundamental noise is intrinsic to the design of the detector.
Fundamental noises include quantum uncertainty, thermal noise in the optics, seismic noise, and Newtonian noise.
These often cannot be improved without major upgrades to the detector, like increasing the arm length or replacing the optical coatings.

Technical noise is not intrinsic to the detector design, but can limit the performance of the detectors.
Technical noises are wide in variety, and most detector work is dedicated to eliminating it.
Important examples of technical noise include
\begin{itemize}
    \item length and angular controls noise
    \item laser frequency and intensity noise
    \item scattered light
    \item residual gas noise
    \item photodetector dark noise
    \item electromagnetic noise
\end{itemize}
All of these noise sources and more are considered carefully in \cite{Buikema2020},
but controls noise is the main technical limit to gravitational wave detectors at low frequencies.

\begin{figure}[H]
    \includegraphics[width=12 cm]{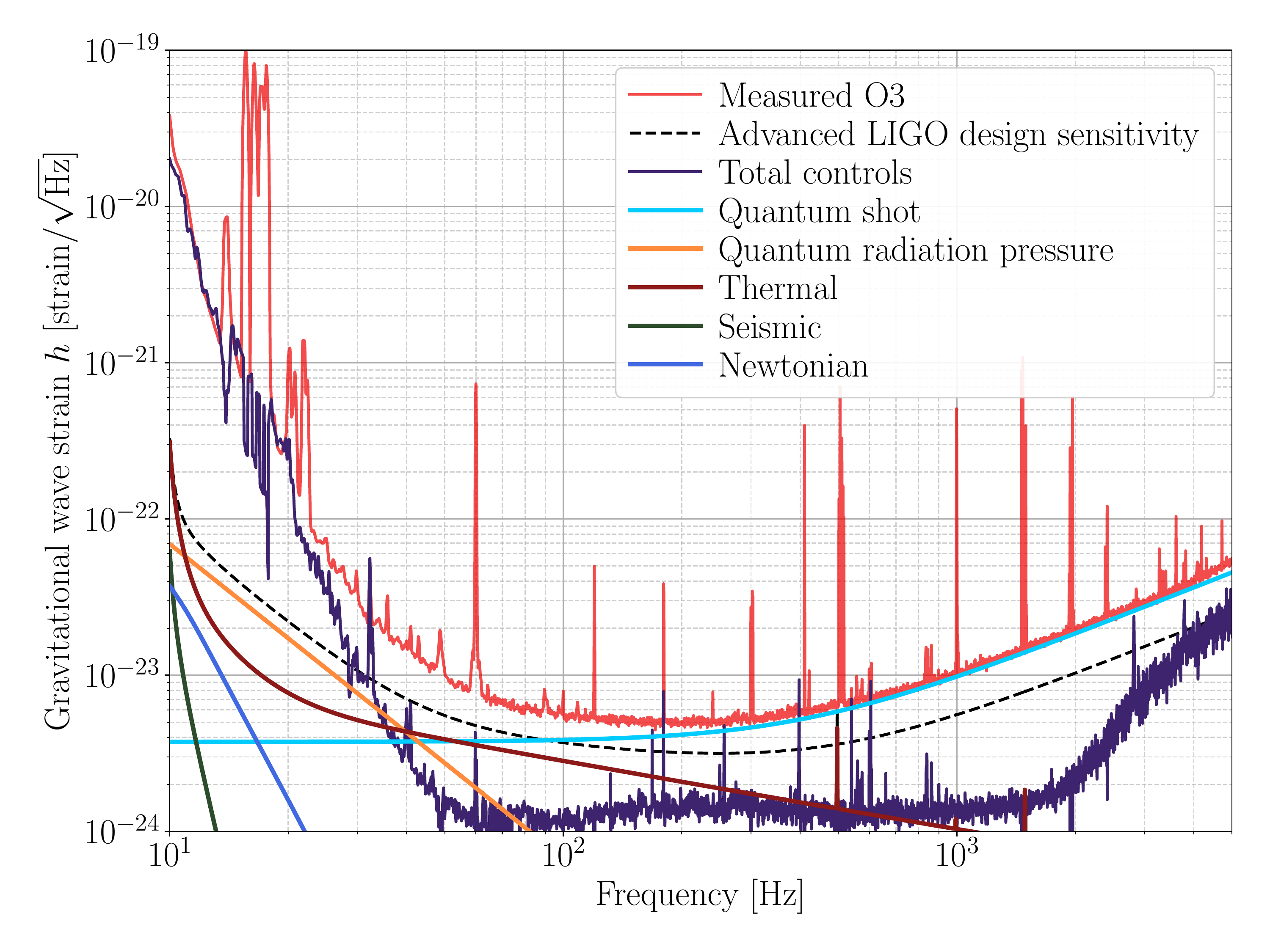}
    \caption{
        O3 LIGO Hanford gravitational wave noise budget, simplified \cite{Buikema2020}.
        The measured curve in red represents the calibrated sensitivity to GWs.
        The dashed black curve is the Advanced LIGO design sensitivity curve, with 125~watts of input power and no squeezing \cite{AdvLIGOPaper, aligoDesignSensitivity2018}.
        All other curves are measured or estimated contributions to the total measured noise.
        Here we have only included significant known noise sources for simplicity.
    }
    \label{fig:noise_budget}
\end{figure}

A \textit{noise budget} is a way of quantifying the contributions to the measured noise curve \cite{Weiss1972}.
Figure~\ref{fig:noise_budget} shows a simplified noise budget of the O3 LIGO Hanford interferometer.
To produce the measured noise curve in Figure~\ref{fig:noise_budget},
the time series of the detector output is Fourier-transformed to represent the frequency content of the noise as an amplitude spectral density (ASD),
then calibrated into units of gravitational wave strain $h$.


\subsection{Design sensitivity}
\label{subsec:design_sensitivity}

\textit{Design sensitivity} is the ultimate sensitivity Advanced LIGO is expected to achieve given the detector configuration and estimated performance.
It is formed by the sum of all fundamental noises.
Design sensitivity has not yet been achieved by the current detectors, due to technical noises and insufficient power buildup.

Figure~\ref{fig:noise_budget} shows the Advanced LIGO design sensitivity as the black dashed curve \cite{AdvLIGOPaper, aligoDesignSensitivity2018}.
No squeezing is included in the design sensitivity curve.
For design sensitivity, 125~watts of laser input power is assumed, resulting in $750~\mathrm{kW}$ of circulating power in the arm cavities.
The O3 input power was much lower, $\sim 34~\mathrm{watts}$ at Hanford and $\sim 38~\mathrm{watts}$ at Livingston.
This affects the quantum noise, lowering the design's strain-referenced quantum shot noise but increasing the quantum radiation pressure noise (Section~\ref{subsec:quantum_noise}).

\subsection{Quantum noise}
\label{subsec:quantum_noise}

Fluctuations of the vacuum electric field at the interferometer readout port impose the main fundamental limit to the interferometer sensitivity \cite{Caves1980, Caves1981, braginsky_khalili_thorne_1992, Buonanno2001}.
Quantum noise appears as \textit{shot noise} and \textit{quantum radiation pressure noise}.

\subsubsection{Shot noise}
\label{subsubsec:shot_noise}

In general, shot noise arises from Poisson fluctuations in the arrival time of discrete objects.
In the case of interferometers, the discrete objects are photons at the detection port.

Heisenberg uncertainty in the measured laser amplitude at the detection port cannot be distinguished from actual mirror motion due to GWs.
The power detected on the photodetector is made up of a finite number of photons which arrive randomly and independently of one another,
leading to a detected white noise amplitude spectral density $\sqrt{ S_\mathrm{shot} }$ proportional to the total power $P_{dc}$ on the photodetector:
\begin{align}
    \label{eq:quantum_shot_noise}
    \sqrt{ S_\mathrm{shot} } = \sqrt{ 2 \hbar \omega_0 P_{dc} }.
\end{align}
where $\omega_0$ is the fundamental angular frequency of the laser.

Eq.~\ref{eq:quantum_shot_noise} is white noise in units of $\mathrm{watts}/\sqrt{\mathrm{Hz}}$, meaning it has no frequency dependence.
However, when this noise is referenced against the GW detector response to yield units $\mathrm{strain}/\sqrt{\mathrm{Hz}}$, it begins to rise above the detector bandwidth of around 450~Hz.
In Figure~\ref{fig:noise_budget}, the cyan curve shows the estimated GW-referenced quantum shot noise.

\subsubsection{Radiation pressure noise}
\label{subsubsec:radiation_pressure_noise}

Quantum radiation pressure noise (QRPN) is displacement noise arising from amplitude fluctuations of the electric field in the arms.
These amplitude fluctuations, again due to Heisenberg uncertainty, mean the arm power is spontaneously increasing and decreasing.
This induces forces on the optics via radiation pressure, moving the optics in the arm \cite{Caves1980}.

For a Michelson interferometer with Fabry-P\'erot arms, the displacement amplitude spectral density due to QRPN $\sqrt{ S_{\mathrm{QRPN}} }$ can be described \cite{KLMTV}
\begin{align}
    \label{eq:quantum_radiation_pressure_noise}
    \sqrt{ S_{\mathrm{QRPN}} }(f) & = \dfrac{1}{m L (2 \pi f)^2} \sqrt{ \dfrac{32 P_\mathrm{bs} \hbar \omega_0}{\omega_c^2 + (2 \pi f)^2} }
\end{align}
where $P_\mathrm{bs}$ is power on the beamsplitter,
$m$ is the mirror mass,
$L$ is the arm length,
$\omega_0$ is the laser frequency,
$\omega_c$ is the arm pole describing the number of reflections inside the Fabry-Perot cavity, and
$f$ is the GW signal frequency.
Eq.~\ref{eq:quantum_radiation_pressure_noise} is plotted in Figure~\ref{fig:noise_budget} as the orange curve.

\subsection{Thermal noise}
\label{subsec:thermal_noise}

Thermal noise refers to the actual displacement in the mirrors induced by thermal fluctuations in the atoms making up
the test mass suspension, substrate, and optical coatings \cite{Braginsky2003, Levin1998, Hong2013, Yam2015}.
In general, thermal noise increases with mechanical loss in the materials making up the optics,
as described by the fluctuation-dissipation theorem~\cite{Callen1952, Kubo1966, Saulson1990}.

For LIGO test masses, the fluctuating observable is the optic displacement due to dissipation from thermal excitations.
The dominant source of mechanical loss, and thus thermal noise, is the optical coatings deposited on the optical surface.
For a single coating with thickness $d$, the dissipated power and coating displacement noise $\sqrt{ S_x }$ due to thermal fluctuations can be calculated \cite{Nakagawa2002, Chalermsongsak2014}:
\begin{align}
    \label{eq:single_coating_thermal_noise}
    \sqrt{ S_x }(f) = \sqrt{ \dfrac{8 k_B T (1 + \sigma)(1 - 2 \sigma) d}{\pi w^2 E} \dfrac{\phi}{2 \pi f} }
\end{align}
where $T$ is temperature,
$\sigma$ is the coating Poisson ratio,
$d$ is the coating thickness,
$E$ is the Young's modulus,
$w$ is the beam radius, and
$\phi$ is the mechanical loss angle of the coating.

The actual LIGO coatings have more than a single layer, and their thermal noise properties are measured directly in the lab \cite{Gras2017, Gras2018}.
Other thermal noise contributions include thermal noise vibrations in the optic suspensions \cite{Saulson1990, Cagnoli2000}.
The total thermal noise estimate, largely from coatings at high frequency and suspensions at low frequency, are plotted as the maroon line in Figure~\ref{fig:noise_budget}.

\subsection{Seismic noise}
\label{subsec:seismic_noise}

Seismic noise is optical displacement due to the motion of the Earth physically shaking the mirrors resting on the Earth's surface.
Unmitigated, the vibrations of the Earth are much larger than LIGO optics can tolerate.
Enormous effort is put into isolating the core optics from the ground vibrations, particularly in the GW sensitive range.
Additionally, earthquakes and windy conditions can make holding the detector lock impossible.

The main LIGO optics are suspended from a quadruple-stage pendulum chain to passively isolate from ground motion \cite{Aston2012}.
These pendulums are suspended from active seismic isolation platforms \cite{Matichard2015}
which themselves are supported by hydraulically actuated pre-isolation structures \cite{HEPI2014}.

Ultimately, the linear coupling due to seismic motion is largely suppressed in the GW detection band, as seen by the green trace in Figure~\ref{fig:noise_budget}.
This curve comes from the measured differential displacement of the seismic isolation platforms, multiplied by the isolation of the suspensions which hang from those platforms.

However, at very low frequencies $(<1~\mathrm{Hz})$, below the pendulum resonance frequencies, seismic motion still dominates.
Worse, high motion at very low frequencies can couple to higher frequencies via bilinear or nonlinear coupling mechanisms.
Work continues to suppress seismic motion further, via monitoring systems and more advanced control schemes \cite{Biscans2018}.

\subsection{Newtonian noise}
\label{subsec:newtonian_noise}

Newtonian noise, or gravity-gradient noise, is from fluctuations in the ground creating changes in the local gravitational potential around the optics, moving them \cite{Saulson1984, Hughes1998, Harms2015}.
Newtonian noise is related to seismic noise, but the coupling mechanism is not from ground motion propagating down a pendulum chain, but changes in ground density due to seismic activity.
Therefore, Newtonian noise cannot be isolated away with longer pendulums, but can be monitored and actively subtracted \cite{Driggers2012newtonian, Coughlin2016, Coughlin2018}.

Upper limits have been placed on Newtonian noise contributions to LIGO, but have never been directly observed \cite{PhysRevD.101.102002}.
The blue trace in Figure~\ref{fig:noise_budget} represents an estimate of Newtonian noise coupling to the GW spectrum given local seismic activity, but is highly uncertain.

\subsection{Controls noise}
\label{subsec:controls_noise}

Controls noise is the noise associated with the sensor and feedback system required to hold the interferometer optics on resonance.
This includes both length control loops, which manage the optic's position \cite{Regehr1995, Sigg1998, Fritschel2001, Strain2003, Hild2009, Fricke2012, Izumi2017, Cahillane2021},
and angular control loops, which point the optics at each other \cite{Anderson1984, Morrison1994, Mavalvala1998, Fritschel1998, Sidles2006, Hirose2010, Barsotti2010, Dooley2013, Enomoto2016}.

Controls noise is the result of multiple effects.
The control loops are required to suppress real motion in the optics, known as \textit{displacement noise}.
The controllers must be strong enough to hold the optics in place.
To hold the optics in place, electromagnetic coil actuators or electrostatic drives are employed \cite{Aston2012, Carbone2012}.

However, the controllers must know where to hold the optics.
The Pound-Drever-Hall error signals hold the information about where each optic must be held \cite{Drever1983, AdvLIGOFinalDesign, AdvLIGOPaper}.
The PDH error signals are detected with radio-frequency photodetectors (RFPDs).
These RFPD sensors are limited by sources of \textit{sensor noise}, largely shot noise, but also potentially ``dark'' noise and analog-to-digital conversion noise.
Sensor noise is indistinguishable from actual displacement noise, and dominates most control loop noise floors above $\sim 50~\mathrm{Hz}$.

\subsubsection{Tradeoff}
\label{subsubsec:tradeoff}

This sets up the fundamental tradeoff involved in LIGO control loop design.
Make the controllers too strong, and sensor noise will be re-injected into the controller actuators, creating true, unintended displacement.
Make the controllers too weak, and excess displacement noise will pollute the spectrum and make it difficult to hold the interferometer on resonance.

LIGO controllers are designed to be overtly strong during the locking phase, to hold the optics strongly on resonance and avoid locklosses.
This injects excess sensor noise.
Near the end of the locking process, the loop bandwidths are reduced, weakening the hold but avoiding sensor noise injection.

\subsubsection{Feedforward}
\label{subsubsec:feedforward}

Even with loop bandwidth reduction, optical cross-couplings cause controllers to inject noise into other loops \cite{Izumi2017}.
In fact, sensor noise in the auxiliary length degrees of freedom is injected into the GW spectrum, causing a major limitation.

However, we have information on this sensor noise, since we are constantly measuring it.
Therefore, we can \textit{feedforward} the sensor noise, with a negative sign, to the interferometer optics \cite{Driggers2016}.
This creates an optic displacement which counteracts the displacement caused by auxiliary length sensor noise.

\subsubsection{Results}
\label{subsubsec:results}

Controls noise dominates the GW spectrum at low frequencies, as seen by the purple curve in Figure~\ref{fig:noise_budget}.
This curve represents the sum of the noise measured via excess power injection into each length, angular, and laser control loop \cite{Buikema2020}.

The problem is more difficult than summarized above.
Optical cross-couplings affect not just the GW spectrum, but all degrees of freedom.
Bilinear and nonlinear couplings can be hard to fully quantify, even with excess power injections.
Damping loops on suspensions and seismic isolation platforms that reduce large low-frequency seismic motion can leak noise into higher frequencies \cite{Adhikari2014}.

Work to lower controls noise further is the highest priority of the commissioning team.
A more advanced feedforward scheme is under consideration, to reduce angle-to-length coupling.
Multiple-input multiple-output controls models are under development.
Controls noise for the output mode cleaner are being analyzed.
Efforts to better quantify important interferometer parameters, such as optical losses and beam mode-matching, are being implemented.

\section{Current performance of the Advanced LIGO detectors}
\label{sec:current_performance}

\begin{figure}[H]
    \includegraphics[width=12 cm]{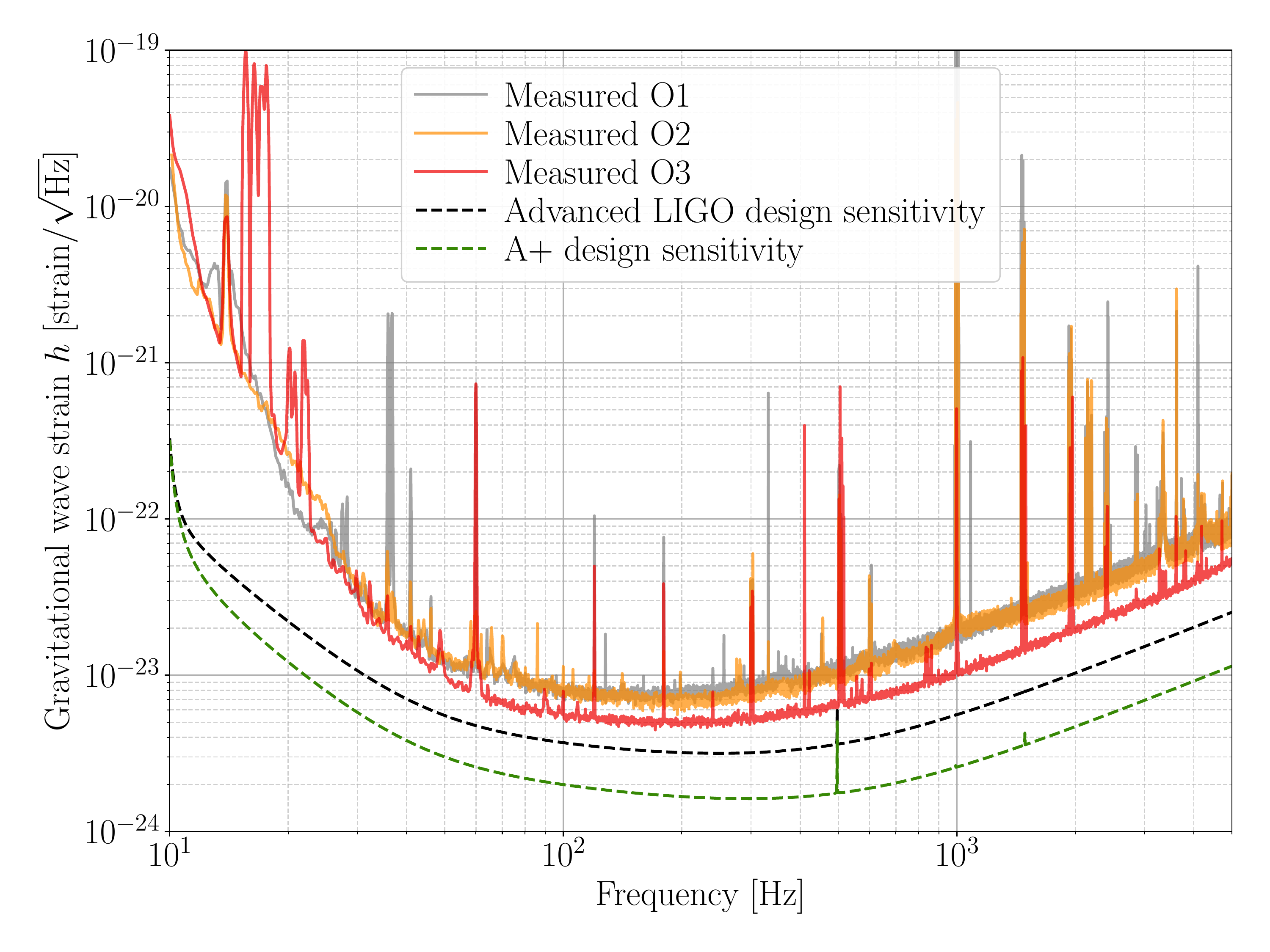}
    \caption{
        Comparison of LIGO Hanford gravitational wave noise spectra from O1 \cite{Martynov2016}, O2 \cite{Driggers2019}, and O3 \cite{Buikema2020}.
        Only the measured O3 spectra includes squeezed light injection.
        The Advanced LIGO design sensitivity assumes no squeezing, 125~watts laser power input yielding $750~\mathrm{kW}$ circulating arm power \cite{AdvLIGOPaper, aligoDesignSensitivity2018}.
        The A$+$ design sensitivity includes $12~\mathrm{dB}$ squeezing and a factor of two lower coatings thermal noise \cite{Barsotti2018, SkyLocation2020, InstrumentScienceWhitePaper2021}.
    }
    \label{fig:run_noise_comparison}
\end{figure}

The performance of the detectors is their overall sensitivity and uptime to astrophysical gravitational wave events.
There are currently 90 gravitational wave candidates, consisting of binary mergers across the universe from the first three observing runs (O1, O2, and O3) \cite{ThirdCatalogPaper}.
Figure~\ref{fig:run_noise_comparison} shows representative GW sensitivity spectra for each observing run.

In this section, we will review the astrophysical range and duty cycle of the detectors,
discuss the power budget and major technical limitations involved with high power interferometry,
the status of squeezing in LIGO interferometers,
and environmental disturbance sources and mitigation efforts.

\subsection{Astrophysical range}
\label{subsec:astrophysical_range}

A common performance gauge for the LIGO interferometers is the \textit{binary neutron star inspiral range}, or simply \textit{range} \cite{Finn1993, Finn1996, Chen2021}.
The range refers to the luminosity distance at which the detector is sensitive to a $1.4~\mathrm{M}_\odot $ neutron star merger with an SNR of 8, averaged over sky location and GW polarization.

During O3, the LIGO Livingston Observatory achieved a median binary neutron star range of 134~Mpc, while the LIGO Hanford Observatory achieved a median range of 111 Mpc \cite{Buikema2020}.
The detector sensitivity to heavier binary black holes extends much further than binary neutron stars, since the signals produced by black hole mergers are stronger.

\subsection{Duty cycle}
\label{subsec:duty_cycle}

The duty cycle is the uptime of the detectors, in other words, how often the detectors are sensitive to GWs.
Also important is coincident observation time, where both the Hanford and Livingston detectors are online at the same time.
Coincident observation is critical for extracting maximum information from detections, particularly verifying the detection is not a false positive and generating sky location data \cite{SkyLocation2020}.

During O3 both detectors were operational a greater percentage of the time compared to the previous two observing runs,
with Hanford and Livingston achieving observation duty cycles of 74.6\% and 77.0\%, respectively,
with coincident observation 62.2\% of the time \cite{Buikema2020}.

\subsection{High power interferometry}
\label{subsec:high_power_interferometry}

Gravitational wave observatories are dedicated to pushing the limits on maximum circulating laser power.
In Equation~\ref{eq:gw_signal_to_noise_approx}, we see that as power in the arms $P_\mathrm{arms}$ increases, so does the detector response to gravitational waves.

The biggest limit on power buildup in the interferometer is imposed by the presence of small, absorbing defects in the test mass optical coatings, known as point absorbers \cite{Brooks2021}.
In this section we will review how we quantify the power budget in the interferometer,
what point absorbers are and how we can mitigate their effect,
and other technical problems with high power, including parametric instability and radiation pressure-based optomechanics.

\subsubsection{Power budget}
\label{subsubsec:power_budget}

Laser power must be conserved inside the interferometer, i.e. $P_\mathrm{in} = P_\mathrm{out}$.
A \textit{power budget} is a record of where the input power of an interferometer is lost in the interferometer steady-state at full lock.
Losses can be from absorption by the optics themselves, or scattered off imperfections in the mirror surfaces into higher order modes \cite{Vajente2014, Isogai2013}.

For the carrier laser light, thanks to the common-mode rejection of the Michelson, we can model the interferometer power buildup using a simple plane-wave \textit{coupled cavity}, i.e. three-mirror cavity \cite{Fritschel1992, AndoThesis}.
The three mirrors are the power-recycling mirror, an input test mass, and an end test mass.

The laser resonates inside the coupled-cavity, building up power until losses exactly equal the input power.
In the coupled-cavity model, there are two main sources of power losses: power-recycling losses $\mathcal{L}_P$ and round-trip arm cavity losses $\mathcal{L}_A$.
There is also the promptly reflected power $P_\mathrm{refl}$ and transmitted power through the arms $P_\mathrm{trans}$.
The transmission of the power-recycling mirror $T_P$ must be strategically selected to be \textit{impedance-matched} with the overall interferometer losses,
or else too much carrier light will be promptly reflected \cite{Meers1988, Fritschel1992, AndoThesis, AdvLIGOPaper, AdvLIGOFinalDesign}.

A convenient proxy for power buildup in the interferometer is the \textit{power-recycling gain}, or PRG.
The power-recycling gain $G_P$ can be expressed as \cite{Brooks2021},
\begin{align}
    \label{eq:power_recycling_gain}
    G_P = \left( \dfrac{t_p}{1 - r_p \left( 1 - \dfrac{G_A (t_e^2 + \mathcal{L}_A) + \mathcal{L}_P}{2} \right)} \right)^2
\end{align}
where $t_p$ is the power-recycling mirror amplitude transmission,
$r_p$ is the power-recycling mirror amplitude reflectivity,
$t_e$ is the end test mass amplitude transmission,
$G_A$ is the arm power gain $\sim 265$,
$\mathcal{L}_A$ is the round-trip arm cavity loss,
$\mathcal{L}_P$ is the power-recycling cavity loss.
Equation~\ref{eq:power_recycling_gain} shows how losses in the arms matter more than losses in the power-recycling cavity,
because the arm gain $G_A$ scales the arm loss to account for the multiple reflections inside the arm cavity.

For LIGO Hanford in mid-2021, preliminary results suggest that interferometer round-trip losses $\mathcal{L}_A$ may now be lower than assumed for Advanced LIGO design \cite{alog58772, alog58794, AdvLIGOPaper}.
There are metrics other than the PRG for measuring the power buildup in the interferometer,
including the total interferometer reflection measurement and direct radiation-pressure based arm power measurement \cite{Buikema2020, alog59142}.
These provide more information on the true power budget,
and indicate that the simple plane-wave coupled-cavity model is not sufficient to fully explain power in the interferometer.




\subsubsection{Point absorbers}
\label{subsubsec:point_absorbers}

When the beam circulating in the interferometer encounters the test mass it uniformly heats an area of the test mass across its Gaussian profile.
The absorber power in this uniform cross section is determined by the power in the beam and the properties of the fused silica and the optical coating.

Point absorbers are sub-millimeter points of non-uniform, anomalously high optical absorption found on the test masses.
A full discussion of the point absorbers in LIGO can be found in Brooks et. al. \cite{Brooks2021}.
Point absorbers have been identified on multiple test masses in LIGO during the first three observing runs.
Their deformation is visible on Hartmann wavefront sensors, an auxiliary sensing system which measures the test mass surface and substrate deformation during power up.
Point absorbers have also been imaged on test masses which have been removed from the interferometers, using dark field microscopy.

The origin of point absorbers is under active investigation.
Point absorbers have been observed in the lab on spare test masses which have never been exposed to high optical power.
They appear to be embedded in the optical coating and are thought to originate during the coating deposition process.
Initial elemental analysis of some point absorbers show high concentrations of aluminum.

The circulating optical power in the interferometer heats up a point absorber, causing light to be scattered out of the arm cavity or into higher order cavity modes.
Depending on the geometry of the cavity, the higher order modes may be resonantly enhanced, causing additional loss to the fundamental mode and coupling unwanted modes to the GW detection port.
As the input power to the interferometer increases, the power scattered into higher order modes increases, as does the optical loss.
Point absorbers on the input test mass affect the arm cavity gain and power-recycling cavity gain, while the end test mass point absorbers only affect the arm cavity.
The thermal timescale of loss due to point absorbers are roughly an order or magnitude faster than that of uniform absorption.
Point absorbers have been observed to increase coupling of scattered light, laser frequency noise, and laser intensity noise to the gravitational wave readout \cite{Buikema2020}.

To avoid the negative effects of point absorbers, the interferometer alignment can be adjusted such that the beam spot overlap with the point absorbers is minimized.
In O3, both Hanford and Livingston operated with the spot position offset from center by $\sim 10$s of millimeters for certain test masses with known point absorbers \cite{alog58794llo, Brooks2021}.
However, beam spot position offsets cannot exceed the size of the optic itself ($34~\mathrm{cm}$ diameter for test masses),
and even small offsets risk scattering excess light out of the fringe of the main Gaussian beam ($\sim 12~\mathrm{cm}$ diameter).

\subsubsection{Parametric instability}
\label{subsubsec:parametric_instability}

Parametric instabilities (PIs) are mechanical modes of the test masses that sap energy from the fundamental mode, putting it into higher order optical modes of the resonating laser \cite{Evans2010, Green2017}.
As the mechanical modes begin to oscillate, it scatters more laser light is scattered into the higher order optical mode, further increasing the mechanical oscillations in runaway positive feedback loop.

Mitigation of PIs is essential to avoid runaway mechanical oscillations causing locklosses.
Ring-heaters on the core optics can change the radius of curvature of the optic, which in turn changes the eigenfrequency of the cavity higher order modes, eliminating the overlap between the optical and mechanical mode frequencies.
The electrostatic drive can be used to directly damp parametric instabilities that ring up at high circulating power \cite{Miller2011, Blair2017, HardwickThesis}.

During O1 and O2, the ring-heaters and electrostatic drive damping were the only ways to combat PI ringups.
However, with the higher circulating laser power in O3, dozens to hundreds of potential PIs were expected to something more was needed to combat multiple PIs constantly being triggered.

Therefore, acoustic mode dampers (AMDs) were installed on all core test masses \cite{Biscans2019}.
AMDs are mechanical tuned mass dampers glued to the sides of the test mass, with a piezo-electric transducer and shunt resistor designed to dissipate mechanical energy by converting it to electric charge and running it through the resistor.
The AMDs passively damp the mechanical oscillations, making it far more difficult for PIs to ring up.

Parametric instabilities have been observed even with the AMDs installed, as expected, but the overall parametric gain of all PIs is drastically reduced.
This makes the problem far more tractable to solve with the ring-heaters and electrostatic drive damping.
This work enables the higher circulating laser power seen in O3 and beyond (Section~\ref{subsubsec:power_budget}).

\subsubsection{Radiation pressure optomechanics}
\label{subsubsec:radiation_pressure_optomechanics}

With the advent of high-power interferometry comes a new era of radiation pressure based optomechanics,
particularly \textit{optical springs} and \textit{optical torques} for suspended optical cavities.
The laser power in the interferometer arm cavities couples the optic suspensions together such that the length and angular degrees of freedom must be considered together.
As circulating laser power increases, the dynamics of the optomechanical plants also change, presenting an additional challenge to high-powered interferometry.

Optical springs occur when the radiation pressure in a cavity creates a non-negligible restoring force (or non-restoring force) that affects the usual pendulum response \cite{Buonanno2002, Sheard2004, Aspelmeyer2014, Bond2017}.
The differential arm degree of freedom at Hanford has exhibited an optical spring effect since O1 \cite{CalUncPaper, HallThesis, Sun2020, Vitale2021}.

Optical torques refer to radiation pressure causing additional torques on the mirrors that affect the usual pendulum angular response \cite{Sidles2006}.
The \textit{hard modes} refer to the angular modes where the torsional stiffness increases due to the laser power torques.
Likewise, the \textit{soft modes} are angular modes where the torsional stiffness decreases due to the laser power torques.
One set of these mode must be statically unstable.

The radius of curvature of the test masses plays a strong role in governing the stability of the angular modes.
In Advanced LIGO, the soft modes have been chosen to be statically unstable, as they have a lower stiffness parameter and are easier to manage with a feedback control loop.
Optical torques have been observed in initial LIGO \cite{Hirose2010, Dooley2013}, and accounted for in Advanced LIGO design \cite{Barsotti2010}.

\subsection{Quantum squeezing}
\label{subsec:quantum_squeezing}

As explained in Section~\ref{subsubsec:squeezer}, quantum shot noise limits the sensitivity of interferometers to GWs.
A squeezed light source reflected off the output port of the interferometer can decrease quantum shot noise, increasing the detector sensitivity.

Frequency-independent squeezing was injected into both LIGO detectors for the duration of O3,
reducing quantum shot noise and increasing the expected gravitational wave detection rate by the cube of the range increase (40\% at LIGO Hanford Observatory and 50\% at LIGO Livingston Observatory) \cite{Tse2019}.
Though the squeezer subsystems at both LIGO sites are identical, small differences in the optical loss in their beam paths result in slightly different squeezer performance.
During O3 LIGO Hanford Observatory measured 2.0~dB of shot noise improvement from squeezing and LIGO Livingston Observatory measured 2.7~dB in the $1.1 - 1.4~\mathrm{kHz}$ regime \cite{Tse2019}.
At LIGO Livingston additional squeezing injection was possible, but would cause measurable increase to radiation pressure noise, and worse detector range.
In addition to improved range, squeezed state injection facilitated interesting investigations into the quantum nature of the LIGO interferometers \cite{Yu2020, Whittle2021, McCuller2021}.

A filter cavity is being installed at both LIGO sites to improve the quantum noise limit.
The filter cavity is discussed in Section~\ref{subsec:filter_cavity}.
A full review of the current status of squeezing in LIGO detectors will be published alongside this review \cite{Dwyer2022}.

\subsection{Environmental disturbances}
\label{subsec:environmental_disturbances}

Every effort is made to isolate the LIGO optics from environmental noise.
Environmental disturbances such as ground motion, acoustic noise, and magnetic noise can couple to the interferometer and cause excess noise in the gravitational-wave readout, masking gravitational-wave signals and limiting sensitivity.
The physical environment monitoring system includes of a variety of sensors and noise injection tools around the main interferometer and is used to monitor environmental noise, and characterize coupling to gravitational readout \cite{Nguyen2021}.
Sensors include seismometers, accelerometers, thermometers, microphones, magnetometers, electrometers, radio receivers, infrasound microphones, tilt meters, anemometers, voltage monitors, and hygrometers.
In this section we discuss some egregious environmental disturbances and their coupling to the interferometer.

\subsubsection{Earthquakes}
\label{subsubsec:earthquakes}

In general the gravitational wave detector sensitivity is not limited by seismic noise, see Figure \ref{fig:noise_budget} and Section \ref{subsec:seismic_noise}.
However, when the seismic waves generated by an earthquake pass through the detector site,
the ground motion can become so high that the active seismic control system \cite{HEPI2014, Matichard2015} cannot sufficiently suppress the motion.

Earthquakes and large seismic motion accounted for 5\% of the unplanned downtime of the LIGO detectors during O1 and O2.
For O3, a new seismic controls mode was implemented during earthquakes, aimed at reducing actuator saturation and gain peaking to maintain interferometer lock during earthquakes \cite{Schwartz2020}.
When the detector is taken to earthquake mode, two major seismic configuration changes occur:
the seismic loops are set to have reduced gain in the 50-60~mHz band,
and the common motion measured by seismometers in the corner and end stations is subtracted from the feedback signal.
Earthquake mode has allowed the LIGO detectors to remain locked through large earthquakes causing ground velocities up to 3.9~$\mu \mathrm{m} / \mathrm{s}$ RMS.

\subsubsection{Wind}
\label{subsubsec:wind}

High velocity wind can cause the corner and end station buildings to tilt, confusing seismometers and making it difficult (or impossible) to maintain lock.
This has been a problem particularly at LIGO Hanford, where gusts over 60~mph are measured.
Between O1 and O2 tilt meters (or ground rotation sensors) were installed at LIGO Hanford, and used to subtract ground tilt from seismometer signals \cite{Ross2020}.
More recently wind fences have been installed at LIGO Hanford, and their effectiveness is under study.

\subsubsection{Anthropogenic noise}
\label{subsubsec:anthropogenic_noise}

Anthropogenic ground motion, caused by human activity near the site, typically occurs in the 1 - 5~Hz frequency band and is particularly problematic at LIGO Livingston observatory.
At LIGO Livingston trains passing near the Y-arm, as well as elevated anthropogenic noise during the daytime cause scattering noise to be visible in the gravitational wave sensitivity in the 10-50~Hz band (see Section \ref{subsubsec:scattered_light}).

\subsubsection{Scattered light}
\label{subsubsec:scattered_light}

The sources of ground motion discussed above can couple to the gravitational wave readout through scattered light.
Tiny imperfections in the main and auxiliary optics can cause light to scatter out of the main interferometer beam.
This scattered light can then reflect off surfaces in the vacuum system, such as the suspension cages, chamber walls, or viewports.
If the light then couples back into the main interferometer beam, it will carry phase modulation from ground motion, and inject noise into the gravitational wave readout.
The characteristic morphology of scattering noise in the gravitational wave stain data is arches in the time-frequency domain - see examples in References \cite{Soni2021} and \cite{Nguyen2021}.
Low-frequency ground motion caused by earthquakes (0.03 - 0.1~Hz) or microseism (0.1 - 0.3~Hz) cause excess noise in the gravitational wave readout in the 10 - 100~Hz band.
In O3 scattered light coupling to the gravitational-wave readout was improved compared to previous runs, thanks to a suite of stray light baffles, and improved vibration isolation in the pre-stabilized laser room, improved test mass suspension control techniques.

\section{Upgrades for observing run four}
\label{sec:upgrades}

Observing run three ended in March 2020, and observing run four is scheduled to begin in December 2022.
Between these runs, there are several key upgrades being made to the LIGO detectors to improve sensitivity as part of the A$+$ upgrades \cite{InstrumentScienceWhitePaper2021, SkyLocation2020}.


\subsection{Y-arm input test mass replacement at Hanford}
\label{subsec:yarm_input_test_mass_replacement}

During O3, a significant point absorber was identified on the Y-arm input test mass (ITMY) at LIGO Hanford (see Section~\ref{subsubsec:point_absorbers}).
In December 2020, during the first part of the break between O3 and O4, the old ITMY was removed and a new test mass installed.
Preliminary results from the mid-2021 commissioning period showed no significant absorbers on the new ITMY and improvement in power buildups (see Section~\ref{subsubsec:power_budget}).

\subsection{End test mass replacements at Livingston}
\label{subsec:end_test_mass_replacement}

The end test masses at Livingston exhibit strong point absorbers limiting the overall circulating power.
In order to achieve the O4 circulating power goal of 400~kW, new end test masses (ETMs) are planned to be installed at LIGO Livingston in 2022.

LIGO Hanford also has exhibited point absorbers on its end test masses, but preliminary results from mid-2021 locking suggest these do not limit power buildup as much as those at Livingston.
The possibility for replacing Hanford's ETMs remains open, depending on commissioning results with the higher input powers from the PSL (see Section~\ref{subsec:higher_input_laser_power}).

\subsection{Filter cavity}
\label{subsec:filter_cavity}

One side effect of frequency-independent squeezing is that quantum radiation pressure noise increases \cite{Yu2020, KLMTV, Buonanno2001}.
This can be mitigated by reflecting the squeezed light off of a \textit{filter cavity} \cite{McCuller2020, filtercavitydesign, Evans2013, Kwee2014, Whittle2020, Kentaro2020} before injecting it into the interferometer, as shown in Figure~\ref{fig:ligo_o4_layout}.

In short, the filter cavity rotates the quadrature of the squeezed light for frequencies below the filter cavity pole frequency, but leaves unrotated the squeezed light quadrature above the filter cavity pole frequency.
This is known as \textit{frequency-dependent} squeezing, since the squeezing uncertainty ellipse rotates about the filter cavity pole frequency.
This can lower quantum uncertainty across the entire detector bandwidth, not just the shot noise dominated regime.

Currently, the facilities infrastructure for the filter cavity is under construction at both Hanford and Livingston.
The filter cavity is expected to be fully built and commissioned in time for O4, starting in December 2022.

\subsection{Higher input laser power}
\label{subsec:higher_input_laser_power}

While commissioning the LIGO detectors, the input laser power is maximized to reduce shot noise, while maintaining stable operation.
During O3 the pre-stabilized laser (PSL) was able to generate $\sim 50~\mathrm{W}$ of optical power at the input mode cleaner, and the detectors operated with 34-38~W input power \cite{Buikema2020}.
Higher power operations were limited by point absorbers.

The goal for O4 is to double the power in the interferometer, from $\sim 200~\mathrm{kW}$ to $400~\mathrm{kW}$ in the arms.
Higher input powers are expected to be possible, especially after the test mass replacements.
Therefore, the pre-stabilized laser is being upgraded to produce more power.

The new PSL configuration features two NeoLASE NeoVAN-4s-HP solid state amplifiers in series, amplifying a seed beam from 2~W to 125~W \cite{Bode2020}.
The seed laser is the same as previous runs: a non-planar ring oscillator (NPRO) Nd-YAG 1064~nm infrared laser \cite{Kane1985}.

\subsection{Output path active mode matching}
\label{subsec:output_path_active_mode_matching}

When the squeezed vacuum state encounters any imperfect optical interface, the squeezed state is degraded slightly.
Additionally, \textit{mode mismatch} is when the spacial mode of the squeezed state is not exactly matched to the spacial mode of the interferometer.
The squeezing improvement to quantum noise is limited by total optical losses from mode mismatch and imperfect optical surfaces.

Of the budgeted sources of loss in O3, mode mismatch was the largest, estimated at 10\% loss \cite{Tse2019}.
Further modeling and analysis of the squeezing over the full detection band has revealed that mode mismatch within the detector itself also induces frequency-dependent loss \cite{McCuller2021}.
For O4, active mode matching elements are being installed between the elements of the squeezer and the interferometer to improve the mode matching.

In the O4 interferometer layout, problematic mode mismatch can occur in several locations.
The squeezer mode must match the cavity modes of the output mode cleaner, the filter cavity and the interferometer mode (i.e. the mode circulating in the signal-recycling cavity).
While the layout is designed such that these cavity modes are matched, in reality there is some uncertainty in the signal-recycling cavity mode, as well as uncertainty in the placement of optics as they are installed in vacuum.
The active mode matching elements will allow changes to the spacial modes propagating between these optical cavities.

Two types of active mode matching element are being installed.
A thermally-actuated mirror with large actuation range (200~millidiopters) and slower response time is being installed on the path between the interferometer and the output mode cleaner \cite{Cao2020}.
Three piezo-electric deformable mirrors, with reduced actuation range (120~millidiopters) but faster response time, are being installed between the squeezer and the filter cavity, and between the filter cavity and the interferometer \cite{Srivastava2021}.

\section{Conclusions}
\label{sec:conclusions}

Powerful black hole and neutron star mergers are now revealed by their imprint on spacetime itself,
traveling to the Earth from the distant past,
carrying a wealth of information about the events that created them.
Advanced LIGO has already revolutionized our understanding of astrophysics and astronomy with 90 detections of gravitational waves.
More detections, and higher SNR detections, from a more sensitive detector will make new results in astrophysics, general relativity, and cosmology possible.

The Advanced LIGO detectors switch between periods of upgrades and installs, commissioning those upgrades to work as intended, and observation runs.
O4 is scheduled to run for one year of coincident observation time between the LIGO detectors, starting in December 2022 with $400~\mathrm{kW}$ circulating power and $\sim 175~\mathrm{Mpc}$ binary neutron star range.
O5 is when we plan to achieve the A$+$ sensitivity shown in Figure~\ref{fig:run_noise_comparison}.

Advanced LIGO's success would not have been possible if not for the lessons learned and support from the first generation of long-baseline interferometers from around the world,
including initial LIGO \cite{Sigg2008, Abbott2009, Aasi2015},
Virgo \cite{Caron1997, Acernese2008},
GEO600 \cite{Grote2005, Grote2010},
and TAMA \cite{Ando2002}.

Advanced Virgo \cite{Acernese2015} currently runs alongside Advanced LIGO, and is the only detector other than LIGO Hanford and LIGO Livingston to have sensed gravitational waves.
KAGRA is anticipated to join O4 in observations \cite{Somiya2012, Aso2013, Akutsu2021}.
LIGO India is expected to begin constructing a new observatory soon \cite{Padma2019}.

More ambitious upgrades to the current facilities are possible for LIGO Voyager, including a new laser wavelength ($2~\mu \mathrm{m}$) and cryogenically cooled optics \cite{Adhikari2020}.
Third generation detector designs are currently being proposed based on results and designs from Advanced LIGO and A$+$,
including Einstein Telescope \cite{EinsteinTelescope2011} in Europe and Cosmic Explorer in the United States \cite{cehorizonstudy2021}.
Space-based interferometers with very long baselines are also being designed and constructed.
The LISA mission is anticipated to detect much lower frequency GWs than LIGO \cite{Baker2019},
with the initial LISA Pathfinder results being extremely promising \cite{Sumner2017}.

The technological achievements made with Advanced LIGO will reverberate into the future,
just as the knowledge gained by first generation detectors paved the way for Advanced LIGO.
Every step toward design sensitivity brings the furthest reaches of the universe closer.







\funding{This research was funded by the National Science Foundation grant number PHY-1764464 and PHY-1834382.}

\acknowledgments{The authors acknowledge the vast amount of work that goes into designing, building, operating, and maintaining the LIGO Laboratory and facilities.
    For locking the interferometer, bringing it to its lowest-noise state, and characterizing the noise sources, we acknowledge the 2021 LIGO Hanford commissioning team, including Sheila Dwyer, Jenne Driggers, Anamaria Effler, Valary Frolov, Keita Kawabe, Jeff Kissel, Robert Schofield, Daniel Sigg, and Varun Srivastava.
    We acknowledge the calibration working group for producing the infrastructure to calibrate the interferometer data.
    We acknowledge the LIGO lab operations teams for locking, running, and managing the detector.
    We acknowledge the LIGO lab detector engineers for fabricating and installing the new Y-arm input test mass that was critical for removing point absorbers from the interferometer core optics.
    We acknowledge the LIGO facilities crew for building the facility, including the new filter cavity infrastrucuture.

    This material is based upon work supported by NSF's LIGO Laboratory which is a major facility fully funded by the National Science Foundation.
}

\conflictsofinterest{The authors declare no conflict of interest.}



\abbreviations{Abbreviations}{The following abbreviations are used in this manuscript:\\

    \noindent
    \begin{tabular}{@{}ll}
        GW   & gravitational waves                                 \\
        LIGO & laser interferometer gravitational wave observatory \\
        O1   & observing run one                                   \\
        O2   & observing run two                                   \\
        O3   & observing run three                                 \\
        O4   & observing run four                                  \\
        SNR  & signal-to-noise ratio                               \\
        PDH  & Pound-Drever-Hall                                   \\
        RF   & radio-frequency                                     \\
        PSL  & pre-stabilized laser                                \\
        ALS  & arm length stabilization                            \\
        DRMI & dual-recycled Michelson interferometer              \\
        DARM & differential arm (length)                           \\
        ASD  & amplitude spectral density                          \\
        QRPN & quantum radiation pressure noise                    \\
        MICH & Michelson length                                    \\
        PRCL & power-recycling cavity length                       \\
        SRCL & signal-recycling cavity length                      \\
        RFPD & radio-frequency photodetectors                      \\
        PRG  & power-recycling gain                                \\
        PI   & parametric instability                              \\
        ITMY & input test mass (Y-arm)                             \\
        ETM  & end test mass                                       \\
    \end{tabular}}

\appendixtitles{yes} 
\appendixstart
\appendix

\section{Michelson interferometer}
\label{sec:michelson_interferometer}

Gravitational waves produce differential motion in orthogonal directions of spacetime (Section~\ref{sec:gravitational_waves}).
Michelson interferometers were originally created to precisely measure differential light velocity in each arm \cite{Michelson1887}.
Today, a Michelson interferometer forms the core of the Advanced LIGO detector, and is used to detect differential motion in the arms \cite{Bond2017}.
This section will overview how a Michelson is sensitive to differential motion.

First, we will assume the plane-wave approximation is valid, so all electric fields will be simplified into a single complex number $E_0 = |E_0| e^{i \phi}$.
Second, we will assume our mirrors are thin mirrors with no losses, so $r^2 + t^2 = R + T = 1$ where $r$ and $t$ are the amplitude reflection and transmission coefficients of the mirrors and $R$ and $T$ are the power reflection and transmission coefficients of the mirrors.
Third, we will use the ``$+/-$'' mirror reflection convention based on the Fresnel relations, which states that a beam reflected off the back of the mirror suffers a $180^\circ$ phase flip, but a beam reflected off the front suffers no phase flip.

\subsection{Basics}
\label{subsec:michelson_basics}

\begin{figure}[h]
    \includegraphics[width=10.5 cm]{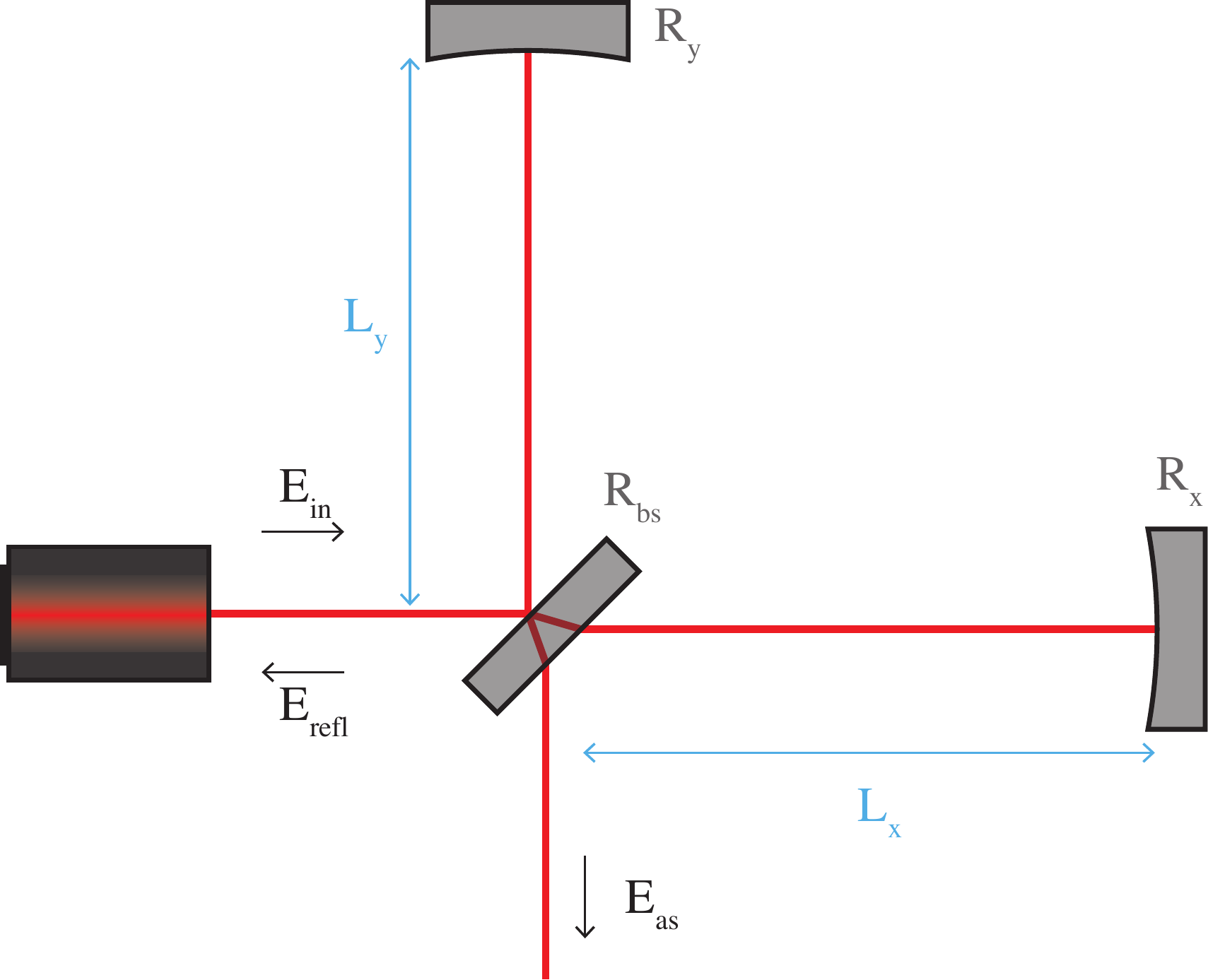}
    \caption{
        Optical layout of a Michelson interferometer.
        The input laser electric field is $E_\mathrm{in}$,
        the field reflected from the Michelson is $E_\mathrm{refl}$,
        and the field transmitter through the Michelson antisymmetric port is $E_\mathrm{as}$.
        The 50:50 beamsplitter $R_\mathrm{bs} = r_\mathrm{bs}^2 = 1 / 2$ reflects half the laser power to the Y-arm and transmits half to the X-arm.
        The highly reflecting end mirrors $R_\mathrm{x} \approx R_\mathrm{y} \approx 1$ send most of the light directly back to the beamsplitter.
    }
    \label{fig:michelson}
\end{figure}

Figure~\ref{fig:michelson} shows a Michelson interferometer, which consists of an input laser, a 50:50 beamsplitter, and two ``arms'' of laser light with highly reflective mirrors at the end.
The laser input amplitude $E_\mathrm{in}$ is split into the arms equally by the beamsplitter $r_\mathrm{bs} = t_\mathrm{bs} = 1/\sqrt{2}$.
The light in each arm $E_\mathrm{x}, E_\mathrm{y}$ propagates the length of its arm $L_\mathrm{x}, L_\mathrm{y}$, is reflected off the end mirrors $r_\mathrm{x} = r_\mathrm{y} = 1$ accruing different amounts of phase $\phi_\mathrm{x}, \phi_\mathrm{y}$:
\begin{align}
    \label{eq:mich_arm_fields_1}
    E_\mathrm{x} & = r_\mathrm{x} t_\mathrm{bs} E_\mathrm{in} e^{i \phi_\mathrm{x}}, & \quad   E_\mathrm{y} & = r_\mathrm{y} r_\mathrm{bs} E_\mathrm{in} e^{i \phi_\mathrm{y}} \\
    \label{eq:mich_arm_fields_2}
    E_\mathrm{x} & = \dfrac{1}{\sqrt{2}} E_\mathrm{in} e^{i 2 k L_\mathrm{x}},       & \quad   E_\mathrm{y} & = \dfrac{1}{\sqrt{2}} E_\mathrm{in} e^{i 2 k L_\mathrm{y}}
\end{align}

The light from each arm is then recombined at the beamsplitter, producing the reflected beam $E_\mathrm{refl}$ and transmitted, or antisymmetric, beam $E_\mathrm{as}$:
\begin{align}
    \label{eq:mich_refl_as_fields_1}
    E_\mathrm{refl} & = t_\mathrm{bs} E_\mathrm{x} + r_\mathrm{bs} E_\mathrm{y},                                   & \quad E_\mathrm{as} & = -r_\mathrm{bs} E_\mathrm{x} + t_\mathrm{bs} E_\mathrm{y}                                   \\
    \label{eq:mich_refl_as_fields_2}
    E_\mathrm{refl} & = \dfrac{1}{2} E_\mathrm{in} \left( e^{i 2 k L_\mathrm{x}} + e^{i 2 k L_\mathrm{y}} \right), & \quad E_\mathrm{as} & = \dfrac{1}{2} E_\mathrm{in} \left( -e^{i 2 k L_\mathrm{x}} + e^{i 2 k L_\mathrm{y}} \right) \\
    \label{eq:mich_refl_as_fields_3}
    E_\mathrm{refl} & = E_\mathrm{in} e^{i 2 k L} \cos(2 k \Delta L),                                              & \quad E_\mathrm{as} & = -i E_\mathrm{in} e^{i 2 k L} \sin(2 k \Delta L)
\end{align}
where between Eqs.~\ref{eq:mich_refl_as_fields_2} and \ref{eq:mich_refl_as_fields_3} we have defined the common length $L = (L_\mathrm{x} + L_\mathrm{y})/2$ and differential length $\Delta L = (L_\mathrm{x} - L_\mathrm{y})/2$.

Calculating the power at the reflected and antisymmetric ports $P_\mathrm{refl}, P_\mathrm{as}$:
\begin{align}
    \label{eq:mich_refl_power}
    P_\mathrm{refl} & = |E_\mathrm{refl}|^2 = P_\mathrm{in} \cos(2 k \Delta L)^2 \\
    \label{eq:mich_as_power}
    P_\mathrm{as}   & = |E_\mathrm{as}|^2 = P_\mathrm{in} \sin(2 k \Delta L)^2
\end{align}
where $P_\mathrm{in} = |E_\mathrm{in}|^2$ is the input power.
The power at the antisymmetric port in Eq.~\ref{eq:mich_as_power} depends on the static differential length $\Delta L$.

\subsection{Transfer function}
\label{subsec:michelson_transfer_function}

Suppose we inject a small differential length oscillation $\Delta x \cos(\omega t)$ into the Michelson, such that $\Delta L = \Delta L_0 + \Delta x \cos(\omega t)$.
The \textit{transfer function} from the differential length to antisymmetric power at the frequency of injection $\omega$ is
\begin{align}
    \label{eq:mich_as_power_transfer_function}
    \dfrac{P_\mathrm{as}}{\Delta x}(\omega) & = k P_\mathrm{in} \sin(4 k \Delta L_0).
\end{align}
The transfer function Eq.~\ref{eq:mich_as_power_transfer_function} defines the frequency response of antisymmetric power to length motion of the Michelson interferometer.
In this case, the transfer function is flat for all $\omega$.

The easiest way to see the effect of the length oscillation $\Delta x$ on $P_\mathrm{as}$ is to think about the derivative of Eq.~\ref{eq:mich_as_power} with respect to $\Delta L$.
The small oscillation will vary $P_\mathrm{as}$ at the same frequency as $\Delta x \cos(\omega t)$, $\Delta P_\mathrm{as}(\omega)$.
The derivative of Eq.~\ref{eq:mich_as_power} is a slight simplification, as it would be missing a factor of two compared to Eq.~\ref{eq:mich_as_power_transfer_function}.

The transfer function of a gravitational wave $h$ to power at the antisymmetric port is more complicated, as seen in \cite{Saulson1994}:
\begin{align}
    \label{eq:mich_as_power_transfer_function_gw}
    \dfrac{P_\mathrm{as}}{h}(\omega) & = k P_\mathrm{in} L \sin(4 k \Delta L_0) \, \mathrm{sinc} \left( \dfrac{\omega L}{c} \right) e^{- \frac{i \omega L}{c}}
\end{align}
Eq.~\ref{eq:mich_as_power_transfer_function_gw} is not flat: when the signal $\omega = 2 \pi \,\mathrm{FSR}$
where the \textit{free spectral range} $\mathrm{FSR} = c / 2 L$, the transfer function dips to zero.
This is from the laser in the Michelson integrating over one full period of the gravitational wave from Eq.~\ref{eq:gw_metric},
yielding zero overall phase change at that frequency.
For the full derivation, see either \cite{Saulson1994, Bond2017} or Appendix B of \cite{CahillaneThesis}.

\section{Fabry-P\'erot interferometer}
\label{sec:fabry_perot_interferometer}

The Fabry-P\'erot interferometer forms a core optomechanical technology in LIGO, with its resonantly-enhanced sensitivity to length motion.
The beam reflected from the Fabry-P\'erot $E_\mathrm{refl}$ has a phase strongly dependent on the cavity length $L$.
Combined with a Michelson interferometer, the Fabry-P\'erot phase shift can be preferentially ``picked-off'' and sent out the antisymmetric port,
enhancing a normal Michelson's sensitivity to differential motion, and gravitational waves.
In the below sections, we will make the same assumptions as Appendix~\ref{sec:michelson_interferometer}.

\subsection{Basics}
\label{subsec:fabry_perot_basics}

\begin{figure}[H]
    \includegraphics[width=10.5 cm]{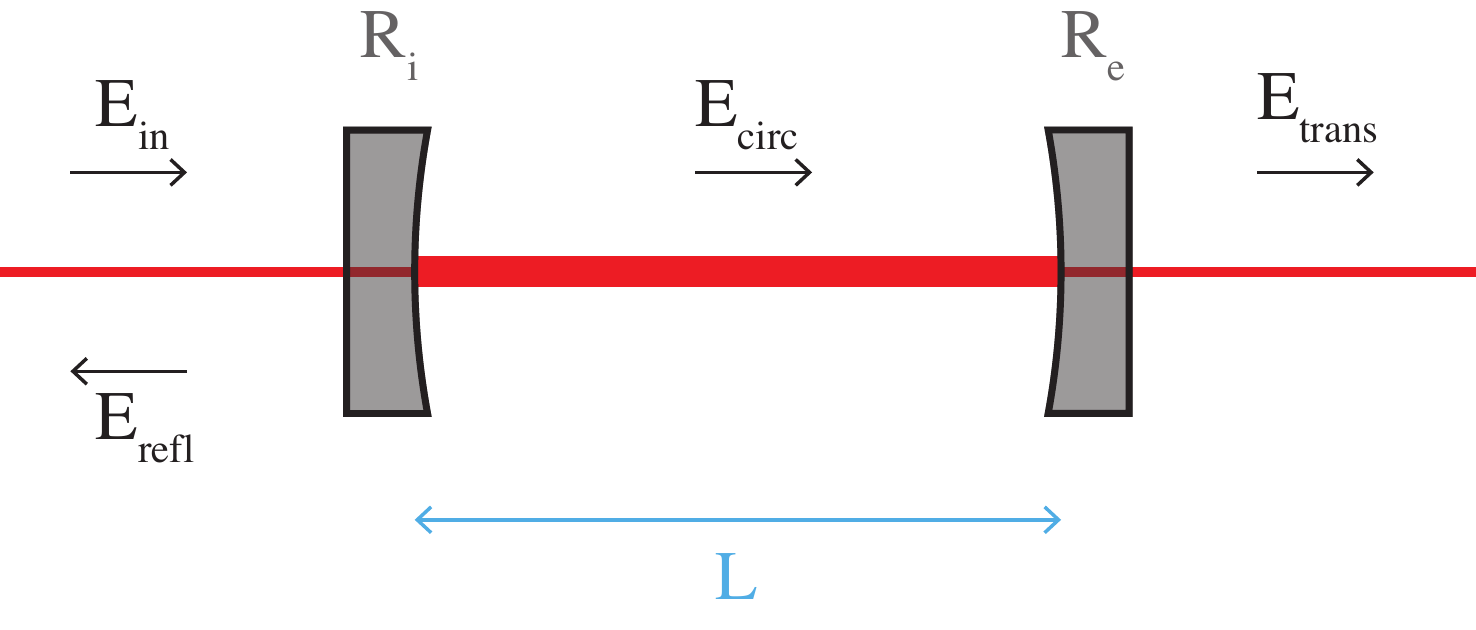}
    \caption{
        Optical layout of a Fabry-P\'erot interferometer.
    }
    \label{fig:fabry_perot}
\end{figure}

Figure~\ref{fig:fabry_perot} shows a Fabry-P\'erot interferometer, which is just a two-mirror aligned optical cavity.
The input mirror has reflectivity $R_i$, and the end mirror has reflectivity $R_e$.

The input beam $E_\mathrm{in}$ is partially reflected into $E_\mathrm{refl}$, and partially transmitted into the cavity $E_\mathrm{circ}$.
The circulating beam $E_\mathrm{circ}$ makes round-trips in the cavity with length $L$, accruing a phase $e^{i 2 k L}$ with every round-trip,
but also partially transmits out the end mirror $E_\mathrm{trans}$ and back out the input mirror $E_\mathrm{refl}$.

We can write out the equations for the plane-wave Fabry-P\'erot beams as
\begin{align}
    \label{eq:fabry_perot_circ_1}
    E_\mathrm{circ}  & = t_i E_\mathrm{in} + r_i r_e e^{i 2 k L} E_\mathrm{circ}  \\
    \label{eq:fabry_perot_refl_1}
    E_\mathrm{refl}  & = -r_i E_\mathrm{in} + t_i r_e e^{i 2 k L} E_\mathrm{circ} \\
    \label{eq:fabry_perot_trans_1}
    E_\mathrm{trans} & = t_e E_\mathrm{circ}
\end{align}
where $t_i, r_i$ are the amplitude transmission and reflection coefficients of the input mirror,
$t_e, r_e$ are the same for the end mirror,
and $k$ is the laser wavenumber.
In Eq.~\ref{eq:fabry_perot_refl_1}, $r_i$ has a negative sign due to the ``$+/-$'' convention, where we have chosen the left side of the input mirror to suffer a phase flip.

Using Eq.~\ref{eq:fabry_perot_circ_1}, we can solve for $E_\mathrm{circ}$, and use that result for Eqs.~\ref{eq:fabry_perot_refl_1} and \ref{eq:fabry_perot_trans_1}:
\begin{align}
    \label{eq:fabry_perot_circ_2}
    E_\mathrm{circ}  & = \dfrac{t_i}{1 - r_i r_e e^{i 2 k L} } E_\mathrm{in}                                    \\
    \label{eq:fabry_perot_refl_2}
    E_\mathrm{refl}  & = \dfrac{-r_i + (r_i^2 + t_i^2) r_e e^{i 2 k L}}{1 - r_i r_e e^{i 2 k L} } E_\mathrm{in} \\
    \label{eq:fabry_perot_trans_2}
    E_\mathrm{trans} & = \dfrac{t_i t_e e^{i k L}}{1 - r_i r_e e^{i 2 k L} } E_\mathrm{in}
\end{align}

From Eq.~\ref{eq:fabry_perot_circ_2}, we can examine how the resonant buildup works for $E_\mathrm{circ}$.
Normally, the product $r_i r_e$ in the denominator is selected to be very close to one.
This leaves the phasor $e^{i 2 k L}$, which can make the resonant power very large when it equals one.
The phasor can equal one when $2 k L = n 2 \pi$, where $n$ is any positive integer.
Simplifying yields the \textit{resonance condition} for the cavity length $L$:
\begin{align}
    \label{eq:resonant_condition}
    L = \dfrac{\lambda}{2} n
\end{align}
where $\lambda$ is the laser wavelength.

One important quantity is the \textit{cavity pole} $f_p$,
which can be derived by setting the denominator of Eqs.~\ref{eq:fabry_perot_circ_2}, \ref{eq:fabry_perot_refl_2}, \ref{eq:fabry_perot_trans_2} to zero:
\begin{align}
    \label{eq:cavity_pole}
    i f_p = - \dfrac{1}{2 \pi} \dfrac{c}{2 L} \log \left( r_i r_e \right)
\end{align}
The cavity pole is the frequency at which the Fabry-P\'erot frequency response to length motion falls by half, see Eq.~\ref{eq:fabry_perot_refl_transfer_function_2}.

Now we examine Eq.~\ref{eq:fabry_perot_refl_2}, the reflection from the Fabry-P\'erot.
First, we'll assume $r_i^2 + t_i^2 = 1$, which is true for a lossless mirror.
Next, we'll assume that we are near resonance except for a small deviation $\Delta L \ll 1$ such that $L \rightarrow L + \Delta L$,
then $e^{i 2 k (L + \Delta L)} \approx 1 + i 2 k \Delta L$.
A first-order series expansion of Eq.~\ref{eq:fabry_perot_refl_2} about $\Delta L$ yields
\begin{align}
    \label{eq:fabry_perot_refl_3}
    E_\mathrm{refl} & \approx E_\mathrm{in} \left[ \dfrac{-r_i + r_e}{ 1 - r_i r_e } - i k \Delta L \dfrac{t_i^2 r_e}{ (1 - r_i r_e)^2 } \right]
\end{align}
The key of Eq.~\ref{eq:fabry_perot_refl_3} is the term that depends on $\Delta L$ is entirely imaginary and very large.
This means that the phase of the reflected light $E_\mathrm{refl}$ is strongly responding to small changes in length $\Delta L$.

\subsection{Transfer function}
\label{subsec:fabry_perot_transfer_function}

Here, we report the frequency response from end mirror displacement modulation $\Delta x \cos(\omega t)$ to the reflected light $E_\mathrm{refl}$.
Assuming the cavity is resonant per Eq.~\ref{eq:resonant_condition}, we can write the length to reflected field transfer function as
\begin{align}
    \label{eq:fabry_perot_refl_transfer_function_1}
    \dfrac{E_\mathrm{refl}}{\Delta x}(\omega) & = i k E_\mathrm{in} \dfrac{t_i^2 r_e e^{i \omega L / c}}{ (1 - r_i r_e)(1 - r_i r_e e^{i 2 \omega L / c}) } \\
    \label{eq:fabry_perot_refl_transfer_function_2}
                                              & \approx i k E_\mathrm{in} r_e G_\mathrm{cav} \dfrac{1}{1 + i \dfrac{\omega}{2 \pi f_p}}
\end{align}
where $G_\mathrm{cav} = t_i^2/(1 - r_i r_e)^2$ is the ideal cavity power gain, and
$f_p$ is the cavity pole from Eq.~\ref{eq:cavity_pole}.
Between Eqs.~\ref{eq:fabry_perot_refl_transfer_function_1} and \ref{eq:fabry_perot_refl_transfer_function_2} we have assumed the end mirror length modulation $\Delta x$ is small.

Again, it's easiest to think of the frequency response in Eqs.~\ref{eq:fabry_perot_refl_transfer_function_1} and \ref{eq:fabry_perot_refl_transfer_function_2} as the derivative of Eq.~\ref{eq:fabry_perot_refl_2} with respect to $L$.
In reality, we must consider the full audio sideband picture to get the more accurate Eq.\ref{eq:fabry_perot_refl_transfer_function_1}.
More complete derivations of the Fabry-P\'erot frequency response can be found in \cite{Willke1999, Rakhmanov2002, Bondu2007, HallThesis, Bond2017}.






\end{paracol}
\reftitle{References}


\externalbibliography{yes}
\bibliography{galaxies_aligo_detector_review.bib}

\begin{thebibliography}{999}

\bibitem[Abbott \em{et~al.}(2016)Abbott et~al.]{GW150914}
Abbott, B.P.;  et~al.
\newblock Observation of Gravitational Waves from a Binary Black Hole Merger.
\newblock {\em Phys. Rev. Lett.} {\bf 2016}, {\em 116},~061102.
\newblock
  doi:{\changeurlcolor{black}\href{https://doi.org/10.1103/PhysRevLett.116.061102}{\detokenize{10.1103/PhysRevLett.116.061102}}}.

\bibitem[Aasi \em{et~al.}(2015)Aasi, Abbott, Abbott, Abbott, Abernathy, Ackley,
  Adams, Adams, Addesso, Adhikari, Adya, Affeldt, Aggarwal, Aguiar, Ain, Ajith,
  Alemic, Allen, Amariutei, Anderson, Anderson, Arai, Araya, Arceneaux, Areeda,
  Ashton, Ast, Aston, Aufmuth, Aulbert, Aylott, Babak, Baker, Ballmer,
  Barayoga, Barbet, Barclay, Barish, Barker, Barr, Barsotti, Bartlett, Barton,
  Bartos, Bassiri, Batch, Baune, Behnke, Bell, Bell, Benacquista, Bergman,
  Bergmann, Berry, Betzwieser, Bhagwat, Bhandare, Bilenko, Billingsley, Birch,
  Biscans, Biwer, Blackburn, Blackburn, Blair, Blair, Bock, Bodiya, Bojtos,
  Bond, Bork, Born, Bose, Brady, Braginsky, Brau, Bridges, Brinkmann, Brooks,
  Brown, Brown, Brown, Buchman, Buikema, Buonanno, Cadonati, {Calder{\'{o}}n
  Bustillo}, Camp, Cannon, Cao, Capano, Caride, Caudill, Cavagli{\`{a}},
  Cepeda, Chakraborty, Chalermsongsak, Chamberlin, Chao, Charlton, Chen, Cho,
  Cho, Chow, Christensen, Chu, Chung, Ciani, Clara, Clark, Collette, Cominsky,
  Constancio, Cook, Corbitt, Cornish, Corsi, Costa, Coughlin, Countryman,
  Couvares, Coward, Cowart, Coyne, Coyne, Craig, Creighton, Creighton, Cripe,
  Crowder, Cumming, Cunningham, Cutler, Dahl, {Dal Canton}, Damjanic,
  Danilishin, Danzmann, Dartez, Dave, Daveloza, Davies, Daw, DeBra, {Del
  Pozzo}, Denker, Dent, Dergachev, DeRosa, DeSalvo, Dhurandhar, Diaz, {Di
  Palma}, Dojcinoski, Dominguez, Donovan, Dooley, Doravari, Douglas, Downes,
  Driggers, Du, Dwyer, Eberle, Edo, Edwards, Edwards, Effler, Eggenstein,
  Ehrens, Eichholz, Eikenberry, Essick, Etzel, Evans, Evans, Factourovich,
  Fairhurst, Fan, Fang, Farr, Farr, Favata, Fays, Fehrmann, Fejer, Feldbaum,
  Ferreira, Fisher, Frei, Freise, Frey, Fricke, Fritschel, Frolov,
  Fuentes-Tapia, Fulda, Fyffe, Gair, Gaonkar, Gehrels, Gergely, Giaime,
  Giardina, Gleason, Goetz, Goetz, Gondan, Gonz{\'{a}}lez, Gordon, Gorodetsky,
  Gossan, Go{\ss}ler, Gr{\"{a}}f, Graff, Grant, Gras, Gray, Greenhalgh,
  Gretarsson, Grote, Grunewald, Guido, Guo, Gushwa, Gustafson, Gustafson,
  Hacker, Hall, Hammond, Hanke, Hanks, Hanna, Hannam, Hanson, Hardwick, Harry,
  Harry, Hart, Hartman, Haster, Haughian, Hee, Heintze, Heinzel, Hendry, Heng,
  Heptonstall, Heurs, Hewitson, Hild, Hoak, Hodge, Hollitt, Holt, Hopkins,
  Hosken, Hough, Houston, Howell, Hu, Huerta, Hughey, Husa, Huttner, Huynh,
  Huynh-Dinh, Idrisy, Indik, Ingram, Inta, Islas, Isler, Isogai, Iyer, Izumi,
  Jacobson, Jang, Jawahar, Ji, Jim{\'{e}}nez-Forteza, Johnson, Jones, Jones,
  Ju, Haris, Kalogera, Kandhasamy, Kang, Kanner, Katsavounidis, Katzman,
  Kaufer, Kaufer, Kaur, Kawabe, Kawazoe, Keiser, Keitel, Kelley, Kells, Keppel,
  Key, Khalaidovski, Khalili, Khazanov, Kim, Kim, Kim, Kim, Kim, King, King,
  Kinzel, Kissel, Klimenko, Kline, Koehlenbeck, Kokeyama, Kondrashov, Korobko,
  Korth, Kozak, Kringel, Krishnan, Krueger, Kuehn, Kumar, Kumar, Kuo, Landry,
  Lantz, Larson, Lasky, Lazzarini, Lazzaro, Le, Leaci, Leavey, Lebigot, Lee,
  Lee, Lee, Leong, Levin, Levine, Lewis, Li, Libbrecht, Libson, Lin,
  Littenberg, Lockerbie, Lockett, Logue, Lombardi, Lormand, Lough, Lubinski,
  L{\"{u}}ck, Lundgren, Lynch, Ma, MacArthur, MacDonald, MacHenschalk,
  MacInnis, MacLeod, Maga{\~{n}}a-Sandoval, Magee, Mageswaran, Maglione,
  Mailand, Mandel, Mandic, Mangano, Mansell, M{\'{a}}rka, M{\'{a}}rka,
  Markosyan, Maros, Martin, Martin, Martynov, Marx, Mason, Massinger,
  Matichard, Matone, Mavalvala, Mazumder, Mazzolo, McCarthy, McClelland,
  McCormick, McGuire, McIntyre, McIver, McLin, McWilliams, Meadors, Meinders,
  Melatos, Mendell, Mercer, Meshkov, Messenger, Meyers, Miao, Middleton,
  Mikhailov, Miller, Miller, Millhouse, Ming, Mirshekari, Mishra, Mitra,
  Mitrofanov, Mitselmakher, Mittleman, Moe, Mohanty, Mohapatra, Moore, Moraru,
  Moreno, Morriss, Mossavi, Mow-Lowry, Mueller, Mueller, Mukherjee, Mullavey,
  Munch, Murphy, Murray, Mytidis, Nash, Nayak, Necula, Nedkova, Newton, Nguyen,
  Nielsen, Nissanke, Nitz, Nolting, Normandin, Nuttall, Ochsner, O'Dell,
  Oelker, Ogin, Oh, Oh, Ohme, Oppermann, Oram, O'Reilly, Ortega, O'Shaughnessy,
  Osthelder, Ott, Ottaway, Ottens, Overmier, Owen, Padilla, Pai, Pai, Palashov,
  Pal-Singh, Pan, Pankow, Pannarale, Pant, Papa, Paris, Patrick, Pedraza,
  Pekowsky, Pele, Penn, Perreca, Phelps, Pierro, Pinto, Pitkin, Poeld, Post,
  Poteomkin, Powell, Prasad, Predoi, Premachandra, Prestegard, Price, Principe,
  Privitera, Prix, Prokhorov, Puncken, P{\"{u}}rrer, Qin, Quetschke, Quintero,
  Quiroga, Quitzow-James, Raab, Rabeling, Radkins, Raffai, Raja, Rajalakshmi,
  Rakhmanov, Ramirez, Raymond, Reed, Reid, Reitze, Reula, Riles, Robertson,
  Robie, Rollins, Roma, Romano, Romanov, Romie, Rowan, R{\"{u}}diger, Ryan,
  Sachdev, Sadecki, Sadeghian, Saleem, Salemi, Sammut, Sandberg, Sanders,
  Sannibale, Santiago-Prieto, Sathyaprakash, Saulson, Savage, Sawadsky,
  Scheuer, Schilling, Schmidt, Schnabel, Schofield, Schreiber, Schuette,
  Schutz, Scott, Scott, Sellers, Sengupta, Sergeev, Serna, Sevigny, Shaddock,
  Shahriar, Shaltev, Shao, Shapiro, Shawhan, Shoemaker, Sidery, Siemens, Sigg,
  Silva, Simakov, Singer, Singer, Singh, Sintes, Slagmolen, Smith, Smith,
  Smith, Smith-Lefebvre, Son, Sorazu, Souradeep, Staley, Stebbins, Steinke,
  Steinlechner, Steinlechner, Steinmeyer, Stephens, Steplewski, Stevenson,
  Stone, Strain, Strigin, Sturani, Stuver, Summerscales, Sutton, Szczepanczyk,
  Szeifert, Talukder, Tanner, T{\'{a}}pai, Tarabrin, Taracchini, Taylor,
  Tellez, Theeg, Thirugnanasambandam, Thomas, Thomas, Thorne, Thorne, Thrane,
  Tiwari, Tomlinson, Torres, Torrie, Traylor, Tse, Tshilumba, Ugolini,
  Unnikrishnan, Urban, Usman, Vahlbruch, Vajente, Valdes, Vallisneri, {Van
  Veggel}, Vass, Vaulin, Vecchio, Veitch, Veitch, Venkateswara, Vincent-Finley,
  Vitale, Vo, Vorvick, Vousden, Vyatchanin, Wade, Wade, Wade, Walker, Wallace,
  Walsh, Wang, Wang, Wang, Ward, Warner, Was, Weaver, Weinert, Weinstein,
  Weiss, Welborn, Wen, Wessels, Westphal, Wette, Whelan, Whitcomb, White,
  Whiting, Wilkinson, Williams, Williams, Williamson, Willis, Willke, Wimmer,
  Winkler, Wipf, Wittel, Woan, Worden, Xie, Yablon, Yakushin, Yam, Yamamoto,
  Yancey, Yang, Zanolin, Zhang, Zhang, Zhang, Zhang, Zhao, Zhou, Zhu, Zucker,
  Zuraw, and Zweizig]{AdvLIGOPaper}
Aasi, J.; Abbott, B.P.; Abbott, R.; Abbott, T.; Abernathy, M.R.; Ackley, K.;
  Adams, C.; Adams, T.; Addesso, P.; Adhikari, R.X.;  et~al.
\newblock {Advanced LIGO}.
\newblock {\em Classical and Quantum Gravity} {\bf 2015},
  \href{https://arxiv.org/abs/1411.4547}{{\normalfont [1411.4547]}}.
\newblock
  doi:{\changeurlcolor{black}\href{https://doi.org/10.1088/0264-9381/32/7/074001}{\detokenize{10.1088/0264-9381/32/7/074001}}}.

\bibitem[Abbott \em{et~al.}(2010)Abbott, Adhikari, Ballmer, Barsotti, Evans,
  Fritschel, Frolov, Mueller, Slagmolen, and Waldman]{AdvLIGOFinalDesign}
Abbott, R.; Adhikari, R.; Ballmer, S.; Barsotti, L.; Evans, M.; Fritschel, P.;
  Frolov, V.; Mueller, G.; Slagmolen, B.; Waldman, S.
\newblock Advanced LIGO Length Sensing and Control Final Design.
\newblock {\em Tech. rep. LIGO-T1000298-v2} {\bf 2010}.

\bibitem[Martynov \em{et~al.}(2016)Martynov, Hall, Abbott, Abbott, Abbott,
  Adams, Adhikari, Anderson, Anderson, Arai, Arain, Aston, Austin, Ballmer,
  Barbet, Barker, Barr, Barsotti, Bartlett, Barton, Bartos, Batch, Bell,
  Belopolski, Bergman, Betzwieser, Billingsley, Birch, Biscans, Biwer, Black,
  Blair, Bogan, Bork, Bridges, Brooks, Celerier, Ciani, Clara, Cook,
  Countryman, Cowart, Coyne, Cumming, Cunningham, Damjanic, Dannenberg,
  Danzmann, Costa, Daw, Debra, Derosa, Desalvo, Dooley, Doravari, Driggers,
  Dwyer, Effler, Etzel, Evans, Evans, Factourovich, Fair, Feldbaum, Fisher,
  Foley, Frede, Fritschel, Frolov, Fulda, Fyffe, Galdi, Giaime, Giardina,
  Gleason, Goetz, Gras, Gray, Greenhalgh, Grote, Guido, Gushwa, Gustafson,
  Gustafson, Hammond, Hanks, Hanson, Hardwick, Harry, Heefner, Heintze,
  Heptonstall, Hoak, Hough, Ivanov, Izumi, Jacobson, James, Jones, Kandhasamy,
  Karki, Kasprzack, Kaufer, Kawabe, Kells, Kijbunchoo, King, King, Kinzel,
  Kissel, Kokeyama, Korth, Kuehn, Kwee, Landry, Lantz, {Le Roux}, Levine,
  Lewis, Lhuillier, Lockerbie, Lormand, Lubinski, Lundgren, Macdonald,
  Macinnis, Macleod, Mageswaran, Mailand, M{\'{a}}rka, M{\'{a}}rka, Markosyan,
  Maros, Martin, Martin, Marx, Mason, Massinger, Matichard, Mavalvala,
  McCarthy, McClelland, McCormick, McIntyre, McIver, Merilh, Meyer, Meyers,
  Miller, Mittleman, Moreno, Mueller, Mueller, Mullavey, Munch, Nuttall,
  Oberling, O'Dell, Oppermann, Oram, O'Reilly, Osthelder, Ottaway, Overmier,
  Palamos, Paris, Parker, Patrick, Pele, Penn, Phelps, Pickenpack, Pierro,
  Pinto, Poeld, Principe, Prokhorov, Puncken, Quetschke, Quintero, Raab,
  Radkins, Raffai, Ramet, Reed, Reid, Reitze, Robertson, Rollins, Roma, Romie,
  Rowan, Ryan, Sadecki, Sanchez, Sandberg, Sannibale, Savage, Schofield,
  Schultz, Schwinberg, Sellers, Sevigny, Shaddock, Shao, Shapiro, Shawhan,
  Shoemaker, Sigg, Slagmolen, Smith, Smith, Smith-Lefebvre, Sorazu, Staley,
  Stein, Stochino, Strain, Taylor, Thomas, Thomas, Thorne, Thrane, Torrie,
  Traylor, Vajente, Valdes, {Van Veggel}, Vargas, Vecchio, Veitch,
  Venkateswara, Vo, Vorvick, Waldman, Walker, Ward, Warner, Weaver, Weiss,
  Welborn, We{\ss}els, Wilkinson, Willems, Williams, Willke, Winkelmann, Wipf,
  Worden, Wu, Yamamoto, Yancey, Yu, Zhang, Zucker, and Zweizig]{Martynov2016}
Martynov, D.V.; Hall, E.D.; Abbott, B.P.; Abbott, R.; Abbott, T.D.; Adams, C.;
  Adhikari, R.X.; Anderson, R.A.; Anderson, S.B.; Arai, K.;  et~al.
\newblock {Sensitivity of the Advanced LIGO detectors at the beginning of
  gravitational wave astronomy}.
\newblock {\em Physical Review D} {\bf 2016},
  \href{https://arxiv.org/abs/1604.00439}{{\normalfont [1604.00439]}}.
\newblock
  doi:{\changeurlcolor{black}\href{https://doi.org/10.1103/PhysRevD.93.112004}{\detokenize{10.1103/PhysRevD.93.112004}}}.

\bibitem[Abbott \em{et~al.}(2016)Abbott et~al.]{GW150914DetectorPaper}
Abbott, B.P.;  et~al.
\newblock GW150914: The Advanced LIGO Detectors in the Era of First
  Discoveries.
\newblock {\em Phys. Rev. Lett.} {\bf 2016}, {\em 116},~131103.
\newblock
  doi:{\changeurlcolor{black}\href{https://doi.org/10.1103/PhysRevLett.116.131103}{\detokenize{10.1103/PhysRevLett.116.131103}}}.

\bibitem[Abbott \em{et~al.}(2017{\natexlab{a}})Abbott et~al.]{GW170817}
Abbott, B.P.;  et~al.
\newblock GW170817: Observation of Gravitational Waves from a Binary Neutron
  Star Inspiral.
\newblock {\em Phys. Rev. Lett.} {\bf 2017}, {\em 119},~161101.
\newblock
  doi:{\changeurlcolor{black}\href{https://doi.org/10.1103/PhysRevLett.119.161101}{\detokenize{10.1103/PhysRevLett.119.161101}}}.

\bibitem[Abbott \em{et~al.}(2017{\natexlab{b}})Abbott
  et~al.]{GW170817multimessenger}
Abbott, B.P.;  et~al.
\newblock Multi-messenger Observations of a Binary Neutron Star Merger.
\newblock {\em Astrophys. J. Lett.} {\bf 2017}, {\em 848},~L12.
\newblock
  doi:{\changeurlcolor{black}\href{https://doi.org/10.3847/2041-8213/aa91c9}{\detokenize{10.3847/2041-8213/aa91c9}}}.

\bibitem[Buikema \em{et~al.}(2020)Buikema, Cahillane, Mansell, Blair, Abbott,
  Adams, Adhikari, Ananyeva, Appert, Arai, Areeda, Asali, Aston, Austin, Baer,
  Ball, Ballmer, Banagiri, Barker, Barsotti, Bartlett, Berger, Betzwieser,
  Bhattacharjee, Billingsley, Biscans, Blair, Bode, Booker, Bork, Bramley,
  Brooks, Brown, Cannon, Chen, Ciobanu, Clara, Cooper, Corley, Countryman,
  Covas, Coyne, Datrier, Davis, {Di Fronzo}, Dooley, Driggers, Dupej, Dwyer,
  Effler, Etzel, Evans, Evans, Feicht, Fernandez-Galiana, Fritschel, Frolov,
  Fulda, Fyffe, Giaime, Giardina, Godwin, Goetz, Gras, Gray, Gray, Green,
  Gustafson, Gustafson, Hanks, Hanson, Hardwick, Hasskew, Heintze,
  Helmling-Cornell, Holland, Jones, Kandhasamy, Karki, Kasprzack, Kawabe,
  Kijbunchoo, King, Kissel, Kumar, Landry, Lane, Lantz, Laxen, Lecoeuche,
  Leviton, Liu, Lormand, Lundgren, Macas, Macinnis, Macleod, M{\'{a}}rka,
  M{\'{a}}rka, Martynov, Mason, Massinger, Matichard, Mavalvala, McCarthy,
  McClelland, McCormick, McCuller, McIver, McRae, Mendell, Merfeld, Merilh,
  Meylahn, Mistry, Mittleman, Moreno, Mow-Lowry, Mozzon, Mullavey, Nelson,
  Nguyen, Nuttall, Oberling, Oram, O'Reilly, Osthelder, Ottaway, Overmier,
  Palamos, Parker, Payne, Pele, Penhorwood, Perez, Pirello, Radkins, Ramirez,
  Richardson, Riles, Robertson, Rollins, Romel, Romie, Ross, Ryan, Sadecki,
  Sanchez, Sanchez, Saravanan, Savage, Schaetzl, Schnabel, Schofield, Schwartz,
  Sellers, Shaffer, Sigg, Slagmolen, Smith, Soni, Sorazu, Spencer, Strain, Sun,
  Szczepa{\'{n}}czyk, Thomas, Thomas, Thorne, Toland, Torrie, Traylor, Tse,
  Urban, Vajente, Valdes, Vander-Hyde, Veitch, Venkateswara, Venugopalan,
  Viets, Vo, Vorvick, Wade, Ward, Warner, Weaver, Weiss, Whittle, Willke, Wipf,
  Xiao, Yamamoto, Yu, Yu, Zhang, Zucker, and Zweizig]{Buikema2020}
Buikema, A.; Cahillane, C.; Mansell, G.L.; Blair, C.D.; Abbott, R.; Adams, C.;
  Adhikari, R.X.; Ananyeva, A.; Appert, S.; Arai, K.;  et~al.
\newblock {Sensitivity and performance of the Advanced LIGO detectors in the
  third observing run}.
\newblock {\em Physical Review D} {\bf 2020},
  \href{https://arxiv.org/abs/2008.01301}{{\normalfont [2008.01301]}}.
\newblock
  doi:{\changeurlcolor{black}\href{https://doi.org/10.1103/PhysRevD.102.062003}{\detokenize{10.1103/PhysRevD.102.062003}}}.

\bibitem[Abbott \em{et~al.}(2021)Abbott et~al.]{ThirdCatalogPaper}
Abbott, R.;  et~al.
\newblock GWTC-3: Compact Binary Coalescences Observed by LIGO and Virgo During
  the Second Part of the Third Observing Run,  2021,
  \href{https://arxiv.org/abs/2111.03606}{{\normalfont
  [arXiv:gr-qc/2111.03606]}}.

\bibitem[Abbott \em{et~al.}(2020)Abbott et~al.]{SkyLocation2020}
Abbott, B.P.;  et~al.
\newblock {Prospects for observing and localizing gravitational-wave transients
  with Advanced LIGO, Advanced Virgo and KAGRA}.
\newblock {\em Living Reviews in Relativity 2020 23:1} {\bf 2020}, {\em
  23},~1--69,  \href{https://arxiv.org/abs/1304.0670}{{\normalfont
  [1304.0670]}}.
\newblock
  doi:{\changeurlcolor{black}\href{https://doi.org/10.1007/S41114-020-00026-9}{\detokenize{10.1007/S41114-020-00026-9}}}.

\bibitem[Fritschel \em{et~al.}(2021)Fritschel, Reid, Vajente, Hammond, Miao,
  Brown, Quetschke, and Steinlechner]{InstrumentScienceWhitePaper2021}
Fritschel, P.; Reid, S.; Vajente, G.; Hammond, G.; Miao, H.; Brown, D.;
  Quetschke, V.; Steinlechner, J.
\newblock Instrument Science White Paper 2021.
\newblock {\em Tech. rep. LIGO-T2100298} {\bf 2021}.

\bibitem[{Misner} \em{et~al.}(1973){Misner}, {Thorne}, and {Wheeler}]{MTW}
{Misner}, C.W.; {Thorne}, K.S.; {Wheeler}, J.A.
\newblock {\em {Gravitation}};  1973.

\bibitem[Sathyaprakash and Schutz(2009)]{Sathyaprakash2009}
Sathyaprakash, B.S.; Schutz, B.F.
\newblock Physics, Astrophysics and Cosmology with Gravitational Waves {\bf
  2009}.
\newblock {\em 12}.
\newblock
  doi:{\changeurlcolor{black}\href{https://doi.org/10.12942/lrr-2009-2}{\detokenize{10.12942/lrr-2009-2}}}.

\bibitem[Saulson(1994)]{Saulson1994}
Saulson, P.R.
\newblock {\em Fundamentals of Interferometric Gravitational Wave Detectors};
  WORLD SCIENTIFIC,  1994;
  \href{https://arxiv.org/abs/https://www.worldscientific.com/doi/pdf/10.1142/2410}{{\normalfont
  [https://www.worldscientific.com/doi/pdf/10.1142/2410]}}.
\newblock
  doi:{\changeurlcolor{black}\href{https://doi.org/10.1142/2410}{\detokenize{10.1142/2410}}}.

\bibitem[Adhikari(2014)]{Adhikari2014}
Adhikari, R.X.
\newblock Gravitational radiation detection with laser interferometry.
\newblock {\em Rev. Mod. Phys.} {\bf 2014}, {\em 86},~121--151.
\newblock
  doi:{\changeurlcolor{black}\href{https://doi.org/10.1103/RevModPhys.86.121}{\detokenize{10.1103/RevModPhys.86.121}}}.

\bibitem[Rakhmanov \em{et~al.}(2008)Rakhmanov, Romano, and
  Whelan]{Rakhmanov2008}
Rakhmanov, M.; Romano, J.D.; Whelan, J.T.
\newblock High-frequency corrections to the detector response and their effect
  on searches for gravitational waves.
\newblock {\em Classical and Quantum Gravity} {\bf 2008}, {\em 25},~184017.
\newblock
  doi:{\changeurlcolor{black}\href{https://doi.org/10.1088/0264-9381/25/18/184017}{\detokenize{10.1088/0264-9381/25/18/184017}}}.

\bibitem[Drever \em{et~al.}(1983)Drever, Hall, Kowalski, Hough, Ford, Munley,
  and Ward]{Drever1983}
Drever, R.W.; Hall, J.L.; Kowalski, F.V.; Hough, J.; Ford, G.M.; Munley, A.J.;
  Ward, H.
\newblock {Laser phase and frequency stabilization using an optical resonator}.
\newblock {\em Applied Physics B Photophysics and Laser Chemistry} {\bf 1983}.
\newblock
  doi:{\changeurlcolor{black}\href{https://doi.org/10.1007/BF00702605}{\detokenize{10.1007/BF00702605}}}.

\bibitem[Regehr \em{et~al.}(1995)Regehr, Raab, and Whitcomb]{Regehr1995}
Regehr, M.W.; Raab, F.J.; Whitcomb, S.E.
\newblock {Demonstration of a power-recycled Michelson interferometer with
  Fabry–Perot arms by frontal modulation}.
\newblock {\em Optics Letters} {\bf 1995}.
\newblock
  doi:{\changeurlcolor{black}\href{https://doi.org/10.1364/ol.20.001507}{\detokenize{10.1364/ol.20.001507}}}.

\bibitem[Sigg \em{et~al.}(1998)Sigg, Mavalvala, Giaime, Fritschel, and
  Shoemaker]{Sigg1998}
Sigg, D.; Mavalvala, N.; Giaime, J.; Fritschel, P.; Shoemaker, D.
\newblock {Signal extraction in a power-recycled Michelson interferometer with
  Fabry–Perot arm cavities by use of a multiple-carrier frontal modulation
  scheme}.
\newblock {\em Applied Optics} {\bf 1998}.
\newblock
  doi:{\changeurlcolor{black}\href{https://doi.org/10.1364/ao.37.005687}{\detokenize{10.1364/ao.37.005687}}}.

\bibitem[Fritschel \em{et~al.}(2001)Fritschel, Bork, Gonz\'{a}lez, Mavalvala,
  Ouimette, Rong, Sigg, and Zucker]{Fritschel2001}
Fritschel, P.; Bork, R.; Gonz\'{a}lez, G.; Mavalvala, N.; Ouimette, D.; Rong,
  H.; Sigg, D.; Zucker, M.
\newblock Readout and control of a power-recycled interferometric
  gravitational-wave antenna.
\newblock {\em Appl. Opt.} {\bf 2001}, {\em 40},~4988--4998.
\newblock
  doi:{\changeurlcolor{black}\href{https://doi.org/10.1364/AO.40.004988}{\detokenize{10.1364/AO.40.004988}}}.

\bibitem[Strain \em{et~al.}(2003)Strain, M{\"{u}}ller, Delker, Reitze, Tanner,
  Mason, Willems, Shaddock, Gray, Mow-Lowry, and McClelland]{Strain2003}
Strain, K.A.; M{\"{u}}ller, G.; Delker, T.; Reitze, D.H.; Tanner, D.B.; Mason,
  J.E.; Willems, P.A.; Shaddock, D.A.; Gray, M.B.; Mow-Lowry, C.;  et~al.
\newblock {Sensing and control in dual-recycling laser interferometer
  gravitational-wave detectors}.
\newblock {\em Applied Optics} {\bf 2003}.
\newblock
  doi:{\changeurlcolor{black}\href{https://doi.org/10.1364/ao.42.001244}{\detokenize{10.1364/ao.42.001244}}}.

\bibitem[Fricke \em{et~al.}(2012)Fricke, Smith-Lefebvre, Abbott, Adhikari,
  Dooley, Evans, Fritschel, Frolov, Kawabe, Kissel, Slagmolen, and
  Waldman]{Fricke2012}
Fricke, T.T.; Smith-Lefebvre, N.D.; Abbott, R.; Adhikari, R.; Dooley, K.L.;
  Evans, M.; Fritschel, P.; Frolov, V.V.; Kawabe, K.; Kissel, J.S.;  et~al.
\newblock {DC} readout experiment in Enhanced {LIGO}.
\newblock {\em Classical and Quantum Gravity} {\bf 2012}, {\em 29},~065005.
\newblock
  doi:{\changeurlcolor{black}\href{https://doi.org/10.1088/0264-9381/29/6/065005}{\detokenize{10.1088/0264-9381/29/6/065005}}}.

\bibitem[Izumi and Sigg(2017)]{Izumi2017}
Izumi, K.; Sigg, D.
\newblock Advanced LIGO: length sensing and control in a dual recycled
  interferometric gravitational wave antenna.
\newblock {\em Classical and Quantum Gravity} {\bf 2017}, {\em 34},~015001.

\bibitem[Anderson(1984)]{Anderson1984}
Anderson, D.Z.
\newblock {Alignment of resonant optical cavities}.
\newblock {\em Applied Optics} {\bf 1984}.
\newblock
  doi:{\changeurlcolor{black}\href{https://doi.org/10.1364/ao.23.002944}{\detokenize{10.1364/ao.23.002944}}}.

\bibitem[Morrison \em{et~al.}(1994)Morrison, Meers, Robertson, and
  Ward]{Morrison1994}
Morrison, E.; Meers, B.J.; Robertson, D.I.; Ward, H.
\newblock {Automatic alignment of optical interferometers}.
\newblock {\em Applied Optics} {\bf 1994}.
\newblock
  doi:{\changeurlcolor{black}\href{https://doi.org/10.1364/ao.33.005041}{\detokenize{10.1364/ao.33.005041}}}.

\bibitem[Mavalvala \em{et~al.}(1998)Mavalvala, Sigg, and
  Shoemaker]{Mavalvala1998}
Mavalvala, N.; Sigg, D.; Shoemaker, D.
\newblock {Experimental test of an alignment-sensing scheme for a
  gravitational-wave interferometer}.
\newblock {\em Applied Optics} {\bf 1998}.
\newblock
  doi:{\changeurlcolor{black}\href{https://doi.org/10.1364/ao.37.007743}{\detokenize{10.1364/ao.37.007743}}}.

\bibitem[Sidles and Sigg(2006)]{Sidles2006}
Sidles, J.A.; Sigg, D.
\newblock {Optical torques in suspended Fabry–Perot interferometers}.
\newblock {\em Physics Letters A} {\bf 2006}, {\em 354},~167--172.
\newblock
  doi:{\changeurlcolor{black}\href{https://doi.org/10.1016/J.PHYSLETA.2006.01.051}{\detokenize{10.1016/J.PHYSLETA.2006.01.051}}}.

\bibitem[Hirose \em{et~al.}(2010)Hirose, Kawabe, Sigg, Adhikari, and
  Saulson]{Hirose2010}
Hirose, E.; Kawabe, K.; Sigg, D.; Adhikari, R.; Saulson, P.R.
\newblock Angular instability due to radiation pressure in the LIGO
  gravitational-wave detector.
\newblock {\em Appl. Opt.} {\bf 2010}, {\em 49},~3474--3484.
\newblock
  doi:{\changeurlcolor{black}\href{https://doi.org/10.1364/AO.49.003474}{\detokenize{10.1364/AO.49.003474}}}.

\bibitem[Barsotti \em{et~al.}(2010)Barsotti, Evans, and
  Fritschel]{Barsotti2010}
Barsotti, L.; Evans, M.; Fritschel, P.
\newblock {Alignment sensing and control in advanced LIGO}.
\newblock {\em Classical and Quantum Gravity} {\bf 2010}.
\newblock
  doi:{\changeurlcolor{black}\href{https://doi.org/10.1088/0264-9381/27/8/084026}{\detokenize{10.1088/0264-9381/27/8/084026}}}.

\bibitem[Dooley \em{et~al.}(2013)Dooley, Barsotti, Adhikari, Evans, Fricke,
  Fritschel, Frolov, Kawabe, and Smith-Lefebvre]{Dooley2013}
Dooley, K.L.; Barsotti, L.; Adhikari, R.X.; Evans, M.; Fricke, T.T.; Fritschel,
  P.; Frolov, V.; Kawabe, K.; Smith-Lefebvre, N.
\newblock Angular control of optical cavities in a radiation-pressure-dominated
  regime: the Enhanced LIGO case.
\newblock {\em J. Opt. Soc. Am. A} {\bf 2013}, {\em 30},~2618--2626.
\newblock
  doi:{\changeurlcolor{black}\href{https://doi.org/10.1364/JOSAA.30.002618}{\detokenize{10.1364/JOSAA.30.002618}}}.

\bibitem[Enomoto \em{et~al.}(2016)Enomoto, Nagano, and Kawamura]{Enomoto2016}
Enomoto, Y.; Nagano, K.; Kawamura, S.
\newblock {Standard quantum limit of angular motion of a suspended mirror and
  homodyne detection of a ponderomotively squeezed vacuum field}.
\newblock {\em Physical Review A} {\bf 2016},
  \href{https://arxiv.org/abs/1602.05344}{{\normalfont [1602.05344]}}.
\newblock
  doi:{\changeurlcolor{black}\href{https://doi.org/10.1103/PhysRevA.94.012115}{\detokenize{10.1103/PhysRevA.94.012115}}}.

\bibitem[Kwee \em{et~al.}(2012)Kwee, Bogan, Danzmann, Frede, Kim, King,
  P\"{o}ld, Puncken, Savage, Seifert, Wessels, Winkelmann, and
  Willke]{Kwee2012}
Kwee, P.; Bogan, C.; Danzmann, K.; Frede, M.; Kim, H.; King, P.; P\"{o}ld, J.;
  Puncken, O.; Savage, R.L.; Seifert, F.;  et~al.
\newblock Stabilized high-power laser system for the gravitational wave
  detector advanced LIGO.
\newblock {\em Opt. Express} {\bf 2012}, {\em 20},~10617--10634.
\newblock
  doi:{\changeurlcolor{black}\href{https://doi.org/10.1364/OE.20.010617}{\detokenize{10.1364/OE.20.010617}}}.

\bibitem[Seifert \em{et~al.}(2006)Seifert, Kwee, Heurs, Willke, and
  Danzmann]{Seifert2006}
Seifert, F.; Kwee, P.; Heurs, M.; Willke, B.; Danzmann, K.
\newblock {Laser power stabilization for second-generation gravitational wave
  detectors}.
\newblock {\em Optics Letters} {\bf 2006}.
\newblock
  doi:{\changeurlcolor{black}\href{https://doi.org/10.1364/ol.31.002000}{\detokenize{10.1364/ol.31.002000}}}.

\bibitem[Kwee \em{et~al.}(2009)Kwee, Willke, and Danzmann]{Kwee2009}
Kwee, P.; Willke, B.; Danzmann, K.
\newblock {Shot-noise-limited laser power stabilization with a high-power
  photodiode array}.
\newblock {\em Optics Letters} {\bf 2009}.
\newblock
  doi:{\changeurlcolor{black}\href{https://doi.org/10.1364/ol.34.002912}{\detokenize{10.1364/ol.34.002912}}}.

\bibitem[Zucker and Whitcomb(1996)]{Zucker1996}
Zucker, M.E.; Whitcomb, S.E.
\newblock Measurement of optical path fluctuations due to residual gas in the
  {LIGO} 40 meter interferometer.
\newblock  Proceedings of the Seventh Marcel Grossman Meeting on recent
  developments in theoretical and experimental general relativity, gravitation,
  and relativistic field theories,  1996, pp. 1434--1436.

\bibitem[Dolesi \em{et~al.}(2011)Dolesi, Hueller, Nicolodi, Tombolato, Vitale,
  Wass, Weber, Evans, Fritschel, Weiss, et~al.]{Dolesi2011}
Dolesi, R.; Hueller, M.; Nicolodi, D.; Tombolato, D.; Vitale, S.; Wass, P.J.;
  Weber, W.J.; Evans, M.; Fritschel, P.; Weiss, R.;  et~al.
\newblock Brownian force noise from molecular collisions and the sensitivity of
  advanced gravitational wave observatories.
\newblock {\em Phys. Rev. D} {\bf 2011}, {\em 84},~063007.
\newblock
  doi:{\changeurlcolor{black}\href{https://doi.org/10.1103/PhysRevD.84.063007}{\detokenize{10.1103/PhysRevD.84.063007}}}.

\bibitem[Phelps \em{et~al.}(2013)Phelps, Gushwa, and Torrie]{Phelps2013}
Phelps, M.H.; Gushwa, K.E.; Torrie, C.I.
\newblock {Optical contamination control in the Advanced LIGO ultra-high vacuum
  system}.
\newblock {\em https://doi.org/10.1117/12.2047327} {\bf 2013}, {\em
  8885},~314--327.
\newblock
  doi:{\changeurlcolor{black}\href{https://doi.org/10.1117/12.2047327}{\detokenize{10.1117/12.2047327}}}.

\bibitem[Robertson \em{et~al.}(2002)Robertson et~al.]{Robertson2002}
Robertson, N.A.;  et~al.
\newblock Quadruple suspension design for Advanced LIGO.
\newblock {\em Classical and Quantum Gravity} {\bf 2002}, {\em 19},~4043.

\bibitem[Aston \em{et~al.}(2012)Aston et~al.]{Aston2012}
Aston, S.M.;  et~al.
\newblock Update on quadruple suspension design for Advanced LIGO.
\newblock {\em Classical and Quantum Gravity} {\bf 2012}, {\em 29},~235004.

\bibitem[Carbone \em{et~al.}(2012)Carbone et~al.]{Carbone2012}
Carbone, L.;  et~al.
\newblock Sensors and actuators for the Advanced LIGO mirror suspensions.
\newblock {\em Classical and Quantum Gravity} {\bf 2012}, {\em 29},~115005.

\bibitem[Daw \em{et~al.}(2004)Daw, Giaime, Lormand, Lubinski, and
  Zweizig]{Daw2004}
Daw, E.J.; Giaime, J.A.; Lormand, D.; Lubinski, M.; Zweizig, J.
\newblock Long-term study of the seismic environment at {LIGO}.
\newblock {\em Classical and Quantum Gravity} {\bf 2004}, {\em 21},~2255--2273.
\newblock
  doi:{\changeurlcolor{black}\href{https://doi.org/10.1088/0264-9381/21/9/003}{\detokenize{10.1088/0264-9381/21/9/003}}}.

\bibitem[Wen \em{et~al.}(2014)Wen, Mittleman, Mason, Giaime, Abbott, Kern,
  O'Reilly, Bork, Hammond, Hardham, et~al.]{HEPI2014}
Wen, S.; Mittleman, R.; Mason, K.; Giaime, J.; Abbott, R.; Kern, J.; O'Reilly,
  B.; Bork, R.; Hammond, M.; Hardham, C.;  et~al.
\newblock Hydraulic external pre-isolator system for {LIGO}.
\newblock {\em Classical and Quantum Gravity} {\bf 2014}, {\em 31},~235001.
\newblock
  doi:{\changeurlcolor{black}\href{https://doi.org/10.1088/0264-9381/31/23/235001}{\detokenize{10.1088/0264-9381/31/23/235001}}}.

\bibitem[Matichard \em{et~al.}(2015)Matichard, Lantz, Mittleman, Mason, Kissel,
  Abbott, Biscans, McIver, Abbott, Abbott, et~al.]{Matichard2015}
Matichard, F.; Lantz, B.; Mittleman, R.; Mason, K.; Kissel, J.; Abbott, B.;
  Biscans, S.; McIver, J.; Abbott, R.; Abbott, S.;  et~al.
\newblock Seismic isolation of {A}dvanced {LIGO}: Review of strategy,
  instrumentation and performance.
\newblock {\em Classical and Quantum Gravity} {\bf 2015}, {\em 32},~185003.
\newblock
  doi:{\changeurlcolor{black}\href{https://doi.org/10.1088/0264-9381/32/18/185003}{\detokenize{10.1088/0264-9381/32/18/185003}}}.

\bibitem[Biscans \em{et~al.}(2018)Biscans, Warner, Mittleman, Buchanan,
  Coughlin, Evans, Gabbard, Harms, Lantz, Mukund, et~al.]{Biscans2018}
Biscans, S.; Warner, J.; Mittleman, R.; Buchanan, C.; Coughlin, M.; Evans, M.;
  Gabbard, H.; Harms, J.; Lantz, B.; Mukund, N.;  et~al.
\newblock Control strategy to limit duty cycle impact of earthquakes on the
  {LIGO} gravitational-wave detectors.
\newblock {\em Classical and Quantum Gravity} {\bf 2018}, {\em 35},~055004.
\newblock
  doi:{\changeurlcolor{black}\href{https://doi.org/10.1088/1361-6382/aaa4aa}{\detokenize{10.1088/1361-6382/aaa4aa}}}.

\bibitem[Bork \em{et~al.}(2001)Bork, Abbott, Barker, and Heefner]{Bork2001}
Bork, R.; Abbott, R.; Barker, D.; Heefner, J.
\newblock {An Overview of the LIGO Control and Data Acquisition System}.
\newblock {\em arXiv.org} {\bf 2001}, {\em physics.in},~19--23,
  \href{https://arxiv.org/abs/0111077}{{\normalfont [arXiv:physics/0111077]}}.

\bibitem[Bartos \em{et~al.}(2010)Bartos, Bork, Factourovich, Heefner, Mrka,
  Mrka, Raics, Schwinberg, and Sigg]{Bartos2010}
Bartos, I.; Bork, R.; Factourovich, M.; Heefner, J.; Mrka, S.; Mrka, Z.; Raics,
  Z.; Schwinberg, P.; Sigg, D.
\newblock {The Advanced LIGO timing system}.
\newblock {\em Classical and Quantum Gravity} {\bf 2010}, {\em 27},~084025.
\newblock
  doi:{\changeurlcolor{black}\href{https://doi.org/10.1088/0264-9381/27/8/084025}{\detokenize{10.1088/0264-9381/27/8/084025}}}.

\bibitem[Rollins(2016)]{Rollins2016}
Rollins, J.G.
\newblock {Distributed state machine supervision for long-baseline
  gravitational-wave detectors}.
\newblock {\em Review of Scientific Instruments} {\bf 2016}, {\em 87},~094502,
  \href{https://arxiv.org/abs/1604.01456}{{\normalfont [1604.01456]}}.
\newblock
  doi:{\changeurlcolor{black}\href{https://doi.org/10.1063/1.4961665}{\detokenize{10.1063/1.4961665}}}.

\bibitem[Bork \em{et~al.}(2021)Bork, Hanks, Barker, Betzwieser, Rollins,
  Thorne, and von Reis]{Bork2021}
Bork, R.; Hanks, J.; Barker, D.; Betzwieser, J.; Rollins, J.; Thorne, K.; von
  Reis, E.
\newblock {advligorts: The Advanced LIGO real-time digital control and data
  acquisition system}.
\newblock {\em SoftwareX} {\bf 2021}, {\em 13},~100619.
\newblock
  doi:{\changeurlcolor{black}\href{https://doi.org/10.1016/J.SOFTX.2020.100619}{\detokenize{10.1016/J.SOFTX.2020.100619}}}.

\bibitem[Mueller \em{et~al.}(2016)Mueller, Arain, Ciani, Derosa, Effler,
  Feldbaum, Frolov, Fulda, Gleason, Heintze, Kawabe, King, Kokeyama, Korth,
  Martin, Mullavey, Peold, Quetschke, Reitze, Tanner, Vorvick, Williams, and
  Mueller]{Mueller2016}
Mueller, C.L.; Arain, M.A.; Ciani, G.; Derosa, R.T.; Effler, A.; Feldbaum, D.;
  Frolov, V.V.; Fulda, P.; Gleason, J.; Heintze, M.;  et~al.
\newblock {The advanced LIGO input optics}.
\newblock {\em Review of Scientific Instruments} {\bf 2016}.
\newblock
  doi:{\changeurlcolor{black}\href{https://doi.org/10.1063/1.4936974}{\detokenize{10.1063/1.4936974}}}.

\bibitem[Arai \em{et~al.}(2013)Arai, Barnum, Fritschel, Lewis, and
  Waldman]{Arai2013}
Arai, K.; Barnum, S.; Fritschel, P.; Lewis, J.; Waldman, S.
\newblock Output Mode Cleaner Design.
\newblock Technical report,  2013.

\bibitem[Evans \em{et~al.}(2013)Evans, Barsotti, Kwee, Harms, and
  Miao]{Evans2013}
Evans, M.; Barsotti, L.; Kwee, P.; Harms, J.; Miao, H.
\newblock {Realistic filter cavities for advanced gravitational wave
  detectors}.
\newblock {\em Physical Review D - Particles, Fields, Gravitation and
  Cosmology} {\bf 2013},  \href{https://arxiv.org/abs/1305.1599}{{\normalfont
  [1305.1599]}}.
\newblock
  doi:{\changeurlcolor{black}\href{https://doi.org/10.1103/PhysRevD.88.022002}{\detokenize{10.1103/PhysRevD.88.022002}}}.

\bibitem[McCuller \em{et~al.}(2020)McCuller, Whittle, Ganapathy, Komori, Tse,
  Fernandez-Galiana, Barsotti, Fritschel, MacInnis, Matichard, Mason,
  Mavalvala, Mittleman, Yu, Zucker, and Evans]{McCuller2020}
McCuller, L.; Whittle, C.; Ganapathy, D.; Komori, K.; Tse, M.;
  Fernandez-Galiana, A.; Barsotti, L.; Fritschel, P.; MacInnis, M.; Matichard,
  F.;  et~al.
\newblock Frequency-Dependent Squeezing for Advanced LIGO.
\newblock {\em Phys. Rev. Lett.} {\bf 2020}, {\em 124},~171102.
\newblock
  doi:{\changeurlcolor{black}\href{https://doi.org/10.1103/PhysRevLett.124.171102}{\detokenize{10.1103/PhysRevLett.124.171102}}}.

\bibitem[McCuller and Barsotti(2020)]{filtercavitydesign}
McCuller, L.; Barsotti, L.
\newblock Design Requirement Document of the A+ filter cavity and relay optics
  for frequency dependent squeezing.
\newblock Technical report, Massachusetts Institute of Technology,  2020.

\bibitem[Buonanno and Chen(2001)]{Buonanno2001}
Buonanno, A.; Chen, Y.
\newblock Quantum noise in second generation, signal-recycled laser
  interferometric gravitational-wave detectors.
\newblock {\em Physical Review D} {\bf 2001}, {\em 64}.
\newblock
  doi:{\changeurlcolor{black}\href{https://doi.org/10.1103/physrevd.64.042006}{\detokenize{10.1103/physrevd.64.042006}}}.

\bibitem[Kimble \em{et~al.}(2001)Kimble, Levin, Matsko, Thorne, and
  Vyatchanin]{KLMTV}
Kimble, H.J.; Levin, Y.; Matsko, A.B.; Thorne, K.S.; Vyatchanin, S.P.
\newblock Conversion of conventional gravitational-wave interferometers into
  quantum nondemolition interferometers by modifying their input and/or output
  optics.
\newblock {\em Phys. Rev. D} {\bf 2001}, {\em 65},~022002.
\newblock
  doi:{\changeurlcolor{black}\href{https://doi.org/10.1103/PhysRevD.65.022002}{\detokenize{10.1103/PhysRevD.65.022002}}}.

\bibitem[Ward(2010)]{WardThesis}
Ward, R.L.
\newblock PhD thesis, California Institute of Technology,  2010.
\newblock
  doi:{\changeurlcolor{black}\href{https://doi.org/10.7907/20SX-2935}{\detokenize{10.7907/20SX-2935}}}.

\bibitem[Hall(2017)]{HallThesis}
Hall, E.D.
\newblock PhD thesis, California Institute of Technology,  2017.
\newblock
  doi:{\changeurlcolor{black}\href{https://doi.org/10.7907/Z9PG1PQ9}{\detokenize{10.7907/Z9PG1PQ9}}}.

\bibitem[Cahillane(2021)]{CahillaneThesis}
Cahillane, C.
\newblock PhD thesis, California Institute of Technology,  2021.
\newblock
  doi:{\changeurlcolor{black}\href{https://doi.org/10.7907/76jj-mr73}{\detokenize{10.7907/76jj-mr73}}}.

\bibitem[Abbott \em{et~al.}(2020)Abbott, Abbott, Abbott, Abraham, Acernese,
  Ackley, Adams, Adya, Affeldt, Agathos, Agatsuma, Aggarwal, Aguiar, Aiello,
  Ain, Ajith, Alford, Allen, Allocca, Aloy, Altin, Amato, Ananyeva, Anderson,
  Anderson, Angelova, Antier, Appert, Arai, Araya, Areeda, Ar~ne, Arnaud, Arun,
  Ascenzi, Ashton, Aston, Astone, Aubin, Aufmuth, Aultoneal, Austin, Avendano,
  Avila-Alvarez, Babak, Bacon, Badaracco, Bader, Bae, Baker, Baldaccini,
  Ballardin, Ballmer, Banagiri, Barayoga, Barclay, Barish, Barker, Barkett,
  Barnum, Barone, Barr, Barsotti, Barsuglia, Barta, Bartlett, Bartos, Bassiri,
  Basti, Bawaj, Bayley, Bazzan, B{\'{e}}csy, Bejger, Belahcene, Bell, Beniwal,
  Berger, Bergmann, Bernuzzi, Bero, Berry, Bersanetti, Bertolini, Betzwieser,
  Bhandare, Bidler, Bilenko, Bilgili, Billingsley, Birch, Birney, Birnholtz,
  Biscans, Biscoveanu, Bisht, Bitossi, Bizouard, Blackburn, Blair, Blair,
  Blair, Bloemen, Bode, Boer, Boetzel, Bogaert, Bondu, Bonilla, Bonnand,
  Booker, Boom, Booth, Bork, Boschi, Bose, Bossie, Bossilkov, Bosveld,
  Bouffanais, Bozzi, Bradaschia, Brady, Bramley, Branchesi, Brau, Briant,
  Briggs, Brighenti, Brillet, Brinkmann, Brisson, Brockill, Brooks, Brown,
  Brunett, Buikema, Bulik, Bulten, Buonanno, Buskulic, Buy, Byer, Cabero,
  Cadonati, Cagnoli, Cahillane, {Calder{\'{o}}n Bustillo}, Callister, Calloni,
  Camp, Campbell, Canepa, Cannon, Cao, Cao, Capocasa, Carbognani, Caride,
  Carney, Carullo, {Casanueva Diaz}, Casentini, Caudill, Cavagli, Cavalier,
  Cavalieri, Cella, Cerd{\'{a}}-Dur{\'{a}}n, Cerretani, Cesarini, Chaibi,
  Chakravarti, Chamberlin, Chan, Chao, Charlton, Chase, Chassande-Mottin,
  Chatterjee, Chaturvedi, Chatziioannou, Cheeseboro, Chen, Chen, Chen, Cheng,
  Cheong, Chia, Chincarini, Chiummo, Cho, Cho, Cho, Christensen, Chu, Chua,
  Chung, Chung, Ciani, Ciobanu, Ciolfi, Cipriano, Cirone, Clara, Clark,
  Clearwater, Cleva, Cocchieri, Coccia, Cohadon, Cohen, Colgan, Colleoni,
  Collette, Collins, Cominsky, Constancio, Conti, Cooper, Corban, Corbitt,
  Cordero-Carri{\'{o}}n, Corley, Cornish, Corsi, Cortese, Costa, Cotesta,
  Coughlin, Coughlin, Coulon, Countryman, Couvares, Covas, Cowan, Coward,
  Cowart, Coyne, Coyne, Creighton, Creighton, Cripe, Croquette, Crowder,
  Cullen, Cumming, Cunningham, Cuoco, {Dal Canton}, D{\'{a}}lya, Danilishin,
  D'Antonio, Danzmann, Dasgupta, {Da Silva Costa}, Datrier, Dattilo, Dave,
  Davier, Davis, Daw, Debra, Deenadayalan, Degallaix, {De Laurentis},
  Del{\'{e}}glise, {Del Pozzo}, Demarchi, Demos, Dent, {De Pietri}, Derby, {De
  Rosa}, {De Rossi}, Desalvo, {De Varona}, Dhurandhar, D{\'{i}}az, Dietrich,
  {Di Fiore}, {Di Giovanni}, {Di Girolamo}, {Di Lieto}, Ding, {Di Pace}, {Di
  Palma}, {Di Renzo}, Dmitriev, Doctor, Donovan, Dooley, Doravari, Dorrington,
  Downes, Drago, Driggers, Du, Ducoin, Dupej, Dwyer, Easter, Edo, Edwards,
  Effler, Ehrens, Eichholz, Eikenberry, Eisenmann, Eisenstein, Essick,
  Estelles, Estevez, Etienne, Etzel, Evans, Evans, Fafone, Fair, Fairhurst,
  Fan, Farinon, Farr, Farr, Fauchon-Jones, Favata, Fays, Fazio, Fee, Feicht,
  Fejer, Feng, Fernandez-Galiana, Ferrante, Ferreira, Ferreira, Ferrini,
  Fidecaro, Fiori, Fiorucci, Fishbach, Fisher, Fishner, Fitz-Axen, Flaminio,
  Fletcher, Flynn, Fong, Font, Forsyth, Fournier, Frasca, Frasconi, Frei,
  Freise, Frey, Frey, Fritschel, Frolov, Fulda, Fyffe, Gabbard, Gadre, Gaebel,
  Gair, Gammaitoni, Ganija, Gaonkar, Garcia, Garc{\'{i}}a-Quir{\'{o}}s, Garufi,
  Gateley, Gaudio, Gaur, Gayathri, Gemme, Genin, Gennai, George, George,
  Gergely, Germain, Ghonge, Ghosh, Ghosh, Ghosh, Giacomazzo, Giaime, Giardina,
  Giazotto, Gill, Giordano, Glover, Godwin, Goetz, Goetz, Goncharov,
  Gonz{\'{a}}lez, {Gonzalez Castro}, Gopakumar, Gorodetsky, Gossan, Gosselin,
  Gouaty, Grado, Graef, Granata, Grant, Gras, Grassia, Gray, Gray, Greco,
  Green, Green, Gretarsson, Groot, Grote, Grunewald, Gruning, Guidi, Gulati,
  Guo, Gupta, Gupta, Gustafson, Gustafson, Haegel, Halim, Hall, Hall, Hamilton,
  Hammond, Haney, Hanke, Hanks, Hanna, Hannam, Hannuksela, Hanson, Hardwick,
  Haris, Harms, Harry, Harry, Haster, Haughian, Hayes, Healy, Heidmann,
  Heintze, Heitmann, Hello, Hemming, Hendry, Heng, Hennig, Heptonstall,
  Vivanco, Heurs, Hild, Hinderer, Hoak, Hochheim, Hofman, Holgado, Holland,
  Holt, Holz, Hopkins, Horst, Hough, Howell, Hoy, Hreibi, Huerta, Huet, Hughey,
  Hulko, Husa, Huttner, Huynh-Dinh, Idzkowski, Iess, Ingram, Inta, Intini,
  Irwin, Isa, Isac, Isi, Iyer, Izumi, Jacqmin, Jadhav, Jani, Janthalur,
  Jaranowski, Jenkins, Jiang, Johnson, Jones, Jones, Jones, Jonker, Ju, Junker,
  Kalaghatgi, Kalogera, Kamai, Kandhasamy, Kang, Kanner, Kapadia, Karki,
  Karvinen, Kashyap, Kasprzack, Katsanevas, Katsavounidis, Katzman, Kaufer,
  Kawabe, Keerthana, K{\'{e}}f{\'{e}}lian, Keitel, Kennedy, Key, Khalili, Khan,
  Khan, Khan, Khan, Khazanov, Khursheed, Kijbunchoo, Kim, Kim, Kim, Kim, Kim,
  Kim, Kim, Kimball, King, King, Kinley-Hanlon, Kirchhoff, Kissel, Kleybolte,
  Klika, Klimenko, Knowles, Koch, Koehlenbeck, Koekoek, Koley, Kondrashov,
  Kontos, Koper, Korobko, Korth, Kowalska, Kozak, Kringel, Krishnendu,
  Kr{\'{o}}lak, Kuehn, Kumar, Kumar, Kumar, Kumar, Kuo, Kutynia, Kwang, Lackey,
  Lai, Lam, Landry, Lane, Lang, Lange, Lantz, Lanza, Larson, Lartaux-Vollard,
  Lasky, Laxen, Lazzarini, Lazzaro, Leaci, Leavey, Lecoeuche, Lee, Lee, Lee,
  Lee, Lee, Lee, Lehmann, Lenon, Leroy, Letendre, Levin, Li, Li, Li, Li, Lin,
  Linde, Linker, Littenberg, Liu, Liu, Lo, Lockerbie, London, Longo, Lorenzini,
  Loriette, Lormand, Losurdo, Lough, Lousto, Lovelace, Lower, L{\"{u}}ck,
  Lumaca, Lundgren, Lynch, Ma, Macas, Macfoy, Macinnis, Macleod, Macquet,
  Maga{\~{n}}a-Sandoval, {Maga{\~{n}}a Zertuche}, Magee, Majorana, Maksimovic,
  Malik, Man, Mandic, Mangano, Mansell, Manske, Mantovani, Marchesoni, Marion,
  M{\'{a}}rka, M{\'{a}}rka, Markakis, Markosyan, Markowitz, Maros, Marquina,
  Marsat, Martelli, Martin, Martin, Martynov, Mason, Massera, Masserot,
  Massinger, Masso-Reid, Mastrogiovanni, Matas, Matichard, Matone, Mavalvala,
  Mazumder, McCann, McCarthy, McClelland, McCormick, McCuller, McGuire, McIver,
  McManus, McRae, McWilliams, Meacher, Meadors, Mehmet, Mehta, Meidam, Melatos,
  Mendell, Mercer, Mereni, Merilh, Merzougui, Meshkov, Messenger, Messick,
  Metzdorff, Meyers, Miao, Michel, Middleton, Mikhailov, Milano, Miller,
  Miller, Millhouse, Mills, Milovich-Goff, Minazzoli, Minenkov, Mishkin,
  Mishra, Mistry, Mitra, Mitrofanov, Mitselmakher, Mittleman, Mo, Moffa,
  Mogushi, Mohapatra, Montani, Moore, Moraru, Moreno, Morisaki, Mours,
  Mow-Lowry, Mukherjee, Mukherjee, Mukherjee, Mukund, Mullavey, Munch,
  Mu{\~{n}}iz, Muratore, Murray, Nagar, Nardecchia, Naticchioni, Nayak,
  Neilson, Nelemans, Nelson, Nery, Neunzert, Ng, Ng, Nguyen, Nichols, Nissanke,
  Nocera, North, Nuttall, Obergaulinger, Oberling, O'Brien, O'Dea, Ogin, Oh,
  Oh, Ohme, Ohta, Okada, Oliver, Oppermann, Oram, O'Reilly, Ormiston, Ortega,
  O'Shaughnessy, Ossokine, Ottaway, Overmier, Owen, Pace, Pagano, Page, Pai,
  Pai, Palamos, Palashov, Palomba, Pal-Singh, Pan, Pang, Pang, Pankow,
  Pannarale, Pant, Paoletti, Paoli, Parida, Parker, Pascucci, Pasqualetti,
  Passaquieti, Passuello, Patil, Patricelli, Pearlstone, Pedersen, Pedraza,
  Pedurand, Pele, Penn, Perez, Perreca, Pfeiffer, Phelps, Phukon, Piccinni,
  Pichot, Piergiovanni, Pillant, Pinard, Pirello, Pitkin, Poggiani, Pong,
  Ponrathnam, Popolizio, Porter, Powell, Prajapati, Prasad, Prasai, Prasanna,
  Pratten, Prestegard, Privitera, Prodi, Prokhorov, Puncken, Punturo, Puppo,
  P{\"{u}}rrer, Qi, Quetschke, Quinonez, Quintero, Quitzow-James, Raab,
  Radkins, Radulescu, Raffai, Raja, Rajan, Rajbhandari, Rakhmanov, Ramirez,
  Ramos-Buades, Rana, Rao, Rapagnani, Raymond, Razzano, Read, Regimbau, Rei,
  Reid, Reitze, Ren, Ricci, Richardson, Richardson, Ricker, Riles, Rizzo,
  Robertson, Robie, Robinet, Rocchi, Rolland, Rollins, Roma, Romanelli, Romano,
  Romel, Romie, Rose, Rosi{\'{n}}ska, Rosofsky, Ross, Rowan, R{\"{u}}diger,
  Ruggi, Rutins, Ryan, Sachdev, Sadecki, Sakellariadou, Salconi, Saleem,
  Samajdar, Sammut, Sanchez, Sanchez, Sanchis-Gual, Sandberg, Sanders,
  Santiago, Sarin, Sassolas, Sathyaprakash, Saulson, Sauter, Savage, Schale,
  Scheel, Scheuer, Schmidt, Schnabel, Schofield, Sch{\"{o}}nbeck, Schreiber,
  Schulte, Schutz, Schwalbe, Scott, Scott, Seidel, Sellers, Sengupta, Sennett,
  Sentenac, Sequino, Sergeev, Setyawati, Shaddock, Shaffer, Shahriar, Shaner,
  Shao, Sharma, Shawhan, Shen, Shink, Shoemaker, Shoemaker, Shyamsundar,
  Siellez, Sieniawska, Sigg, Silva, Singer, Singh, Singhal, Sintes,
  Sitmukhambetov, Skliris, Slagmolen, Slaven-Blair, Smith, Smith, Somala, Son,
  Sorazu, Sorrentino, Souradeep, Sowell, Spencer, Srivastava, Srivastava,
  Staats, Stachie, Standke, Steer, Steinke, Steinlechner, Steinlechner,
  Steinmeyer, Stevenson, Stocks, Stone, Stops, Strain, Stratta, Strigin,
  Strunk, Sturani, Stuver, Sudhir, Summerscales, Sun, Sunil, Suresh, Sutton,
  Swinkels, Szczepa{\'{n}}czyk, Tacca, Tait, Talbot, Talukder, Tanner,
  T{\'{a}}pai, Taracchini, Tasson, Taylor, Thies, Thomas, Thomas, Thondapu,
  Thorne, Thrane, Tiwari, Tiwari, Tiwari, Toland, Tonelli, Tornasi,
  Torres-Forn{\'{e}}, Torrie, T{\"{o}}yr, Travasso, Traylor, Tringali, Trovato,
  Trozzo, Trudeau, Tsang, Tse, Tso, Tsukada, Tsuna, Tuyenbayev, Ueno, Ugolini,
  Unnikrishnan, Urban, Usman, Vahlbruch, Vajente, Valdes, {Van Bakel}, {Van
  Beuzekom}, {Van Den Brand}, {Van Den Broeck}, Vander-Hyde, {Van Heijningen},
  {Van Der Schaaf}, {Van Veggel}, Vardaro, Varma, Vass, Vas{\'{u}}th, Vecchio,
  Vedovato, Veitch, Veitch, Venkateswara, Venugopalan, Verkindt, Vetrano,
  Vicer{\'{e}}, Viets, Vine, Vinet, Vitale, Vo, Vocca, Vorvick, Vyatchanin,
  Wade, Wade, Wade, Walet, Walker, Wallace, Walsh, Wang, Wang, Wang, Wang,
  Wang, Ward, Warden, Warner, Was, Watchi, Weaver, Wei, Weinert, Weinstein,
  Weiss, Wellmann, Wen, Wessel, We{\ss}els, Westhouse, Wette, Whelan, Whiting,
  Whittle, Wilken, Williams, Williamson, Willis, Willke, Wimmer, Winkler, Wipf,
  Wittel, Woan, Woehler, Wofford, Worden, Wright, Wu, Wysocki, Xiao, Yamamoto,
  Yancey, Yang, Yap, Yazback, Yeeles, Yu, Yu, Yuen, Yvert, Zadro{\.{z}}ny,
  Zanolin, Zelenova, Zendri, Zevin, Zhang, Zhang, Zhang, Zhao, Zhou, Zhou, Zhu,
  Zucker, and Zweizig]{NoiseGuide2020}
Abbott, B.P.; Abbott, R.; Abbott, T.D.; Abraham, S.; Acernese, F.; Ackley, K.;
  Adams, C.; Adya, V.B.; Affeldt, C.; Agathos, M.;  et~al.
\newblock {A guide to LIGO–Virgo detector noise and extraction of transient
  gravitational-wave signals}.
\newblock {\em Classical and Quantum Gravity} {\bf 2020}, {\em 37},~055002,
  \href{https://arxiv.org/abs/1908.11170}{{\normalfont [1908.11170]}}.
\newblock
  doi:{\changeurlcolor{black}\href{https://doi.org/10.1088/1361-6382/AB685E}{\detokenize{10.1088/1361-6382/AB685E}}}.

\bibitem[Allen \em{et~al.}(2012)Allen, Anderson, Brady, Brown, and
  Creighton]{Allen2012}
Allen, B.; Anderson, W.G.; Brady, P.R.; Brown, D.A.; Creighton, J.D.E.
\newblock FINDCHIRP: An algorithm for detection of gravitational waves from
  inspiraling compact binaries.
\newblock {\em Phys. Rev. D} {\bf 2012}, {\em 85},~122006.
\newblock
  doi:{\changeurlcolor{black}\href{https://doi.org/10.1103/PhysRevD.85.122006}{\detokenize{10.1103/PhysRevD.85.122006}}}.

\bibitem[Meers(1988)]{Meers1988}
Meers, B.J.
\newblock {Recycling in laser-interferometric gravitational-wave detectors}.
\newblock {\em Physical Review D} {\bf 1988}.
\newblock
  doi:{\changeurlcolor{black}\href{https://doi.org/10.1103/PhysRevD.38.2317}{\detokenize{10.1103/PhysRevD.38.2317}}}.

\bibitem[Strain and Meers(1991)]{Strain1991}
Strain, K.A.; Meers, B.J.
\newblock {Experimental demonstration of dual recycling for interferometric
  gravitational-wave detectors}.
\newblock {\em Physical Review Letters} {\bf 1991}, {\em 66},~1391.
\newblock
  doi:{\changeurlcolor{black}\href{https://doi.org/10.1103/PhysRevLett.66.1391}{\detokenize{10.1103/PhysRevLett.66.1391}}}.

\bibitem[Heinzel \em{et~al.}(1998)Heinzel, Strain, Mizuno, Skeldon, Willke,
  Winkler, Schilling, R{\"{u}}diger, and Danzmann]{Heinzel1998}
Heinzel, G.; Strain, K.A.; Mizuno, J.; Skeldon, K.D.; Willke, B.; Winkler, W.;
  Schilling, R.; R{\"{u}}diger, A.; Danzmann, K.
\newblock {Experimental demonstration of a suspended dual recycling
  interferometer for gravitational wave detection}.
\newblock {\em Phys. Rev. Lett.} {\bf 1998}, {\em 81},~5493--5496.
\newblock
  doi:{\changeurlcolor{black}\href{https://doi.org/10.1103/physrevlett.81.5493}{\detokenize{10.1103/physrevlett.81.5493}}}.

\bibitem[Grote \em{et~al.}(2004)Grote, Freise, Malec, Heinzel, Willke,
  L{\"{u}}ck, Strain, Hough, and Danzmann]{Grote2004}
Grote, H.; Freise, A.; Malec, M.; Heinzel, G.; Willke, B.; L{\"{u}}ck, H.;
  Strain, K.A.; Hough, J.; Danzmann, K.
\newblock {Dual recycling for GEO 600}.
\newblock {\em Classical and Quantum Gravity} {\bf 2004}, {\em 21},~S473,
  \href{https://arxiv.org/abs/0306053}{{\normalfont [arXiv:gr-qc/0306053]}}.
\newblock
  doi:{\changeurlcolor{black}\href{https://doi.org/10.1088/0264-9381/21/5/013}{\detokenize{10.1088/0264-9381/21/5/013}}}.

\bibitem[Fritschel \em{et~al.}(1992)Fritschel, Shoemaker, and
  Weiss]{Fritschel1992}
Fritschel, P.; Shoemaker, D.; Weiss, R.
\newblock {Demonstration of light recycling in a Michelson interferometer with
  Fabry–Perot cavities}.
\newblock {\em Applied Optics, Vol. 31, Issue 10, pp. 1412-1418} {\bf 1992},
  {\em 31},~1412--1418.
\newblock
  doi:{\changeurlcolor{black}\href{https://doi.org/10.1364/AO.31.001412}{\detokenize{10.1364/AO.31.001412}}}.

\bibitem[Ando(1998)]{AndoThesis}
Ando, M.
\newblock Power recycling for an interferometric gravitational wave detector.
\newblock PhD thesis, University of Tokyo,  1998.

\bibitem[Arain and Mueller(2008)]{Arain2008}
Arain, M.A.; Mueller, G.
\newblock {Design of the Advanced LIGO recycling cavities} {\bf 2008}.

\bibitem[Fritschel \em{et~al.}(1998)Fritschel, Mavalvala, Shoemaker, Sigg,
  Zucker, and Gonz{\'{a}}lez]{Fritschel1998}
Fritschel, P.; Mavalvala, N.; Shoemaker, D.; Sigg, D.; Zucker, M.;
  Gonz{\'{a}}lez, G.
\newblock {Alignment of an interferometric gravitational wave detector}.
\newblock {\em Applied Optics, Vol. 37, Issue 28, pp. 6734-6747} {\bf 1998},
  {\em 37},~6734--6747.
\newblock
  doi:{\changeurlcolor{black}\href{https://doi.org/10.1364/AO.37.006734}{\detokenize{10.1364/AO.37.006734}}}.

\bibitem[Caves(1980)]{Caves1980}
Caves, C.M.
\newblock Quantum-Mechanical Radiation-Pressure Fluctuations in an
  Interferometer.
\newblock {\em Phys. Rev. Lett.} {\bf 1980}, {\em 45},~75--79.
\newblock
  doi:{\changeurlcolor{black}\href{https://doi.org/10.1103/PhysRevLett.45.75}{\detokenize{10.1103/PhysRevLett.45.75}}}.

\bibitem[Caves(1981)]{Caves1981}
Caves, C.M.
\newblock Quantum-mechanical noise in an interferometer.
\newblock {\em Phys. Rev. D} {\bf 1981}, {\em 23},~1693--1708.
\newblock
  doi:{\changeurlcolor{black}\href{https://doi.org/10.1103/PhysRevD.23.1693}{\detokenize{10.1103/PhysRevD.23.1693}}}.

\bibitem[Tse \em{et~al.}(2019)Tse, Yu, Kijbunchoo, Fernandez-Galiana, Dupej,
  Barsotti, Blair, Brown, Dwyer, Effler, Evans, Fritschel, Frolov, Green,
  Mansell, Matichard, Mavalvala, McClelland, McCuller, McRae, Miller, Mullavey,
  Oelker, Phinney, Sigg, Slagmolen, Vo, Ward, Whittle, Abbott, Adams, Adhikari,
  Ananyeva, Appert, Arai, Areeda, Asali, Aston, Austin, Baer, Ball, Ballmer,
  Banagiri, Barker, Bartlett, Berger, Betzwieser, Bhattacharjee, Billingsley,
  Biscans, Blair, Bode, Booker, Bork, Bramley, Brooks, Buikema, Cahillane,
  Cannon, Chen, Ciobanu, Clara, Cooper, Corley, Countryman, Covas, Coyne,
  Datrier, Davis, Fronzo, Driggers, Etzel, Evans, Feicht, Fulda, Fyffe, Giaime,
  Giardina, Godwin, Goetz, Gras, Gray, Gray, Gupta, Gustafson, Gustafson,
  Hanks, Hanson, Hardwick, Hasskew, Heintze, Helmling-Cornell, Holland, Jones,
  Kandhasamy, Karki, Kasprzack, Kawabe, King, Kissel, Kumar, Landry, Lane,
  Lantz, Laxen, Lecoeuche, Leviton, Liu, Lormand, Lundgren, Macas, MacInnis,
  Macleod, M{\'{a}}rka, M{\'{a}}rka, Martynov, Mason, Massinger, McCarthy,
  McCormick, McIver, Mendell, Merfeld, Merilh, Meylahn, Mistry, Mittleman,
  Moreno, Mow-Lowry, Mozzon, Nelson, Nguyen, Nuttall, Oberling, Oram, O'Reilly,
  Osthelder, Ottaway, Overmier, Palamos, Parker, Payne, Pele, Perez, Pirello,
  Radkins, Ramirez, Richardson, Riles, Robertson, Rollins, Romel, Romie, Ross,
  Ryan, Sadecki, Sanchez, Sanchez, Saravanan, Savage, Schaetzl, Schnabel,
  Schofield, Schwartz, Sellers, Shaffer, Smith, Soni, Sorazu, Spencer, Strain,
  Sun, Szczepa{\'{n}}czyk, Thomas, Thomas, Thorne, Toland, Torrie, Traylor,
  Urban, Vajente, Valdes, Vander-Hyde, Veitch, Venkateswara, Venugopalan,
  Viets, Vorvick, Wade, Warner, Weaver, Weiss, Willke, Wipf, Xiao, Yamamoto,
  Yap, Yu, Zhang, Zucker, and Zweizig]{Tse2019}
Tse, M.; Yu, H.; Kijbunchoo, N.; Fernandez-Galiana, A.; Dupej, P.; Barsotti,
  L.; Blair, C.; Brown, D.; Dwyer, S.; Effler, A.;  et~al.
\newblock Quantum-Enhanced Advanced {LIGO} Detectors in the Era of
  Gravitational-Wave Astronomy.
\newblock {\em Physical Review Letters} {\bf 2019}, {\em 123}.
\newblock
  doi:{\changeurlcolor{black}\href{https://doi.org/10.1103/physrevlett.123.231107}{\detokenize{10.1103/physrevlett.123.231107}}}.

\bibitem[Oelker \em{et~al.}(2016)Oelker, Mansell, Tse, Miller, Matichard,
  Barsotti, Fritschel, McClelland, Evans, and Mavalvala]{Oelker2016a}
Oelker, E.; Mansell, G.; Tse, M.; Miller, J.; Matichard, F.; Barsotti, L.;
  Fritschel, P.; McClelland, D.E.; Evans, M.; Mavalvala, N.
\newblock Ultra-low phase noise squeezed vacuum source for gravitational wave
  detectors.
\newblock {\em Optica} {\bf 2016}, {\em 3},~682--685.
\newblock
  doi:{\changeurlcolor{black}\href{https://doi.org/10.1364/OPTICA.3.000682}{\detokenize{10.1364/OPTICA.3.000682}}}.

\bibitem[Yu \em{et~al.}(2020)Yu, McCuller, Tse, Kijbunchoo, Barsotti, and
  Mavalvala]{Yu2020}
Yu, H.; McCuller, L.; Tse, M.; Kijbunchoo, N.; Barsotti, L.; Mavalvala, N.
\newblock Quantum correlations between light and the kilogram-mass mirrors of
  {LIGO}.
\newblock {\em Nature} {\bf 2020}, {\em 583},~43--47.
\newblock
  doi:{\changeurlcolor{black}\href{https://doi.org/10.1038/s41586-020-2420-8}{\detokenize{10.1038/s41586-020-2420-8}}}.

\bibitem[McCuller \em{et~al.}(2021)McCuller, Dwyer, Green, Yu, Kuns, Barsotti,
  Blair, Brown, Effler, Evans, Fernandez-Galiana, Fritschel, Frolov,
  Kijbunchoo, Mansell, Matichard, Mavalvala, McClelland, McRae, Mullavey, Sigg,
  Slagmolen, Tse, Vo, Ward, Whittle, Abbott, Adams, Adhikari, Ananyeva, Appert,
  Arai, Areeda, Asali, Aston, Austin, Baer, Ball, Ballmer, Banagiri, Barker,
  Bartlett, Berger, Betzwieser, Bhattacharjee, Billingsley, Biscans, Blair,
  Bode, Booker, Bork, Bramley, Brooks, Buikema, Cahillane, Cannon, Chen,
  Ciobanu, Clara, Compton, Cooper, Corley, Countryman, Covas, Coyne, Datrier,
  Davis, Di~Fronzo, Dooley, Driggers, Etzel, Evans, Feicht, Fulda, Fyffe,
  Giaime, Giardina, Godwin, Goetz, Gras, Gray, Gray, Gustafson, Gustafson,
  Hanks, Hanson, Hardwick, Hasskew, Heintze, Helmling-Cornell, Holland, Jones,
  Kandhasamy, Karki, Kasprzack, Kawabe, King, Kissel, Kumar, Landry, Lane,
  Lantz, Laxen, Lecoeuche, Leviton, Liu, Lormand, Lundgren, Macas, MacInnis,
  Macleod, M\'arka, M\'arka, Martynov, Mason, Massinger, McCarthy, McCormick,
  McIver, Mendell, Merfeld, Merilh, Meylahn, Mistry, Mittleman, Moreno,
  Mow-Lowry, Mozzon, Nelson, Nguyen, Nuttall, Oberling, Oram, Osthelder,
  Ottaway, Overmier, Palamos, Parker, Payne, Pele, Penhorwood, Perez, Pirello,
  Radkins, Ramirez, Richardson, Riles, Robertson, Rollins, Romel, Romie, Ross,
  Ryan, Sadecki, Sanchez, Sanchez, Saravanan, Savage, Schaetzl, Schnabel,
  Schofield, Schwartz, Sellers, Shaffer, Smith, Soni, Sorazu, Spencer, Strain,
  Sun, Szczepa\ifmmode~\acute{n}\else \'{n}\fi{}czyk, Thomas, Thomas, Thorne,
  Toland, Torrie, Traylor, Urban, Vajente, Valdes, Vander-Hyde, Veitch,
  Venkateswara, Venugopalan, Viets, Vorvick, Wade, Warner, Weaver, Weiss,
  Willke, Wipf, Xiao, Yamamoto, Yu, Zhang, Zucker, and Zweizig]{McCuller2021}
McCuller, L.; Dwyer, S.E.; Green, A.C.; Yu, H.; Kuns, K.; Barsotti, L.; Blair,
  C.D.; Brown, D.D.; Effler, A.; Evans, M.;  et~al.
\newblock LIGO's quantum response to squeezed states.
\newblock {\em Phys. Rev. D} {\bf 2021}, {\em 104},~062006.
\newblock
  doi:{\changeurlcolor{black}\href{https://doi.org/10.1103/PhysRevD.104.062006}{\detokenize{10.1103/PhysRevD.104.062006}}}.

\bibitem[Cahillane \em{et~al.}(2021)Cahillane, Sigg, Mansell, and
  Sigg]{Cahillane2021}
Cahillane, C.; Sigg, D.; Mansell, G.L.; Sigg, D.
\newblock {Laser frequency noise in next generation gravitational-wave
  detectors}.
\newblock {\em Optics Express, Vol. 29, Issue 25, pp. 42144-42161} {\bf 2021},
  {\em 29},~42144--42161,  \href{https://arxiv.org/abs/2107.14349}{{\normalfont
  [2107.14349]}}.
\newblock
  doi:{\changeurlcolor{black}\href{https://doi.org/10.1364/OE.439253}{\detokenize{10.1364/OE.439253}}}.

\bibitem[Abbott \em{et~al.}(2017)Abbott et~al.]{GW150914CalPaper}
Abbott, B.P.;  et~al.
\newblock Calibration of the Advanced LIGO detectors for the discovery of the
  binary black-hole merger GW150914.
\newblock {\em Phys. Rev. D} {\bf 2017}, {\em 95},~062003.
\newblock
  doi:{\changeurlcolor{black}\href{https://doi.org/10.1103/PhysRevD.95.062003}{\detokenize{10.1103/PhysRevD.95.062003}}}.

\bibitem[Cahillane \em{et~al.}(2017)Cahillane, Betzwieser, Brown, Goetz, Hall,
  Izumi, Kandhasamy, Karki, Kissel, Mendell, Savage, Tuyenbayev, Urban, Viets,
  Wade, and Weinstein]{CalUncPaper}
Cahillane, C.; Betzwieser, J.; Brown, D.A.; Goetz, E.; Hall, E.D.; Izumi, K.;
  Kandhasamy, S.; Karki, S.; Kissel, J.S.; Mendell, G.;  et~al.
\newblock Calibration uncertainty for Advanced LIGO's first and second
  observing runs.
\newblock {\em Phys. Rev. D} {\bf 2017}, {\em 96},~102001.
\newblock
  doi:{\changeurlcolor{black}\href{https://doi.org/10.1103/PhysRevD.96.102001}{\detokenize{10.1103/PhysRevD.96.102001}}}.

\bibitem[Sun \em{et~al.}(2020)Sun, Goetz, Kissel, Betzwieser, Karki, Viets,
  Wade, Bhattacharjee, Bossilkov, Covas, Datrier, Gray, Kandhasamy, Lecoeuche,
  Mendell, Mistry, Payne, Savage, Weinstein, Aston, Buikema, Cahillane,
  Driggers, Dwyer, Kumar, and Urban]{Sun2020}
Sun, L.; Goetz, E.; Kissel, J.S.; Betzwieser, J.; Karki, S.; Viets, A.; Wade,
  M.; Bhattacharjee, D.; Bossilkov, V.; Covas, P.B.;  et~al.
\newblock {Characterization of systematic error in Advanced LIGO calibration}.
\newblock {\em Classical and Quantum Gravity} {\bf 2020}.
\newblock
  doi:{\changeurlcolor{black}\href{https://doi.org/10.1088/1361-6382/abb14e}{\detokenize{10.1088/1361-6382/abb14e}}}.

\bibitem[Sun \em{et~al.}(2021)Sun, Goetz, Kissel, Betzwieser, Karki,
  Bhattacharjee, Covas, Datrier, Kandhasamy, Lecoeuche, Mendell, Mistry, Payne,
  Savage, Viets, Wade, Weinstein, Aston, Cahillane, Driggers, Dwyer, and
  Urban]{Sun2021}
Sun, L.; Goetz, E.; Kissel, J.S.; Betzwieser, J.; Karki, S.; Bhattacharjee, D.;
  Covas, P.B.; Datrier, L.E.H.; Kandhasamy, S.; Lecoeuche, Y.K.;  et~al.
\newblock Characterization of systematic error in Advanced LIGO calibration in
  the second half of O3,  2021,
  \href{https://arxiv.org/abs/2107.00129}{{\normalfont
  [arXiv:astro-ph.IM/2107.00129]}}.

\bibitem[Lindblom(2009)]{Lindblom2009}
Lindblom, L.
\newblock Optimal calibration accuracy for gravitational-wave detectors.
\newblock {\em Physical Review D} {\bf 2009}, {\em 80}.
\newblock
  doi:{\changeurlcolor{black}\href{https://doi.org/10.1103/physrevd.80.042005}{\detokenize{10.1103/physrevd.80.042005}}}.

\bibitem[Tuyenbayev \em{et~al.}(2017)Tuyenbayev et~al.]{CALTimeDependence}
Tuyenbayev, D.;  et~al.
\newblock Improving LIGO calibration accuracy by tracking and compensating for
  slow temporal variations.
\newblock {\em Classical and Quantum Gravity} {\bf 2017}, {\em 34},~015002.

\bibitem[Vitale \em{et~al.}(2021)Vitale, Haster, Sun, Farr, Goetz, Kissel, and
  Cahillane]{Vitale2021}
Vitale, S.; Haster, C.J.; Sun, L.; Farr, B.; Goetz, E.; Kissel, J.; Cahillane,
  C.
\newblock Physical approach to the marginalization of LIGO calibration
  uncertainties.
\newblock {\em Phys. Rev. D} {\bf 2021}, {\em 103},~063016.
\newblock
  doi:{\changeurlcolor{black}\href{https://doi.org/10.1103/PhysRevD.103.063016}{\detokenize{10.1103/PhysRevD.103.063016}}}.

\bibitem[Karki \em{et~al.}(2016)Karki et~al.]{aLIGOPCALPaper}
Karki, S.;  et~al.
\newblock The Advanced LIGO photon calibrators.
\newblock {\em Rev. Sci. Instrum.} {\bf 2016}, {\em 87},~114503--114503.

\bibitem[Bhattacharjee \em{et~al.}(2020)Bhattacharjee, Lecoeuche, Karki,
  Betzwieser, Bossilkov, Kandhasamy, Payne, and Savage]{Bhattacharjee2020}
Bhattacharjee, D.; Lecoeuche, Y.; Karki, S.; Betzwieser, J.; Bossilkov, V.;
  Kandhasamy, S.; Payne, E.; Savage, R.L.
\newblock Fiducial displacements with improved accuracy for the global network
  of gravitational wave detectors.
\newblock {\em Classical and Quantum Gravity} {\bf 2020}, {\em 38},~015009.
\newblock
  doi:{\changeurlcolor{black}\href{https://doi.org/10.1088/1361-6382/aba9ed}{\detokenize{10.1088/1361-6382/aba9ed}}}.

\bibitem[Estevez \em{et~al.}(2018)Estevez, Lieunard, Marion, Mours, Rolland,
  and Verkindt]{Estevez2018}
Estevez, D.; Lieunard, B.; Marion, F.; Mours, B.; Rolland, L.; Verkindt, D.
\newblock First tests of a Newtonian calibrator on an interferometric
  gravitational wave detector.
\newblock {\em Classical and Quantum Gravity} {\bf 2018}, {\em 35},~235009.
\newblock
  doi:{\changeurlcolor{black}\href{https://doi.org/10.1088/1361-6382/aae95f}{\detokenize{10.1088/1361-6382/aae95f}}}.

\bibitem[Inoue \em{et~al.}(2018)Inoue, Haino, Kanda, Ogawa, Suzuki, Tomaru,
  Yamanmoto, and Yokozawa]{Inoue2018}
Inoue, Y.; Haino, S.; Kanda, N.; Ogawa, Y.; Suzuki, T.; Tomaru, T.; Yamanmoto,
  T.; Yokozawa, T.
\newblock Improving the absolute accuracy of the gravitational wave detectors
  by combining the photon pressure and gravity field calibrators.
\newblock {\em Physical Review D} {\bf 2018}, {\em 98}.
\newblock
  doi:{\changeurlcolor{black}\href{https://doi.org/10.1103/physrevd.98.022005}{\detokenize{10.1103/physrevd.98.022005}}}.

\bibitem[Schreiner \em{et~al.}(2021)Schreiner, Z{\"{u}}nd, G{\"{u}}nter, al,
  Kraft, Estevez, Mours, and Pradier]{Schreiner2021}
Schreiner, D.; Z{\"{u}}nd, T.; G{\"{u}}nter, F.J.; al.; Kraft, L.; Estevez, D.;
  Mours, B.; Pradier, T.
\newblock {Newtonian calibrator tests during the Virgo O3 data taking}.
\newblock {\em Classical and Quantum Gravity} {\bf 2021}, {\em 38},~075012.
\newblock
  doi:{\changeurlcolor{black}\href{https://doi.org/10.1088/1361-6382/ABE2DA}{\detokenize{10.1088/1361-6382/ABE2DA}}}.

\bibitem[Ross \em{et~al.}(2021)Ross, Mistry, Datrier, Kissel, Venkateswara,
  Weller, Kumar, Hagedorn, Adelberger, Lee, Shaw, Thomas, Barker, Clara,
  Gateley, Guidry, Daw, Hendry, and Gundlach]{Ross2021}
Ross, M.P.; Mistry, T.; Datrier, L.; Kissel, J.; Venkateswara, K.; Weller, C.;
  Kumar, K.; Hagedorn, C.; Adelberger, E.; Lee, J.;  et~al.
\newblock Initial Results from the LIGO Newtonian Calibrator,  2021,
  \href{https://arxiv.org/abs/2107.00141}{{\normalfont
  [arXiv:gr-qc/2107.00141]}}.

\bibitem[Kane and Byer(1985)]{Kane1985}
Kane, T.J.; Byer, R.L.
\newblock Monolithic, unidirectional single-mode Nd:YAG ring laser.
\newblock {\em Opt. Lett.} {\bf 1985}, {\em 10},~65--67.
\newblock
  doi:{\changeurlcolor{black}\href{https://doi.org/10.1364/OL.10.000065}{\detokenize{10.1364/OL.10.000065}}}.

\bibitem[Korth(2019)]{KorthThesis}
Korth, Z.
\newblock PhD thesis, California Institute of Technology,  2019.
\newblock
  doi:{\changeurlcolor{black}\href{https://doi.org/10.7907/4H7V-W213}{\detokenize{10.7907/4H7V-W213}}}.

\bibitem[Hoak(2015)]{HoakThesis}
Hoak, D.
\newblock PhD thesis, University of Massachusetts Amherst,  2015.
\newblock
  doi:{\changeurlcolor{black}\href{https://doi.org/10.7275/7510756.0}{\detokenize{10.7275/7510756.0}}}.

\bibitem[Venugopalan(2021)]{VenugopalanThesis}
Venugopalan, G.
\newblock PhD thesis, California Institute of Technology,  2021.
\newblock
  doi:{\changeurlcolor{black}\href{https://doi.org/10.7907/ttwp-1h12}{\detokenize{10.7907/ttwp-1h12}}}.

\bibitem[Hild \em{et~al.}(2009)Hild, Grote, Degallaix, Chelkowski, Danzmann,
  Freise, Hewitson, Hough, Lück, Prijatelj, Strain, Smith, and
  Willke]{Hild2009}
Hild, S.; Grote, H.; Degallaix, J.; Chelkowski, S.; Danzmann, K.; Freise, A.;
  Hewitson, M.; Hough, J.; Lück, H.; Prijatelj, M.;  et~al.
\newblock {DC}-readout of a signal-recycled gravitational wave detector.
\newblock {\em Classical and Quantum Gravity} {\bf 2009}, {\em 26},~055012.
\newblock
  doi:{\changeurlcolor{black}\href{https://doi.org/10.1088/0264-9381/26/5/055012}{\detokenize{10.1088/0264-9381/26/5/055012}}}.

\bibitem[Martynov(2015)]{MartynovThesis}
Martynov, D.V.
\newblock Lock Acquisition and Sensitivity Analysis of Advanced LIGO
  Interferometers.
\newblock PhD thesis,  2015.
\newblock
  doi:{\changeurlcolor{black}\href{https://doi.org/10.7907/Z9Q81B1F}{\detokenize{10.7907/Z9Q81B1F}}}.

\bibitem[Staley(2015)]{StaleyThesis}
Staley, A.N.
\newblock PhD thesis, Columbia University,  2015.
\newblock
  doi:{\changeurlcolor{black}\href{https://doi.org/https://doi.org/10.7916/D8X34WQ4}{\detokenize{https://doi.org/10.7916/D8X34WQ4}}}.

\bibitem[Staley \em{et~al.}(2014)Staley, Martynov, Abbott, Adhikari, Arai,
  Ballmer, Barsotti, Brooks, Derosa, Dwyer, Effler, Evans, Fritschel, Frolov,
  Gray, Guido, Gustafson, Heintze, Hoak, Izumi, Kawabe, King, Kissel, Kokeyama,
  Landry, McClelland, Miller, Mullavey, Oreilly, Rollins, Sanders, Schofield,
  Sigg, Slagmolen, Smith-Lefebvre, Vajente, Ward, and Wipf]{Staley2014}
Staley, A.; Martynov, D.; Abbott, R.; Adhikari, R.X.; Arai, K.; Ballmer, S.;
  Barsotti, L.; Brooks, A.F.; Derosa, R.T.; Dwyer, S.;  et~al.
\newblock {Achieving resonance in the Advanced LIGO gravitational-wave
  interferometer}.
\newblock {\em Classical and Quantum Gravity} {\bf 2014}.
\newblock
  doi:{\changeurlcolor{black}\href{https://doi.org/10.1088/0264-9381/31/24/245010}{\detokenize{10.1088/0264-9381/31/24/245010}}}.

\bibitem[Izumi \em{et~al.}(2012)Izumi, Arai, Barr, Betzwieser, Brooks, Dahl,
  Doravari, Driggers, Korth, Miao, Rollins, Vass, Yeaton-Massey, and
  Adhikari]{Izumi2012}
Izumi, K.; Arai, K.; Barr, B.; Betzwieser, J.; Brooks, A.; Dahl, K.; Doravari,
  S.; Driggers, J.C.; Korth, W.Z.; Miao, H.;  et~al.
\newblock {Multicolor cavity metrology}.
\newblock {\em Journal of the Optical Society of America A} {\bf 2012}.
\newblock
  doi:{\changeurlcolor{black}\href{https://doi.org/10.1364/josaa.29.002092}{\detokenize{10.1364/josaa.29.002092}}}.

\bibitem[Mullavey \em{et~al.}(2012)Mullavey, Slagmolen, Miller, Evans,
  Fritschel, Sigg, Waldman, Shaddock, and McClelland]{Mullavey2012}
Mullavey, A.J.; Slagmolen, B.J.J.; Miller, J.; Evans, M.; Fritschel, P.; Sigg,
  D.; Waldman, S.J.; Shaddock, D.A.; McClelland, D.E.
\newblock {Arm-length stabilisation for interferometric gravitational-wave
  detectors using frequency-doubled auxiliary lasers}.
\newblock {\em Optics Express} {\bf 2012},
  \href{https://arxiv.org/abs/1112.3118}{{\normalfont [1112.3118]}}.
\newblock
  doi:{\changeurlcolor{black}\href{https://doi.org/10.1364/oe.20.000081}{\detokenize{10.1364/oe.20.000081}}}.

\bibitem[Arai \em{et~al.}(2000)Arai, Ando, Moriwaki, Kawabe, and
  Tsubono]{Arai3f}
Arai, K.; Ando, M.; Moriwaki, S.; Kawabe, K.; Tsubono, K.
\newblock New signal extraction scheme with harmonic demodulation for
  power-recycled {F}abry–{P}erot–{M}ichelson interferometers.
\newblock {\em Physics Letters A} {\bf 2000}, {\em 273},~15--24.
\newblock
  doi:{\changeurlcolor{black}\href{https://doi.org/https://doi.org/10.1016/S0375-9601(00)00467-9}{\detokenize{https://doi.org/10.1016/S0375-9601(00)00467-9}}}.

\bibitem[Cahillane and Sigg()]{alog43119}
Cahillane, C.; Sigg, D.
\newblock Out of loop ALS COMM frequency measurement via IR.
\newblock
  \href{https://alog.ligo-wa.caltech.edu/aLOG/index.php?callRep=43119}{LHO alog
  43119}.

\bibitem[Cahillane \em{et~al.}()Cahillane, Dwyer, and Sigg]{alog43214}
Cahillane, C.; Dwyer, S.; Sigg, D.
\newblock Out of Loop ALS COMM frequency measurement - Take Two.
\newblock
  \href{https://alog.ligo-wa.caltech.edu/aLOG/index.php?callRep=43214}{LHO alog
  43214}.

\bibitem[Vajente(2014)]{Vajente2014}
Vajente, G.
\newblock {In situ correction of mirror surface to reduce round-trip losses in
  Fabry-Perot cavities}.
\newblock {\em Applied Optics, Vol. 53, Issue 7, pp. 1459-1465} {\bf 2014},
  {\em 53},~1459--1465.
\newblock
  doi:{\changeurlcolor{black}\href{https://doi.org/10.1364/AO.53.001459}{\detokenize{10.1364/AO.53.001459}}}.

\bibitem[Brooks \em{et~al.}(2016)Brooks, Heptonstall, Lynch, Cole, Abbott,
  Vorvick, Guido, Ottaway, King, Gustafson, Ciani, Grabeel, Mueller, Munch,
  Mailand, Heintze, Smith, Jacobson, Arain, Veitch, Willems, O'Connor, Vo,
  Sannibale, Kim, and Shao]{Brooks2016}
Brooks, A.F.; Heptonstall, A.; Lynch, A.; Cole, A.; Abbott, B.; Vorvick, C.;
  Guido, C.; Ottaway, D.; King, E.; Gustafson, E.;  et~al.
\newblock {Overview of Advanced LIGO adaptive optics}.
\newblock {\em Applied Optics, Vol. 55, Issue 29, pp. 8256-8265} {\bf 2016},
  {\em 55},~8256--8265,  \href{https://arxiv.org/abs/1608.02934}{{\normalfont
  [1608.02934]}}.
\newblock
  doi:{\changeurlcolor{black}\href{https://doi.org/10.1364/AO.55.008256}{\detokenize{10.1364/AO.55.008256}}}.

\bibitem[Weiss(1972)]{Weiss1972}
Weiss, R.
\newblock Electronically Coupled Broadband Gravitational Antenna.
\newblock {\em Quarterly Progress Report, Research Laboratory of Electronics
  (MIT)} {\bf 1972}, {\em 105}.

\bibitem[Barsotti \em{et~al.}(2018)Barsotti, Fritschel, Evans, and
  Gras]{aligoDesignSensitivity2018}
Barsotti, L.; Fritschel, P.; Evans, M.; Gras, S.
\newblock Updated Advanced LIGO sensitivity design curve.
\newblock {\em Tech. rep. LIGO-T1800044} {\bf 2018}.

\bibitem[Braginsky \em{et~al.}(1992)Braginsky, Khalili, and
  Thorne]{braginsky_khalili_thorne_1992}
Braginsky, V.B.; Khalili, F.Y.; Thorne, K.S.
\newblock {\em Quantum Measurement}; Cambridge University Press,  1992.
\newblock
  doi:{\changeurlcolor{black}\href{https://doi.org/10.1017/CBO9780511622748}{\detokenize{10.1017/CBO9780511622748}}}.

\bibitem[Braginsky and Vyatchanin(2003)]{Braginsky2003}
Braginsky, V.; Vyatchanin, S.
\newblock Thermodynamical fluctuations in optical mirror coatings.
\newblock {\em Physics Letters A} {\bf 2003}, {\em 312},~244 -- 255.
\newblock
  doi:{\changeurlcolor{black}\href{https://doi.org/https://doi.org/10.1016/S0375-9601(03)00473-0}{\detokenize{https://doi.org/10.1016/S0375-9601(03)00473-0}}}.

\bibitem[Levin(1998)]{Levin1998}
Levin, Y.
\newblock Internal thermal noise in the LIGO test masses: A direct approach.
\newblock {\em Phys. Rev. D} {\bf 1998}, {\em 57},~659--663.
\newblock
  doi:{\changeurlcolor{black}\href{https://doi.org/10.1103/PhysRevD.57.659}{\detokenize{10.1103/PhysRevD.57.659}}}.

\bibitem[Hong \em{et~al.}(2013)Hong, Yang, Gustafson, Adhikari, and
  Chen]{Hong2013}
Hong, T.; Yang, H.; Gustafson, E.K.; Adhikari, R.X.; Chen, Y.
\newblock Brownian thermal noise in multilayer coated mirrors.
\newblock {\em Phys. Rev. D} {\bf 2013}, {\em 87},~082001.
\newblock
  doi:{\changeurlcolor{black}\href{https://doi.org/10.1103/PhysRevD.87.082001}{\detokenize{10.1103/PhysRevD.87.082001}}}.

\bibitem[Yam \em{et~al.}(2015)Yam, Gras, and Evans]{Yam2015}
Yam, W.; Gras, S.; Evans, M.
\newblock Multimaterial coatings with reduced thermal noise.
\newblock {\em Phys. Rev. D} {\bf 2015}, {\em 91},~042002.
\newblock
  doi:{\changeurlcolor{black}\href{https://doi.org/10.1103/PhysRevD.91.042002}{\detokenize{10.1103/PhysRevD.91.042002}}}.

\bibitem[Callen and Greene(1952)]{Callen1952}
Callen, H.; Greene, R.
\newblock On a Theorem of Irreversible Thermodynamics.
\newblock {\em Phys. Rev.} {\bf 1952}, {\em 86},~702--710.
\newblock
  doi:{\changeurlcolor{black}\href{https://doi.org/10.1103/PhysRev.86.702}{\detokenize{10.1103/PhysRev.86.702}}}.

\bibitem[Kubo(1966)]{Kubo1966}
Kubo, R.
\newblock The fluctuation-dissipation theorem.
\newblock {\em Reports on Progress in Physics} {\bf 1966}, {\em 29},~255--284.
\newblock
  doi:{\changeurlcolor{black}\href{https://doi.org/10.1088/0034-4885/29/1/306}{\detokenize{10.1088/0034-4885/29/1/306}}}.

\bibitem[Saulson(1990)]{Saulson1990}
Saulson, P.R.
\newblock Thermal noise in mechanical experiments.
\newblock {\em Phys. Rev. D} {\bf 1990}, {\em 42},~2437--2445.
\newblock
  doi:{\changeurlcolor{black}\href{https://doi.org/10.1103/PhysRevD.42.2437}{\detokenize{10.1103/PhysRevD.42.2437}}}.

\bibitem[Nakagawa \em{et~al.}(2002)Nakagawa, Gretarsson, Gustafson, and
  Fejer]{Nakagawa2002}
Nakagawa, N.; Gretarsson, A.M.; Gustafson, E.K.; Fejer, M.M.
\newblock Thermal noise in half-infinite mirrors with nonuniform loss: A slab
  of excess loss in a half-infinite mirror.
\newblock {\em Phys. Rev. D} {\bf 2002}, {\em 65},~102001.
\newblock
  doi:{\changeurlcolor{black}\href{https://doi.org/10.1103/PhysRevD.65.102001}{\detokenize{10.1103/PhysRevD.65.102001}}}.

\bibitem[Chalermsongsak \em{et~al.}(2014)Chalermsongsak, Seifert, Hall, Arai,
  Gustafson, and Adhikari]{Chalermsongsak2014}
Chalermsongsak, T.; Seifert, F.; Hall, E.D.; Arai, K.; Gustafson, E.K.;
  Adhikari, R.X.
\newblock Broadband measurement of coating thermal noise in rigid
  Fabry{\textendash}P{\'{e}}rot cavities.
\newblock {\em Metrologia} {\bf 2014}, {\em 52},~17--30.
\newblock
  doi:{\changeurlcolor{black}\href{https://doi.org/10.1088/0026-1394/52/1/17}{\detokenize{10.1088/0026-1394/52/1/17}}}.

\bibitem[Gras \em{et~al.}(2017)Gras, Yu, Yam, Martynov, and Evans]{Gras2017}
Gras, S.; Yu, H.; Yam, W.; Martynov, D.; Evans, M.
\newblock {Audio-band coating thermal noise measurement for Advanced LIGO with
  a multimode optical resonator}.
\newblock {\em Physical Review D} {\bf 2017},
  \href{https://arxiv.org/abs/1609.05595}{{\normalfont [1609.05595]}}.
\newblock
  doi:{\changeurlcolor{black}\href{https://doi.org/10.1103/PhysRevD.95.022001}{\detokenize{10.1103/PhysRevD.95.022001}}}.

\bibitem[Gras and Evans(2018)]{Gras2018}
Gras, S.; Evans, M.
\newblock Direct measurement of coating thermal noise in optical resonators.
\newblock {\em Physical Review D} {\bf 2018}, {\em 98}.
\newblock
  doi:{\changeurlcolor{black}\href{https://doi.org/10.1103/physrevd.98.122001}{\detokenize{10.1103/physrevd.98.122001}}}.

\bibitem[Cagnoli \em{et~al.}(2000)Cagnoli, Hough, DeBra, Fejer, Gustafson,
  Rowan, and Mitrofanov]{Cagnoli2000}
Cagnoli, G.; Hough, J.; DeBra, D.; Fejer, M.M.; Gustafson, E.; Rowan, S.;
  Mitrofanov, V.
\newblock {Damping dilution factor for a pendulum in an interferometric
  gravitational waves detector}.
\newblock {\em Physics Letters A} {\bf 2000}, {\em 272},~39--45.
\newblock
  doi:{\changeurlcolor{black}\href{https://doi.org/10.1016/S0375-9601(00)00411-4}{\detokenize{10.1016/S0375-9601(00)00411-4}}}.

\bibitem[Saulson(1984)]{Saulson1984}
Saulson, P.R.
\newblock Terrestrial gravitational noise on a gravitational wave antenna.
\newblock {\em Phys. Rev. D} {\bf 1984}, {\em 30},~732--736.
\newblock
  doi:{\changeurlcolor{black}\href{https://doi.org/10.1103/PhysRevD.30.732}{\detokenize{10.1103/PhysRevD.30.732}}}.

\bibitem[Hughes and Thorne(1998)]{Hughes1998}
Hughes, S.A.; Thorne, K.S.
\newblock Seismic gravity-gradient noise in interferometric gravitational-wave
  detectors.
\newblock {\em Phys. Rev. D} {\bf 1998}, {\em 58},~122002.
\newblock
  doi:{\changeurlcolor{black}\href{https://doi.org/10.1103/PhysRevD.58.122002}{\detokenize{10.1103/PhysRevD.58.122002}}}.

\bibitem[Harms(2015)]{Harms2015}
Harms, J.
\newblock Terrestrial Gravity Fluctuations.
\newblock {\em Living Reviews in Relativity} {\bf 2015}, {\em 18},~3.
\newblock
  doi:{\changeurlcolor{black}\href{https://doi.org/10.1007/lrr-2015-3}{\detokenize{10.1007/lrr-2015-3}}}.

\bibitem[Driggers \em{et~al.}(2012)Driggers, Harms, and
  Adhikari]{Driggers2012newtonian}
Driggers, J.C.; Harms, J.; Adhikari, R.X.
\newblock Subtraction of {N}ewtonian noise using optimized sensor arrays.
\newblock {\em Phys. Rev. D} {\bf 2012}, {\em 86},~102001.
\newblock
  doi:{\changeurlcolor{black}\href{https://doi.org/10.1103/PhysRevD.86.102001}{\detokenize{10.1103/PhysRevD.86.102001}}}.

\bibitem[Coughlin \em{et~al.}(2016)Coughlin, Mukund, Harms, Driggers, Adhikari,
  and Mitra]{Coughlin2016}
Coughlin, M.; Mukund, N.; Harms, J.; Driggers, J.; Adhikari, R.; Mitra, S.
\newblock Towards a first design of a {N}ewtonian-noise cancellation system for
  {A}dvanced {LIGO}.
\newblock {\em Classical and Quantum Gravity} {\bf 2016}, {\em 33},~244001.

\bibitem[Coughlin \em{et~al.}(2018)Coughlin, Harms, Driggers, McManus, Mukund,
  Ross, Slagmolen, and Venkateswara]{Coughlin2018}
Coughlin, M.W.; Harms, J.; Driggers, J.; McManus, D.J.; Mukund, N.; Ross, M.P.;
  Slagmolen, B.J.J.; Venkateswara, K.
\newblock Implications of Dedicated Seismometer Measurements on
  {N}ewtonian-Noise Cancellation for {A}dvanced {LIGO}.
\newblock {\em Phys. Rev. Lett.} {\bf 2018}, {\em 121},~221104.
\newblock
  doi:{\changeurlcolor{black}\href{https://doi.org/10.1103/PhysRevLett.121.221104}{\detokenize{10.1103/PhysRevLett.121.221104}}}.

\bibitem[Harms \em{et~al.}(2020)Harms, Bonilla, Coughlin, Driggers, Dwyer,
  McManus, Ross, Slagmolen, and Venkateswara]{PhysRevD.101.102002}
Harms, J.; Bonilla, E.L.; Coughlin, M.W.; Driggers, J.; Dwyer, S.E.; McManus,
  D.J.; Ross, M.P.; Slagmolen, B.J.J.; Venkateswara, K.
\newblock Observation of a potential future sensitivity limitation from ground
  motion at {LIGO} {H}anford.
\newblock {\em Phys. Rev. D} {\bf 2020}, {\em 101},~102002.
\newblock
  doi:{\changeurlcolor{black}\href{https://doi.org/10.1103/PhysRevD.101.102002}{\detokenize{10.1103/PhysRevD.101.102002}}}.

\bibitem[Driggers(2016)]{Driggers2016}
Driggers, J.
\newblock Length feedforward calculations.
\newblock Technical report,  2016.

\bibitem[Driggers \em{et~al.}(2019)Driggers, Vitale, Lundgren, Evans, Kawabe,
  Dwyer, Izumi, Schofield, Effler, Sigg, Fritschel, Drago, Nitz, Abbott,
  Abbott, Abbott, Adams, Adhikari, Adya, Ananyeva, Appert, Arai, Aston, Austin,
  Ballmer, Barker, Barr, Barsotti, Bartlett, Bartos, Batch, Bell, Betzwieser,
  Billingsley, Birch, Biscans, Blair, Blair, Bork, Brooks, Cao, Ciani, Clara,
  Cooper, Corban, Countryman, Covas, Cowart, Coyne, Cumming, Cunningham,
  Danzmann, {Da Silva Costa}, Daw, Debra, Desalvo, Dooley, Doravari, Edo,
  Etzel, Evans, Fair, Fernandez-Galiana, Ferreira, Fisher, Fong, Frey, Frolov,
  Fulda, Fyffe, Gateley, Giaime, Giardina, Goetz, Goetz, Gras, Gray, Grote,
  Gushwa, Gustafson, Gustafson, Hall, Hammond, Hanks, Hanson, Hardwick, Harry,
  Heintze, Heptonstall, Hough, Jones, Kandhasamy, Karki, Kasprzack, Kaufer,
  Kennedy, Kijbunchoo, Kim, King, King, Kissel, Korth, Kuehn, Landry, Lantz,
  Laxen, Liu, Lockerbie, Lormand, Macinnis, Macleod, M{\'{a}}rka, M{\'{a}}rka,
  Markosyan, Maros, Marsh, Martin, Martynov, Mason, Massinger, Matichard,
  Mavalvala, McCarthy, McClelland, McCormick, McCuller, McIver, McManus, McRae,
  Mendell, Merilh, Meyers, Mittleman, Mogushi, Moraru, Moreno, Mow-Lowry,
  Mueller, Mukund, Mullavey, Munch, Nelson, Nguyen, Nuttall, Oberling, Oliver,
  Oppermann, Oram, O'Reilly, Ottaway, Overmier, Palamos, Parker, Pele, Penn,
  Perez, Phelps, Pierro, Pinto, Pirello, Principe, Prokhorov, Puncken,
  Quetschke, Quintero, Radkins, Raffai, Ramirez, Reid, Reitze, Robertson,
  Rollins, Roma, Romel, Romie, Ross, Rowan, Ryan, Sadecki, Sanchez, Sanchez,
  Sandberg, Savage, Sellers, Shaddock, Shaffer, Shapiro, Shoemaker, Slagmolen,
  Smith, Smith, Sorazu, Spencer, Strain, Tanner, Taylor, Thomas, Thomas,
  Thorne, Thrane, Toland, Torrie, Traylor, Tse, Tuyenbayev, Vajente, Valdes,
  {Van Veggel}, Vass, Vecchio, Veitch, Venkateswara, Venugopalan, Vo, Vorvick,
  Walker, Ward, Warner, Weaver, Weiss, We{\ss}els, Willke, Wipf, Worden,
  Yamamoto, Yancey, Yu, Yu, Zhang, Zucker, and Zweizig]{Driggers2019}
Driggers, J.C.; Vitale, S.; Lundgren, A.P.; Evans, M.; Kawabe, K.; Dwyer, S.E.;
  Izumi, K.; Schofield, R.M.; Effler, A.; Sigg, D.;  et~al.
\newblock {Improving astrophysical parameter estimation via offline noise
  subtraction for Advanced LIGO}.
\newblock {\em Physical Review D} {\bf 2019},
  \href{https://arxiv.org/abs/1806.00532}{{\normalfont [1806.00532]}}.
\newblock
  doi:{\changeurlcolor{black}\href{https://doi.org/10.1103/PhysRevD.99.042001}{\detokenize{10.1103/PhysRevD.99.042001}}}.

\bibitem[Barsotti \em{et~al.}(2018)Barsotti, McCuller, Evans, and
  Fritschel]{Barsotti2018}
Barsotti, L.; McCuller, L.; Evans, M.; Fritschel, P.
\newblock {The A+ design curve}.
\newblock Technical report,  2018.

\bibitem[Finn and Chernoff(1993)]{Finn1993}
Finn, L.S.; Chernoff, D.F.
\newblock {Observing binary inspiral in gravitational radiation: One
  interferometer}.
\newblock {\em Physical Review D} {\bf 1993}, {\em 47},~2198,
  \href{https://arxiv.org/abs/9301003}{{\normalfont [arXiv:gr-qc/9301003]}}.
\newblock
  doi:{\changeurlcolor{black}\href{https://doi.org/10.1103/PhysRevD.47.2198}{\detokenize{10.1103/PhysRevD.47.2198}}}.

\bibitem[Finn(1996)]{Finn1996}
Finn, L.S.
\newblock Binary inspiral, gravitational radiation, and cosmology.
\newblock {\em Phys. Rev. D} {\bf 1996}, {\em 53},~2878--2894.
\newblock
  doi:{\changeurlcolor{black}\href{https://doi.org/10.1103/PhysRevD.53.2878}{\detokenize{10.1103/PhysRevD.53.2878}}}.

\bibitem[Chen \em{et~al.}(2021)Chen, Holz, Miller, Evans, Vitale, and
  Creighton]{Chen2021}
Chen, H.Y.; Holz, D.E.; Miller, J.; Evans, M.; Vitale, S.; Creighton, J.
\newblock {Distance measures in gravitational-wave astrophysics and cosmology}.
\newblock {\em Classical and Quantum Gravity} {\bf 2021}, {\em 38},~055010,
  \href{https://arxiv.org/abs/1709.08079}{{\normalfont [1709.08079]}}.
\newblock
  doi:{\changeurlcolor{black}\href{https://doi.org/10.1088/1361-6382/ABD594}{\detokenize{10.1088/1361-6382/ABD594}}}.

\bibitem[Brooks \em{et~al.}(2021)Brooks, Vajente, Yamamoto, Abbott, Adams,
  Adhikari, Ananyeva, Appert, Arai, Areeda, Asali, Aston, Austin, Baer, Ball,
  Ballmer, Banagiri, Barker, Barsotti, Bartlett, Berger, Betzwieser,
  Bhattacharjee, Billingsley, Biscans, Biscans, Blair, Blair, Bode, Bode,
  Booker, Booker, Bork, Bramley, Brown, Buikema, Cahillane, Cannon, Cao, Chen,
  Ciobanu, Clara, Compton, Cooper, Corley, Countryman, Covas, Coyne, Datrier,
  Davis, Difronzo, Dooley, Dooley, Driggers, Dupej, Dwyer, Effler, Etzel,
  Evans, Evans, Feicht, Fernandez-Galiana, Fritschel, Frolov, Fulda, Fyffe,
  Giaime, Giaime, Giardina, Godwin, Goetz, Goetz, Goetz, Goetz, Gras, Gray,
  Gray, Green, Gupta, Gustafson, Gustafson, Hall, Hanks, Hanson, Hardwick,
  Hasskew, Heintze, Helmling-Cornell, Holland, Izmui, Jia, Jones, Kandhasamy,
  Karki, Kasprzack, Kawabe, Kijbunchoo, King, Kissel, Kumar, Landry, Lane,
  Lantz, Laxen, Lecoeuche, Leviton, Jian, Jian, Lormand, Lundgren, Macas,
  Macinnis, Macleod, Mansell, Mansell, Marka, Marka, Martynov, Mason,
  Massinger, Matichard, Matichard, Mavalvala, McCarthy, McClelland, McCormick,
  McCuller, McIver, McIver, McRae, Mendell, Merfeld, Merilh, Meylahn, Meylahn,
  Mistry, Mittleman, Moreno, Mow-Lowry, Mozzon, Mullavey, Nelson, Nguyen,
  Nuttall, Oberling, Oram, Osthelder, Ottaway, Overmier, Palamos, Parker,
  Parker, Payne, Pele, Penhorwood, Perez, Pirello, Radkins, Ramirez,
  Richardson, Riles, Robertson, Robertson, Rollins, Romel, Romie, Ross, Ryan,
  Sadecki, Sanchez, Sanchez, Tiruppatturrajamanikkam, Savage, Schaetzl,
  Schnabel, Schofield, Schwartz, Sellers, Shaffer, Sigg, Slagmolen, Smith,
  Soni, Sorazu, Spencer, Strain, Sun, Szczepanczyk, Thomas, Thomas, Thorne,
  Toland, Torrie, Traylor, Tse, Urban, Valdes, Vander-Hyde, Veitch,
  Venkateswara, Venugopalan, Viets, Vo, Vorvick, Wade, Ward, Warner, Weaver,
  Weiss, Whittle, Willke, Willke, Wipf, Xiao, Yu, Yu, Zhang, Zucker, Zucker,
  and Zweizig]{Brooks2021}
Brooks, A.F.; Vajente, G.; Yamamoto, H.; Abbott, R.; Adams, C.; Adhikari, R.X.;
  Ananyeva, A.; Appert, S.; Arai, K.; Areeda, J.S.;  et~al.
\newblock {Point absorbers in Advanced LIGO}.
\newblock {\em Applied Optics, Vol. 60, Issue 13, pp. 4047-4063} {\bf 2021},
  {\em 60},~4047--4063,  \href{https://arxiv.org/abs/2101.05828}{{\normalfont
  [2101.05828]}}.
\newblock
  doi:{\changeurlcolor{black}\href{https://doi.org/10.1364/AO.419689}{\detokenize{10.1364/AO.419689}}}.

\bibitem[Isogai \em{et~al.}(2013)Isogai, Miller, Kwee, Barsotti, Evans,
  Hinkley, Sherman, Phillips, Schioppo, Lemke, Beloy, Pizzocaro, Oates, Ludlow,
  Kessler, Hagemann, Grebing, Legero, Sterr, Riehle, Martin, Chen, and
  Ye]{Isogai2013}
Isogai, T.; Miller, J.; Kwee, P.; Barsotti, L.; Evans, M.; Hinkley, N.;
  Sherman, J.A.; Phillips, N.B.; Schioppo, M.; Lemke, N.D.;  et~al.
\newblock {Loss in long-storage-time optical cavities}.
\newblock {\em Optics Express, Vol. 21, Issue 24, pp. 30114-30125} {\bf 2013},
  {\em 21},~30114--30125,  \href{https://arxiv.org/abs/1310.1820}{{\normalfont
  [1310.1820]}}.
\newblock
  doi:{\changeurlcolor{black}\href{https://doi.org/10.1364/OE.21.030114}{\detokenize{10.1364/OE.21.030114}}}.

\bibitem[Cahillane({\natexlab{a}})]{alog58772}
Cahillane, C.
\newblock Power trend of latest high power lock.
\newblock
  \href{https://alog.ligo-wa.caltech.edu/aLOG/index.php?callRep=58772}{LHO alog
  58772}.

\bibitem[Cahillane({\natexlab{b}})]{alog58794}
Cahillane, C.
\newblock Power trends for a lock today.
\newblock
  \href{https://alog.ligo-wa.caltech.edu/aLOG/index.php?callRep=58794}{LHO alog
  58794}.

\bibitem[Cahillane({\natexlab{c}})]{alog59142}
Cahillane, C.
\newblock Power in the interferometer - Pre-O4.
\newblock
  \href{https://alog.ligo-wa.caltech.edu/aLOG/index.php?callRep=59142}{LHO alog
  59142}.

\bibitem[Martynov()]{alog58794llo}
Martynov, D.
\newblock {PRG and optical gain increase}.
\newblock
  \href{https://alog.ligo-la.caltech.edu/aLOG/index.php?callRep=43121}{LLO alog
  43121}.

\bibitem[Evans \em{et~al.}(2010)Evans, Barsotti, and Fritschel]{Evans2010}
Evans, M.; Barsotti, L.; Fritschel, P.
\newblock {A general approach to optomechanical parametric instabilities}.
\newblock {\em Physics Letters A} {\bf 2010}, {\em 374},~665--671,
  \href{https://arxiv.org/abs/0910.2716}{{\normalfont [0910.2716]}}.
\newblock
  doi:{\changeurlcolor{black}\href{https://doi.org/10.1016/J.PHYSLETA.2009.11.023}{\detokenize{10.1016/J.PHYSLETA.2009.11.023}}}.

\bibitem[Green \em{et~al.}(2017)Green, Brown, Dovale-Alvarez, Collins, Miao,
  Mow-Lowry, and Freise]{Green2017}
Green, A.C.; Brown, D.D.; Dovale-Alvarez, M.; Collins, C.; Miao, H.; Mow-Lowry,
  C.M.; Freise, A.
\newblock {The influence of dual-recycling on parametric instabilities at
  Advanced LIGO}.
\newblock {\em Classical and Quantum Gravity} {\bf 2017}, {\em 34},
  \href{https://arxiv.org/abs/1704.08595}{{\normalfont [1704.08595]}}.
\newblock
  doi:{\changeurlcolor{black}\href{https://doi.org/10.1088/1361-6382/AA8AF8}{\detokenize{10.1088/1361-6382/AA8AF8}}}.

\bibitem[Miller \em{et~al.}(2011)Miller, Evans, Barsotti, Fritschel, MacInnis,
  Mittleman, Shapiro, Soto, and Torrie]{Miller2011}
Miller, J.; Evans, M.; Barsotti, L.; Fritschel, P.; MacInnis, M.; Mittleman,
  R.; Shapiro, B.; Soto, J.; Torrie, C.
\newblock {Damping parametric instabilities in future gravitational wave
  detectors by means of electrostatic actuators}.
\newblock {\em Physics Letters A} {\bf 2011}, {\em 375},~788--794,
  \href{https://arxiv.org/abs/1704.03587}{{\normalfont [1704.03587]}}.
\newblock
  doi:{\changeurlcolor{black}\href{https://doi.org/10.1016/J.PHYSLETA.2010.12.032}{\detokenize{10.1016/J.PHYSLETA.2010.12.032}}}.

\bibitem[Blair \em{et~al.}(2017)Blair, Gras, Abbott, Aston, Betzwieser, Blair,
  Derosa, Evans, Frolov, Fritschel, Grote, Hardwick, Liu, Lormand, Miller,
  Mullavey, O'Reilly, Zhao, Abbott, Abbott, Adams, Adhikari, Anderson,
  Ananyeva, Appert, Arai, Ballmer, Barker, Barr, Barsotti, Bartlett, Bartos,
  Batch, Bell, Billingsley, Birch, Biscans, Biwer, Bork, Brooks, Ciani, Clara,
  Countryman, Cowart, Coyne, Cumming, Cunningham, Danzmann, {Da Silva Costa},
  Daw, Debra, Desalvo, Dooley, Doravari, Driggers, Dwyer, Effler, Etzel, Evans,
  Factourovich, Fair, {Fern{\'{a}}ndez Galiana}, Fisher, Fulda, Fyffe, Giaime,
  Giardina, Goetz, Goetz, Gray, Gushwa, Gustafson, Gustafson, Hall, Hammond,
  Hanks, Hanson, Harry, Heintze, Heptonstall, Hough, Izumi, Jones, Kandhasamy,
  Karki, Kasprzack, Kaufer, Kawabe, Kijbunchoo, King, King, Kissel, Korth,
  Kuehn, Landry, Lantz, Lockerbie, Lundgren, MacInnis, Macleod, M{\'{a}}rka,
  M{\'{a}}rka, Markosyan, Maros, Martin, Martynov, Mason, Massinger, Matichard,
  Mavalvala, McCarthy, McClelland, McCormick, McIntyre, McIver, Mendell,
  Merilh, Meyers, Mittleman, Moreno, Mueller, Munch, Nuttall, Oberling,
  Oppermann, Oram, Ottaway, Overmier, Palamos, Paris, Parker, Pele, Penn,
  Phelps, Pierro, Pinto, Principe, Prokhorov, Puncken, Quetschke, Quintero,
  Raab, Radkins, Raffai, Reid, Reitze, Robertson, Rollins, Roma, Romie, Rowan,
  Ryan, Sadecki, Sanchez, Sandberg, Savage, Schofield, Sellers, Shaddock,
  Shaffer, Shapiro, Shawhan, Shoemaker, Sigg, Slagmolen, Smith, Smith, Sorazu,
  Staley, Strain, Tanner, Taylor, Thomas, Thomas, Thorne, Thrane, Torrie,
  Traylor, Vajente, Valdes, {Van Veggel}, Vecchio, Veitch, Venkateswara, Vo,
  Vorvick, Walker, Ward, Warner, Weaver, Weiss, We{\ss}els, Willke, Wipf,
  Worden, Wu, Yamamoto, Yancey, Yu, Yu, Zhang, Zucker, and Zweizig]{Blair2017}
Blair, C.; Gras, S.; Abbott, R.; Aston, S.; Betzwieser, J.; Blair, D.; Derosa,
  R.; Evans, M.; Frolov, V.; Fritschel, P.;  et~al.
\newblock {First Demonstration of Electrostatic Damping of Parametric
  Instability at Advanced LIGO}.
\newblock {\em Physical Review Letters} {\bf 2017}, {\em 118},~151102,
  \href{https://arxiv.org/abs/1611.08997}{{\normalfont [1611.08997]}}.
\newblock
  doi:{\changeurlcolor{black}\href{https://doi.org/10.1103/PHYSREVLETT.118.151102/FIGURES/5/MEDIUM}{\detokenize{10.1103/PHYSREVLETT.118.151102/FIGURES/5/MEDIUM}}}.

\bibitem[Hardwick(2019)]{HardwickThesis}
Hardwick, T.
\newblock High Power and Optomechanics in Advanced LIGO Detectors.
\newblock PhD thesis, Louisiana State University,  2019.

\bibitem[Biscans \em{et~al.}(2019)Biscans, Gras, Blair, Driggers, Evans,
  Fritschel, Hardwick, and Mansell]{Biscans2019}
Biscans, S.; Gras, S.; Blair, C.D.; Driggers, J.; Evans, M.; Fritschel, P.;
  Hardwick, T.; Mansell, G.
\newblock Suppressing parametric instabilities in {LIGO} using low-noise
  acoustic mode dampers.
\newblock {\em Phys. Rev. D} {\bf 2019}, {\em 100},~122003.
\newblock
  doi:{\changeurlcolor{black}\href{https://doi.org/10.1103/PhysRevD.100.122003}{\detokenize{10.1103/PhysRevD.100.122003}}}.

\bibitem[Buonanno and Chen(2002)]{Buonanno2002}
Buonanno, A.; Chen, Y.
\newblock Signal recycled laser-interferometer gravitational-wave detectors as
  optical springs.
\newblock {\em Phys. Rev. D} {\bf 2002}, {\em 65},~042001.
\newblock
  doi:{\changeurlcolor{black}\href{https://doi.org/10.1103/PhysRevD.65.042001}{\detokenize{10.1103/PhysRevD.65.042001}}}.

\bibitem[Sheard \em{et~al.}(2004)Sheard, Gray, Mow-Lowry, McClelland, and
  Whitcomb]{Sheard2004}
Sheard, B.S.; Gray, M.B.; Mow-Lowry, C.M.; McClelland, D.E.; Whitcomb, S.E.
\newblock Observation and characterization of an optical spring.
\newblock {\em Phys. Rev. A} {\bf 2004}, {\em 69},~051801.
\newblock
  doi:{\changeurlcolor{black}\href{https://doi.org/10.1103/PhysRevA.69.051801}{\detokenize{10.1103/PhysRevA.69.051801}}}.

\bibitem[Aspelmeyer \em{et~al.}(2014)Aspelmeyer, Kippenberg, and
  Marquardt]{Aspelmeyer2014}
Aspelmeyer, M.; Kippenberg, T.J.; Marquardt, F.
\newblock {Cavity optomechanics}.
\newblock {\em Reviews of Modern Physics} {\bf 2014}, {\em 86},~1391--1452,
  \href{https://arxiv.org/abs/1303.0733}{{\normalfont [1303.0733]}}.
\newblock
  doi:{\changeurlcolor{black}\href{https://doi.org/10.1103/REVMODPHYS.86.1391/FIGURES/46/MEDIUM}{\detokenize{10.1103/REVMODPHYS.86.1391/FIGURES/46/MEDIUM}}}.

\bibitem[Bond \em{et~al.}(2017)Bond, Brown, Freise, and Strain]{Bond2017}
Bond, C.; Brown, D.; Freise, A.; Strain, K.A.
\newblock {Interferometer techniques for gravitational-wave detection}.
\newblock {\em Living Reviews in Relativity 2016 19:1} {\bf 2017}, {\em
  19},~1--217.
\newblock
  doi:{\changeurlcolor{black}\href{https://doi.org/10.1007/S41114-016-0002-8}{\detokenize{10.1007/S41114-016-0002-8}}}.

\bibitem[Whittle \em{et~al.}(2021)Whittle, Hall, Dwyer, Mavalvala, Sudhir,
  Abbott, Ananyeva, Austin, Barsotti, Betzwieser, Blair, Brooks, Brown,
  Buikema, Cahillane, Driggers, Effler, Fernandez-Galiana, Fritschel, Frolov,
  Hardwick, Kasprzack, Kawabe, Kijbunchoo, Kissel, Mansell, Matichard,
  McCuller, McRae, Mullavey, Pele, Schofield, Sigg, Tse, Vajente, Vander-Hyde,
  Yu, Yu, Adams, Adhikari, Appert, Arai, Areeda, Asali, Aston, Baer, Ball,
  Ballmer, Banagiri, Barker, Bartlett, Berger, Bhattacharjee, Billingsley,
  Biscans, Blair, Bode, Booker, Bork, Bramley, Cannon, Chen, Ciobanu, Clara,
  Compton, Cooper, Corley, Countryman, Covas, Coyne, Datrier, Davis, {Di
  Fronzo}, Dooley, Dupej, Etzel, Evans, Evans, Feicht, Fulda, Fyffe, Giaime,
  Giardina, Godwin, Goetz, Gras, Gray, Gray, Green, Gustafson, Gustafson,
  Hanks, Hanson, Hasskew, Heintze, Helmling-Cornell, Holland, Jones,
  Kandhasamy, Karki, King, Kumar, Landry, Lane, Lantz, Laxen, Lecoeuche,
  Leviton, Liu, Lormand, Lundgren, Macas, MacInnis, Macleod, M{\'{a}}rka,
  M{\'{a}}rka, Martynov, Mason, Massinger, McCarthy, McClelland, McCormick,
  McIver, Mendell, Merfeld, Merilh, Meylahn, Mistry, Mittleman, Moreno,
  Mow-Lowry, Mozzon, Nelson, Nguyen, Nuttall, Oberling, Oram, Osthelder,
  Ottaway, Overmier, Palamos, Parker, Payne, Penhorwood, Perez, Pirello,
  Radkins, Ramirez, Richardson, Riles, Robertson, Rollins, Romel, Romie, Ross,
  Ryan, Sadecki, Sanchez, Sanchez, Saravanan, Savage, Schaetz, Schnabel,
  Schwartz, Sellers, Shaffer, Slagmolen, Smith, Soni, Sorazu, Spencer, Strain,
  Sun, Szczepa~czyk, Thomas, Thomas, Thorne, Toland, Torrie, Traylor, Urban,
  Valdes, Veitch, Venkateswara, Venugopalan, Viets, Vo, Vorvick, Wade, Ward,
  Warner, Weaver, Weiss, Willke, Wipf, Xiao, Yamamoto, Zhang, Zucker, and
  Zweizig]{Whittle2021}
Whittle, C.; Hall, E.D.; Dwyer, S.; Mavalvala, N.; Sudhir, V.; Abbott, R.;
  Ananyeva, A.; Austin, C.; Barsotti, L.; Betzwieser, J.;  et~al.
\newblock {Approaching the motional ground state of a 10-kg object}.
\newblock {\em Science} {\bf 2021}, {\em 372},~1333--1336,
  \href{https://arxiv.org/abs/2102.12665}{{\normalfont [2102.12665]}}.
\newblock
  doi:{\changeurlcolor{black}\href{https://doi.org/10.1126/SCIENCE.ABH2634/SUPPL_FILE/ABH2634-WHITTLE-SM.PDF}{\detokenize{10.1126/SCIENCE.ABH2634/SUPPL_FILE/ABH2634-WHITTLE-SM.PDF}}}.

\bibitem[Dwyer \em{et~al.}(2022)Dwyer, Mansell, and McCuller]{Dwyer2022}
Dwyer, S.; Mansell, G.; McCuller, L.
\newblock {Squeezing in gravitational wave detectors}.
\newblock Submitted to Galaxies special issue.

\bibitem[Nguyen \em{et~al.}(2021)Nguyen, Schofield, Effler, Austin, Adya, Ball,
  Banagiri, Banowetz, Billman, Blair, Buikema, Cahillane, Clara, Covas, Dalya,
  Daniel, Dawes, Derosa, Dwyer, Frey, Frolov, Ghirado, Goetz, Hardwick,
  Helmling-Cornell, Hollows, Kijbunchoo, Kruk, Laxen, Maaske, Mansell,
  McCarthy, Merfeld, Neunzert, Palamos, Parker, Pearlstone, Pele, Radkins,
  Roma, Savage, Schale, Shoemaker, Shoemaker, Soni, Talukder, Tse, Valdes,
  Vidreo, Vorvick, Abbott, Adams, Adhikari, Ananyeva, Appert, Arai, Areeda,
  Asali, Aston, Baer, Ballmer, Barker, Barsotti, Bartlett, Berger, Betzwieser,
  Bhattacharjee, Billingsley, Biscans, Blair, Bode, Booker, Bork, Bramley,
  Brooks, Brown, Cannon, Chen, Ciobanu, Cooper, Compton, Corley, Countryman,
  Coyne, Datrier, Davis, {Di Fronzo}, Dooley, Driggers, Dupej, Etzel, Evans,
  Evans, Feicht, Fernandez-Galiana, Fritschel, Fulda, Fyffe, Giaime, Giardina,
  Godwin, Gras, Gray, Gray, Green, Gustafson, Gustafson, Hanks, Hanson,
  Hasskew, Heintze, Holland, Jones, Kandhasamy, Karki, Kasprzack, Kawabe, King,
  Kissel, Kumar, Landry, Lane, Lantz, Lecoeuche, Leviton, Liu, Lormand,
  Lundgren, Macas, Macinnis, Macleod, M{\'{a}}rka, M{\'{a}}rka, Martynov,
  Mason, Massinger, Matichard, Mavalvala, McClelland, McCormick, McCuller,
  McIver, McRae, Mendell, Merilh, Meylahn, Meyers, Mistry, Mittleman, Moreno,
  Mow-Lowry, Mozzon, Mullavey, Nelson, Nuttall, Oberling, Oram, Osthelder,
  Ottaway, Overmier, Payne, Penhorwood, Perez, Pirello, Ramirez, Richardson,
  Riles, Robertson, Rollins, Romel, Romie, Ross, Ryan, Sadecki, Sanchez,
  Sanchez, Saravanan, Schaetzl, Schnabel, Schwartz, Sellers, Shaffer, Sigg,
  Slagmolen, Smith, Sorazu, Spencer, Strain, Sun, Szczepa{\'{n}}czyk, Thomas,
  Thomas, Thorne, Toland, Torrie, Traylor, Urban, Vajente, Vander-Hyde, Veitch,
  Venkateswara, Venugopalan, Viets, Vo, Wade, Ward, Warner, Weaver, Weiss,
  Whittle, Willke, Wipf, Xiao, Yamamoto, Yu, Yu, Zhang, Zucker, and
  Zweizig]{Nguyen2021}
Nguyen, P.; Schofield, R.M.; Effler, A.; Austin, C.; Adya, V.; Ball, M.;
  Banagiri, S.; Banowetz, K.; Billman, C.; Blair, C.D.;  et~al.
\newblock {Environmental noise in advanced LIGO detectors}.
\newblock {\em Classical and Quantum Gravity} {\bf 2021}, {\em 38},~145001,
  \href{https://arxiv.org/abs/2101.09935}{{\normalfont [2101.09935]}}.
\newblock
  doi:{\changeurlcolor{black}\href{https://doi.org/10.1088/1361-6382/AC011A}{\detokenize{10.1088/1361-6382/AC011A}}}.

\bibitem[Schwartz \em{et~al.}(2020)Schwartz, Pele, Warner, Lantz, Betzwieser,
  Dooley, Biscans, Coughlin, Mukund, Abbott, Adams, Adhikari, Ananyeva, Appert,
  Arai, Areeda, Asali, Aston, Austin, Baer, Ball, Ballmer, Banagiri, Barker,
  Barsotti, Bartlett, Berger, Bhattacharjee, Billingsley, Blair, Blair, Bode,
  Booker, Bork, Bramley, Brooks, Brown, Buikema, Cahillane, Cannon, Chen,
  Ciobanu, Clara, Cooper, Corley, Countryman, Covas, Coyne, Datrier, Davis,
  Fronzo, Driggers, Dupej, Dwyer, Effler, Etzel, Evans, Evans, Feicht,
  Fernandez-Galiana, Fritschel, Frolov, Fulda, Fyffe, Giaime, Giardina, Godwin,
  Goetz, Gras, Gray, Gray, Green, Gupta, Gustafson, Gustafson, Hanks, Hanson,
  Hardwick, Hasskew, Heintze, Helmling-Cornell, Holland, Jones, Kandhasamy,
  Karki, Kasprzack, Kawabe, Kijbunchoo, King, Kissel, Kumar, Landry, Lane,
  Laxen, Lecoeuche, Leviton, Liu, Lormand, Lundgren, Macas, MacInnis, Macleod,
  Mansell, M{\'{a}}rka, M{\'{a}}rka, Martynov, Mason, Massinger, Matichard,
  Mavalvala, McCarthy, McClelland, McCormick, McCuller, McIver, McRae, Mendell,
  Merfeld, Merilh, Meylahn, Mistry, Mittleman, Moreno, Mow-Lowry, Mozzon,
  Mullavey, Nelson, Nguyen, Nuttall, Oberling, Oram, Osthelder, Ottaway,
  Overmier, Palamos, Parker, Payne, Perez, Pirello, Radkins, Ramirez,
  Richardson, Riles, Robertson, Rollins, Romel, Romie, Ross, Ryan, Sadecki,
  Sanchez, Sanchez, Saravanan, Savage, Schaetzl, Schnabel, Schofield, Sellers,
  Shaffer, Sigg, Slagmolen, Smith, Soni, Sorazu, Spencer, Strain, Sun,
  Szczepa{\'{n}}czyk, Thomas, Thomas, Thorne, Toland, Torrie, Traylor, Tse,
  Urban, Vajente, Valdes, Vander-Hyde, Veitch, Venkateswara, Venugopalan,
  Viets, Vo, Vorvick, Wade, Ward, Weaver, Weiss, Whittle, Willke, Wipf, Xiao,
  Yamamoto, Yu, Yu, Zhang, Zucker, and Zweizig]{Schwartz2020}
Schwartz, E.; Pele, A.; Warner, J.; Lantz, B.; Betzwieser, J.; Dooley, K.L.;
  Biscans, S.; Coughlin, M.; Mukund, N.; Abbott, R.;  et~al.
\newblock Improving the robustness of the advanced {LIGO} detectors to
  earthquakes.
\newblock {\em Classical and Quantum Gravity} {\bf 2020}, {\em 37},~235007.
\newblock
  doi:{\changeurlcolor{black}\href{https://doi.org/10.1088/1361-6382/abbc8c}{\detokenize{10.1088/1361-6382/abbc8c}}}.

\bibitem[Ross \em{et~al.}(2020)Ross, Venkateswara, Mow-Lowry, Cooper, Warner,
  Lantz, Kissel, Radkins, Shaffer, Mittleman, Pele, and Gundlach]{Ross2020}
Ross, M.P.; Venkateswara, K.; Mow-Lowry, C.; Cooper, S.; Warner, J.; Lantz, B.;
  Kissel, J.; Radkins, H.; Shaffer, T.; Mittleman, R.;  et~al.
\newblock Towards windproofing {LIGO}: reducing the effect of wind-driven floor
  tilt by using rotation sensors in active seismic isolation.
\newblock {\em Classical and Quantum Gravity} {\bf 2020}, {\em 37},~185018.
\newblock
  doi:{\changeurlcolor{black}\href{https://doi.org/10.1088/1361-6382/ab9d5c}{\detokenize{10.1088/1361-6382/ab9d5c}}}.

\bibitem[Soni \em{et~al.}(2021)Soni, Austin, Effler, Schofield, Gonz{\'{a}}lez,
  Frolov, Driggers, Pele, Urban, Valdes, Abbott, Adams, Adhikari, Ananyeva,
  Appert, Arai, Areeda, Asali, Aston, Baer, Ball, Ballmer, Banagiri, Barker,
  Barsotti, Bartlett, Berger, Betzwieser, Bhattacharjee, Billingsley, Biscans,
  Blair, Blair, Bode, Booker, Bork, Bramley, Brooks, Brown, Buikema, Cahillane,
  Cannon, Chen, Ciobanu, Clara, Cooper, Corley, Countryman, Covas, Coyne,
  Datrier, Davis, Fronzo, Dooley, Dupej, Dwyer, Etzel, Evans, Evans, Feicht,
  Fernandez-Galiana, Fritschel, Fulda, Fyffe, Giaime, Giardina, Godwin, Goetz,
  Gras, Gray, Gray, Green, Gustafson, Gustafson, Hanks, Hanson, Hardwick,
  Hasskew, Heintze, Helmling-Cornell, Holland, Jones, Kandhasamy, Karki,
  Kasprzack, Kawabe, Kijbunchoo, King, Kissel, Kumar, Landry, Lane, Lantz,
  Laxen, Lecoeuche, Leviton, Liu, Lormand, Lundgren, Macas, MacInnis, Macleod,
  Mansell, M{\'{a}}rka, M{\'{a}}rka, Martynov, Mason, Massinger, Matichard,
  Mavalvala, McCarthy, McClelland, McCormick, McCuller, McIver, McRae, Mendell,
  Merfeld, Merilh, Meylahn, Mistry, Mittleman, Moreno, Mow-Lowry, Mozzon,
  Mullavey, Nelson, Nguyen, Nuttall, Oberling, Oram, Osthelder, Ottaway,
  Overmier, Palamos, Parker, Payne, Penhorwood, Perez, Pirello, Radkins,
  Ramirez, Richardson, Riles, Robertson, Rollins, Romel, Romie, Ross, Ryan,
  Sadecki, Sanchez, Sanchez, Saravanan, Savage, Schaetzl, Schnabel, Schwartz,
  Sellers, Shaffer, Sigg, Slagmolen, Smith, Sorazu, Spencer, Strain, Sun,
  Szczepa{\'{n}}czyk, Thomas, Thomas, Thorne, Toland, Torrie, Traylor, Tse,
  Vajente, Vander-Hyde, Veitch, Venkateswara, Venugopalan, Viets, Vo, Vorvick,
  Wade, Ward, Warner, Weaver, Weiss, Whittle, Willke, Wipf, Xiao, Yamamoto, Yu,
  Yu, Zhang, Zucker, Zweizig, and Collaboration)]{Soni2021}
Soni, S.; Austin, C.; Effler, A.; Schofield, R.M.S.; Gonz{\'{a}}lez, G.;
  Frolov, V.V.; Driggers, J.C.; Pele, A.; Urban, A.L.; Valdes, G.;  et~al.
\newblock Reducing scattered light in {LIGO}'s third observing run.
\newblock {\em Classical and Quantum Gravity} {\bf 2021}, {\em 38},~025016.
\newblock
  doi:{\changeurlcolor{black}\href{https://doi.org/10.1088/1361-6382/abc906}{\detokenize{10.1088/1361-6382/abc906}}}.

\bibitem[Kwee \em{et~al.}(2014)Kwee, Miller, Isogai, Barsotti, and
  Evans]{Kwee2014}
Kwee, P.; Miller, J.; Isogai, T.; Barsotti, L.; Evans, M.
\newblock {Decoherence and degradation of squeezed states in quantum filter
  cavities}.
\newblock {\em Physical Review D - Particles, Fields, Gravitation and
  Cosmology} {\bf 2014},  \href{https://arxiv.org/abs/1704.03531}{{\normalfont
  [1704.03531]}}.
\newblock
  doi:{\changeurlcolor{black}\href{https://doi.org/10.1103/PhysRevD.90.062006}{\detokenize{10.1103/PhysRevD.90.062006}}}.

\bibitem[Whittle \em{et~al.}(2020)Whittle, Komori, Ganapathy, McCuller,
  Barsotti, Mavalvala, and Evans]{Whittle2020}
Whittle, C.; Komori, K.; Ganapathy, D.; McCuller, L.; Barsotti, L.; Mavalvala,
  N.; Evans, M.
\newblock Optimal detuning for quantum filter cavities.
\newblock {\em Phys. Rev. D} {\bf 2020}, {\em 102},~102002.
\newblock
  doi:{\changeurlcolor{black}\href{https://doi.org/10.1103/PhysRevD.102.102002}{\detokenize{10.1103/PhysRevD.102.102002}}}.

\bibitem[Komori \em{et~al.}(2020)Komori, Ganapathy, Whittle, McCuller,
  Barsotti, Mavalvala, and Evans]{Kentaro2020}
Komori, K.; Ganapathy, D.; Whittle, C.; McCuller, L.; Barsotti, L.; Mavalvala,
  N.; Evans, M.
\newblock Demonstration of an amplitude filter cavity at gravitational-wave
  frequencies.
\newblock {\em Phys. Rev. D} {\bf 2020}, {\em 102},~102003.
\newblock
  doi:{\changeurlcolor{black}\href{https://doi.org/10.1103/PhysRevD.102.102003}{\detokenize{10.1103/PhysRevD.102.102003}}}.

\bibitem[Bode \em{et~al.}(2020)Bode, Meylahn, and Willke]{Bode2020}
Bode, N.; Meylahn, F.; Willke, B.
\newblock Sequential high power laser amplifiers for gravitational wave
  detection.
\newblock {\em Opt. Express} {\bf 2020}, {\em 28},~29469--29478.
\newblock
  doi:{\changeurlcolor{black}\href{https://doi.org/10.1364/OE.401826}{\detokenize{10.1364/OE.401826}}}.

\bibitem[Cao \em{et~al.}(2020)Cao, Brooks, Ng, Ottaway, Perreca, Richardson,
  Chaderjian, and Veitch]{Cao2020}
Cao, H.T.; Brooks, A.; Ng, S.W.S.; Ottaway, D.; Perreca, A.; Richardson, J.W.;
  Chaderjian, A.; Veitch, P.J.
\newblock High dynamic range thermally actuated bimorph mirror for
  gravitational wave detectors.
\newblock {\em Appl. Opt.} {\bf 2020}, {\em 59},~2784--2790.
\newblock
  doi:{\changeurlcolor{black}\href{https://doi.org/10.1364/AO.376764}{\detokenize{10.1364/AO.376764}}}.

\bibitem[Srivastava \em{et~al.}(submitted 2021)Srivastava, Mansell, Makarem,
  Noh, Abbott, Ballmer, Billingsley, Brooks, Cao, Fritschel, Griffith, Jia,
  Kasprzack, MacInnis, Ng, Sanchez, Torrie, Veitch, and
  Matichard]{Srivastava2021}
Srivastava, V.; Mansell, G.; Makarem, C.; Noh, M.; Abbott, R.; Ballmer, S.;
  Billingsley, G.; Brooks, A.; Cao, H.T.; Fritschel, P.;  et~al.
\newblock Piezo-deformable Mirrors for Active Mode Matching in Advanced LIGO.
\newblock {\em Optics Express} {\bf submitted 2021},
  \href{https://arxiv.org/abs/2110.00674}{{\normalfont
  [arXiv:astro-ph.IM/2110.00674]}}.

\bibitem[Sigg(2008)]{Sigg2008}
Sigg, D.
\newblock {Status of the LIGO detectors}.
\newblock {\em Classical and Quantum Gravity} {\bf 2008}, {\em 25},~114041.
\newblock
  doi:{\changeurlcolor{black}\href{https://doi.org/10.1088/0264-9381/25/11/114041}{\detokenize{10.1088/0264-9381/25/11/114041}}}.

\bibitem[Abbott \em{et~al.}(2009)Abbott, Abbott, Adhikari, Ajith, Allen, Allen,
  Amin, Anderson, Anderson, Arain, Araya, Armandula, Armor, Aso, Aston,
  Aufmuth, Aulbert, Babak, Baker, Ballmer, Barker, Barker, Barr, Barriga,
  Barsotti, Barton, Bartos, Bassiri, Bastarrika, Behnke, Benacquista,
  Betzwieser, Beyersdorf, Bilenko, Billingsley, Biswas, Black, Blackburn,
  Blackburn, Blair, Bland, Bodiya, Bogue, Bork, Boschi, Bose, Brady, Braginsky,
  Brau, Bridges, Brinkmann, Brooks, Brown, Brummit, Brunet, Bullington,
  Buonanno, Burmeister, Byer, Cadonati, Camp, Cannizzo, Cannon, Cao, Cardenas,
  Caride, Castaldi, Caudill, Cavagli{\`{a}}, Cepeda, Chalermsongsak, Chalkley,
  Charlton, Chatterji, Chelkowski, Chen, Christensen, Chung, Clark, Clark,
  Clayton, Cokelaer, Colacino, Conte, Cook, Corbitt, Cornish, Coward, Coyne,
  Creighton, Creighton, Cruise, Culter, Cumming, Cunningham, Danilishin,
  Danzmann, Daudert, Davies, Daw, DeBra, Degallaix, Dergachev, Desai, DeSalvo,
  Dhurandhar, D{\'{\i}}az, Dietz, Donovan, Dooley, Doomes, Drever, Dueck, Duke,
  Dumas, Dwyer, Echols, Edgar, Effler, Ehrens, Espinoza, Etzel, Evans, Evans,
  Fairhurst, Faltas, Fan, Fazi, Fehrmenn, Finn, Flasch, Foley, Forrest,
  Fotopoulos, Franzen, Frede, Frei, Frei, Freise, Frey, Fricke, Fritschel,
  Frolov, Fyffe, Galdi, Garofoli, Gholami, Giaime, Giampanis, Giardina, Goda,
  Goetz, Goggin, Gonz{\'{a}}lez, Gorodetsky, Go{\ss}ler, Gouaty, Grant, Gras,
  Gray, Gray, Greenhalgh, Gretarsson, Grimaldi, Grosso, Grote, Grunewald,
  Guenther, Gustafson, Gustafson, Hage, Hallam, Hammer, Hammond, Hanna, Hanson,
  Harms, Harry, Harry, Harstad, Haughian, Hayama, Heefner, Heng, Heptonstall,
  Hewitson, Hild, Hirose, Hoak, Hodge, Holt, Hosken, Hough, Hoyland, Hughey,
  Huttner, Ingram, Isogai, Ito, Ivanov, Johnson, Johnson, Jones, Jones, Jones,
  Ju, Kalmus, Kalogera, Kandhasamy, Kanner, Kasprzyk, Katsavounidis, Kawabe,
  Kawamura, Kawazoe, Kells, Keppel, Khalaidovski, Khalili, Khan, Khazanov,
  King, Kissel, Klimenko, Kokeyama, Kondrashov, Kopparapu, Koranda, Kozak,
  Krishnan, Kumar, Kwee, Lam, Landry, Lantz, Lazzarini, Lei, Lei, Leindecker,
  Leonor, Li, Lin, Lindquist, Littenberg, Lockerbie, Lodhia, Longo, Lormand,
  Lu, Lubinski, Lucianetti, L\"{u}ck, Machenschalk, MacInnis, Mageswaran,
  Mailand, Mandel, Mandic, M{\'{a}}rka, M{\'{a}}rka, Markosyan, Markowitz,
  Maros, Martin, Martin, Marx, Mason, Matichard, Matone, Matzner, Mavalvala,
  McCarthy, McClelland, McGuire, McHugh, McIntyre, McKechan, McKenzie, Mehmet,
  Melatos, Melissinos, Men{\'{e}}ndez, Mendell, Mercer, Meshkov, Messenger,
  Meyer, Miller, Minelli, Mino, Mitrofanov, Mitselmakher, Mittleman, Miyakawa,
  Moe, Mohanty, Mohapatra, Moreno, Morioka, Mors, Mossavi, MowLowry, Mueller,
  M\"{u}ller-Ebhardt, Muhammad, Mukherjee, Mukhopadhyay, Mullavey, Munch,
  Murray, Myers, Myers, Nash, Nelson, Newton, Nishizawa, Numata,
  O{\textquotesingle}Dell, O{\textquotesingle}Reilly,
  O{\textquotesingle}Shaughnessy, Ochsner, Ogin, Ottaway, Ottens, Overmier,
  Owen, Pan, Pankow, Papa, Parameshwaraiah, Patel, Pedraza, Penn, Perraca,
  Pierro, Pinto, Pitkin, Pletsch, Plissi, Postiglione, Principe, Prix,
  Prokhorov, Punken, Quetschke, Raab, Rabeling, Radkins, Raffai, Raics, Rainer,
  Rakhmanov, Raymond, Reed, Reed, Rehbein, Reid, Reitze, Riesen, Riles, Rivera,
  Roberts, Robertson, Robinson, Robinson, Roddy, R\"{o}ver, Rollins, Romano,
  Romie, Rowan, R\"{u}diger, Russell, Ryan, Sakata, de~la Jordana, Sandberg,
  Sannibale, Santamar{\'{\i}}a, Saraf, Sarin, Sathyaprakash, Sato,
  Satterthwaite, Saulson, Savage, Savov, Scanlan, Schilling, Schnabel,
  Schofield, Schulz, Schutz, Schwinberg, Scott, Scott, Searle, Sears, Seifert,
  Sellers, Sengupta, Sergeev, Shapiro, Shawhan, Shoemaker, Sibley, Siemens,
  Sigg, Sinha, Sintes, Slagmolen, Slutsky, Smith, Smith, Smith, Somiya, Sorazu,
  Stein, Stein, Steplewski, Stochino, Stone, Strain, Strigin, Stroeer, Stuver,
  Summerscales, Sun, Sung, Sutton, Szokoly, Talukder, Tang, Tanner, Tarabrin,
  Taylor, Taylor, Thacker, Thorne, Th\"{u}ring, Tokmakov, Torres, Torrie,
  Traylor, Trias, Ugolini, Ulmen, Urbanek, Vahlbruch, Vallisneri, Broeck,
  van~der Sluys, van Veggel, Vass, Vaulin, Vecchio, Veitch, Veitch, Veltkamp,
  Villar, Vorvick, Vyachanin, Waldman, Wallace, Ward, Weidner, Weinert,
  Weinstein, Weiss, Wen, Wen, Wette, Whelan, Whitcomb, Whiting, Wilkinson,
  Willems, Williams, Williams, Willke, Wilmut, Winkelmann, Winkler, Wipf,
  Wiseman, Woan, Wooley, Worden, Wu, Yakushin, Yamamoto, Yan, Yoshida, Zanolin,
  Zhang, Zhang, Zhao, Zotov, Zucker, zur M\"{u}hlen, and Zweizig]{Abbott2009}
Abbott, B.P.; Abbott, R.; Adhikari, R.; Ajith, P.; Allen, B.; Allen, G.; Amin,
  R.S.; Anderson, S.B.; Anderson, W.G.; Arain, M.A.;  et~al.
\newblock {LIGO}: the Laser Interferometer Gravitational-Wave Observatory.
\newblock {\em Reports on Progress in Physics} {\bf 2009}, {\em 72},~076901.
\newblock
  doi:{\changeurlcolor{black}\href{https://doi.org/10.1088/0034-4885/72/7/076901}{\detokenize{10.1088/0034-4885/72/7/076901}}}.

\bibitem[Aasi \em{et~al.}(2015)Aasi, Abadie, Abbott, Abbott, Abbott, Abernathy,
  Accadia, Acernese, Adams, Adams, Adhikari, Affeldt, Agathos, Aggarwal,
  Aguiar, Ajith, Allen, Allocca, Ceron, Amariutei, Anderson, Anderson,
  Anderson, Arai, Araya, Arceneaux, Areeda, Ast, Aston, Astone, Aufmuth,
  Aulbert, Austin, Aylott, Babak, Baker, Ballardin, Ballmer, Barayoga, Barker,
  Barnum, Barone, Barr, Barsotti, Barsuglia, Barton, Bartos, Bassiri, Basti,
  Batch, Bauchrowitz, Bauer, Bebronne, Behnke, Bejger, Beker, Bell, Bell,
  Belopolski, Bergmann, Berliner, Bertolini, Bessis, Betzwieser, Beyersdorf,
  Bhadbhade, Bilenko, Billingsley, Birch, Bitossi, Bizouard, Black, Blackburn,
  Blackburn, Blair, Blom, Bock, Bodiya, Boer, Bogan, Bond, Bondu, Bonelli,
  Bonnand, Bork, Born, Bose, Bosi, Bowers, Bradaschia, Brady, Braginsky,
  Branchesi, Brannen, Brau, Breyer, Briant, Bridges, Brillet, Brinkmann,
  Brisson, Britzger, Brooks, Brown, Brown, Br\"{u}ckner, Bulik, Bulten,
  Buonanno, Buskulic, Buy, Byer, Cadonati, Cagnoli, Bustillo, Calloni, Camp,
  Campsie, Cannon, Canuel, Cao, Capano, Carbognani, Carbone, Caride, Castiglia,
  Caudill, Cavagli{\'{a}}, Cavalier, Cavalieri, Cella, Cepeda, Cesarini,
  Chakraborty, Chalermsongsak, Chao, Charlton, Chassande-Mottin, Chen, Chen,
  Chincarini, Chiummo, Cho, Chow, Christensen, Chu, Chua, Chung, Ciani, Clara,
  Clark, Clark, Cleva, Coccia, Cohadon, Colla, Colombini, Jr, Conte, Conte,
  Cook, Corbitt, Cordier, Cornish, Corsi, Costa, Coughlin, Coulon, Countryman,
  Couvares, Coward, Cowart, Coyne, Craig, Creighton, Creighton, Crowder,
  Cumming, Cunningham, Cuoco, Dahl, Canton, Damjanic, Danilishin, D'Antonio,
  Danzmann, Dattilo, Daudert, Daveloza, Davier, Davies, Daw, Day, Dayanga,
  Debreczeni, Degallaix, Deleeuw, Del{\'{e}}glise, Pozzo, Denker, Dent, Dereli,
  Dergachev, Rosa, DeRosa, DeSalvo, Dhurandhar, az, Dietz, Fiore, Lieto, Palma,
  Virgilio, Dmitry, Donovan, Dooley, Doravari, Drago, Drever, Driggers, Du,
  Dumas, Dwyer, Eberle, Edwards, Effler, Ehrens, Eichholz, Eikenberry,
  Endr\"{o}czi, Essick, Etzel, Evans, Evans, Evans, Factourovich, Fafone,
  Fairhurst, Fang, Farr, Farr, Favata, Fazi, Fehrmann, Feldbaum, Ferrante,
  Ferrini, Fidecaro, Finn, Fiori, Fisher, Flaminio, Foley, Foley, Forsi, Forte,
  Fotopoulos, Fournier, Franco, Frasca, Frasconi, Frede, Frei, Frei, Freise,
  Frey, Fricke, Fritschel, Frolov, Fujimoto, Fulda, Fyffe, Gair, Gammaitoni,
  Garcia, Garufi, Gehrels, Gemme, Genin, Gennai, Gergely, Ghosh, Giaime,
  Giampanis, Giardina, Giazotto, Gil-Casanova, Gill, Gleason, Goetz, Goetz,
  Gondan, Gonz{\'{a}}lez, Gordon, Gorodetsky, Gossan, Go{\ss}ler, Gouaty,
  Graef, Graff, Granata, Grant, Gras, Gray, Greenhalgh, Gretarsson, Griffo,
  Grote, Grover, Grunewald, Guidi, Guido, Gushwa, Gustafson, Gustafson, Hall,
  Hall, Hammer, Hammond, Hanke, Hanks, Hanna, Hanson, Harms, Harry, Harry,
  Harstad, Hartman, Haughian, Hayama, Heefner, Heidmann, Heintze, Heitmann,
  Hello, Hemming, Hendry, Heng, Heptonstall, Heurs, Hild, Hoak, Hodge, Holt,
  Hong, Hooper, Horrom, Hosken, Hough, Howell, Hu, Hua, Huang, Huerta, Hughey,
  Husa, Huttner, Huynh, Huynh-Dinh, Iafrate, Ingram, Inta, Isogai, Ivanov,
  Iyer, Izumi, Jacobson, James, Jang, Jang, Jaranowski, Jim{\'{e}}nez-Forteza,
  Johnson, Jones, Jones, Jones, Jonker, Ju, K, Kalmus, Kalogera, Kandhasamy,
  Kang, Kanner, Kasprzack, Kasturi, Katsavounidis, Katzman, Kaufer, Kaufman,
  Kawabe, Kawamura, Kawazoe, K{\'{e}}f{\'{e}}lian, Keitel, Kelley, Kells,
  Keppel, Khalaidovski, Khalili, Khazanov, Kim, Kim, Kim, Kim, Kim, Kim, King,
  King, Kinzel, Kissel, Klimenko, Kline, Koehlenbeck, Kokeyama, Kondrashov,
  Koranda, Korth, Kowalska, Kozak, Kremin, Kringel, Krishnan, Kr{\'{o}}lak,
  Kucharczyk, Kudla, Kuehn, Kumar, Kumar, Kumar, Kumar, Kurdyumov, Kwee,
  Landry, Lantz, Larson, Lasky, Lawrie, Lazzarini, Leaci, Lebigot, Lee, Lee,
  Lee, Lee, Lee, Leonardi, Leong, Roux, Leroy, Letendre, Levine, Lewis,
  Lhuillier, Li, Lin, Littenberg, Litvine, Liu, Liu, Liu, Liu, Lloyd,
  Lockerbie, Lockett, Lodhia, Loew, Logue, Lombardi, Lorenzini, Loriette,
  Lormand, Losurdo, Lough, Luan, Lubinski, L\"{u}ck, Lundgren, Macarthur,
  Macdonald, Machenschalk, MacInnis, Macleod, Magana-Sandoval, Mageswaran,
  Mailand, Majorana, Maksimovic, Malvezzi, Man, Manca, Mandel, Mandic, Mangano,
  Mantovani, Marchesoni, Marion, M{\'{a}}rka, M{\'{a}}rka, Markosyan, Maros,
  Marque, Martelli, Martellini, Martin, Martin, Martynov, Marx, Mason,
  Masserot, Massinger, Matichard, Matone, Matzner, Mavalvala, May, Mazumder,
  Mazzolo, McCarthy, McClelland, McGuire, McIntyre, McIver, Meacher, Meadors,
  Mehmet, Meidam, Meier, Melatos, Mendell, Mercer, Meshkov, Messenger, Meyer,
  Miao, Michel, Mikhailov, Milano, Miller, Minenkov, Mingarelli, Mitra,
  Mitrofanov, Mitselmakher, Mittleman, Moe, Mohan, Mohapatra, Mokler, Moraru,
  Moreno, Morgado, Mori, Morriss, Mossavi, Mours, Mow-Lowry, Mueller, Mueller,
  Mukherjee, Mullavey, Munch, Murphy, Murray, Mytidis, Nagy, Nardecchia, Nash,
  Naticchioni, Nayak, Necula, Neri, Newton, Nguyen, Nishida, Nishizawa, Nitz,
  Nocera, Nolting, Normandin, Nuttall, Ochsner, O'Dell, Oelker, Ogin, Oh, Oh,
  Ohme, Oppermann, O'Reilly, Larcher, O'Shaughnessy, Osthelder, Ott, Ottaway,
  Ottens, Ou, Overmier, Owen, Padilla, Pai, Palomba, Pan, Pankow, Paoletti,
  Paoletti, Papa, Paris, Pasqualetti, Passaquieti, Passuello, Pedraza, Peiris,
  Penn, Perreca, Phelps, Pichot, Pickenpack, Piergiovanni, Pierro, Pinard,
  Pindor, Pinto, Pitkin, Poeld, Poggiani, Poole, Poux, Predoi, Prestegard,
  Price, Prijatelj, Principe, Privitera, Prodi, Prokhorov, Puncken, Punturo,
  Puppo, Quetschke, Quintero, Quitzow-James, Raab, Rabeling, R{\'{a}}cz,
  Radkins, Raffai, Raja, Rajalakshmi, Rakhmanov, Ramet, Rapagnani, Raymond, Re,
  Reed, Reed, Regimbau, Reid, Reitze, Ricci, Riesen, Riles, Robertson, Robinet,
  Rocchi, Roddy, Rodriguez, Rodruck, Roever, Rolland, Rollins, Romano, Romanov,
  Romie, Rosi{\'{n}}ska, Rowan, R\"{u}diger, Ruggi, Ryan, Salemi, Sammut,
  Sandberg, Sanders, Sannibale, Santiago-Prieto, Saracco, Sassolas,
  Sathyaprakash, Saulson, Savage, Schilling, Schnabel, Schofield, Schreiber,
  Schuette, Schulz, Schutz, Schwinberg, Scott, Scott, Seifert, Sellers,
  Sengupta, Sentenac, Sergeev, Shaddock, Shah, Shahriar, Shaltev, Shapiro,
  Shawhan, Shoemaker, Sidery, Siellez, Siemens, Sigg, Simakov, Singer, Singer,
  Sintes, Skelton, Slagmolen, Slutsky, Smith, Smith, Smith, Smith-Lefebvre,
  Soden, Son, Sorazu, Souradeep, Sperandio, Staley, Steinert, Steinlechner,
  Steinlechner, Steplewski, Stevens, Stochino, Stone, Strain, Strigin, Stroeer,
  Sturani, Stuver, Summerscales, Susmithan, Sutton, Swinkels, Szeifert, Tacca,
  Talukder, Tang, Tanner, Tarabrin, Taylor, ter Braack, Thirugnanasambandam,
  Thomas, Thomas, Thorne, Thorne, Thrane, Tiwari, Tokmakov, Tomlinson,
  Toncelli, Tonelli, Torre, Torres, Torrie, Travasso, Traylor, Tse, Ugolini,
  Unnikrishnan, Vahlbruch, Vajente, Vallisneri, van~den Brand, Broeck, van~der
  Putten, van~der Sluys, van Heijningen, van Veggel, Vass, Vas{\'{u}}th,
  Vaulin, Vecchio, Vedovato, Veitch, Veitch, Venkateswara, Verkindt, Verma,
  Vetrano, Vicer{\'{e}}, Vincent-Finley, Vinet, Vitale, Vlcek, Vo, Vocca,
  Vorvick, Vousden, Vrinceanu, Vyachanin, Wade, Wade, Wade, Waldman, Walker,
  Wallace, Wan, Wang, Wang, Wang, Wanner, Ward, Was, Weaver, Wei, Weinert,
  Weinstein, Weiss, Welborn, Wen, Wessels, West, Westphal, Wette, Whelan,
  Whitcomb, White, Whiting, Wibowo, Wiesner, Wilkinson, Williams, Williams,
  Williams, Willis, Willke, Wimmer, Winkelmann, Winkler, Wipf, Wittel, Woan,
  Worden, Yablon, Yakushin, Yamamoto, Yancey, Yang, Yeaton-Massey, Yoshida,
  Yum, Yvert, Zadro{\.{z}}ny, Zanolin, Zendri, Zhang, Zhang, Zhao, Zhu, Zhu,
  Zotov, Zucker, and Zweizig]{Aasi2015}
Aasi, J.; Abadie, J.; Abbott, B.P.; Abbott, R.; Abbott, T.; Abernathy, M.R.;
  Accadia, T.; Acernese, F.; Adams, C.; Adams, T.;  et~al.
\newblock Characterization of the {LIGO} detectors during their sixth science
  run.
\newblock {\em Classical and Quantum Gravity} {\bf 2015}, {\em 32},~115012.
\newblock
  doi:{\changeurlcolor{black}\href{https://doi.org/10.1088/0264-9381/32/11/115012}{\detokenize{10.1088/0264-9381/32/11/115012}}}.

\bibitem[Caron \em{et~al.}(1997)Caron, Dominjon, Drezen, Flaminio, Grave,
  Marion, Massonnet, Mehmel, Morand, Mours, Sannibale, Yvert, Babusci,
  Bellucci, Candusso, Giordano, Matone, Mackowski, Pinard, Barone, Calloni, {Di
  Fiore}, Flagiello, Garuti, Grado, Longo, Lops, Marano, Milano, Solimeno,
  Brisson, Cavalier, Davier, Hello, Heusse, Mann, Acker, Barsuglia, Bhawal,
  Bondu, Brillet, Heitmann, Innocent, Latrach, Man, Pham-Tu, Tournier,
  Taubmann, Vinet, Boccara, Gleyzes, Loriette, Roger, Cagnoli, Gammaitoni,
  Kovalik, Marchesoni, Punturo, Beccaria, Bernardini, Bougleux, Braccini,
  Bradaschia, Cella, Ciampa, Cuoco, Curci, {Del Fabbro}, {De Salvo}, {Di
  Virgilio}, Enard, Ferrante, Fidecaro, Giassi, Giazotto, Holloway, {La Penna},
  Losurdo, Mancini, Mazzoni, Palla, Pan, Passuello, Pelfer, Poggiani, Stanga,
  Vicere', Zhang, Ferrari, Majorana, Puppo, Rapagnani, and Ricci]{Caron1997}
Caron, B.; Dominjon, A.; Drezen, C.; Flaminio, R.; Grave, X.; Marion, F.;
  Massonnet, L.; Mehmel, C.; Morand, R.; Mours, B.;  et~al.
\newblock {The VIRGO interferometer for gravitational wave detection}.
\newblock {\em Nuclear Physics B - Proceedings Supplements} {\bf 1997}.
\newblock
  doi:{\changeurlcolor{black}\href{https://doi.org/10.1016/S0920-5632(97)00109-6}{\detokenize{10.1016/S0920-5632(97)00109-6}}}.

\bibitem[Acernese \em{et~al.}(2008)Acernese, Alshourbagy, Amico, Antonucci,
  Aoudia, Astone, Avino, Baggio, Ballardin, Barone, Barsotti, Barsuglia, Bauer,
  Bigotta, Birindelli, Bizouard, Boccara, Bondu, Bosi, Braccini, Bradaschia,
  Brillet, Brisson, Buskulic, Cagnoli, Calloni, Campagna, Carbognani, Cavalier,
  Cavalieri, Cella, Cesarini, Chassande-Mottin, Clapson, Cleva, Coccia, Corda,
  Corsi, Cottone, Coulon, Cuoco, D'Antonio, Dari, Dattilo, Davier, {De Rosa},
  DelPrete, {Di Fiore}, {Di Lieto}, {Di Paolo Emilio}, {Di Virgilio}, Evans,
  Fafone, Ferrante, Fidecaro, Fiori, Flaminio, Fournier, Frasca, Frasconi,
  Gammaitoni, Garufi, Genin, Gennai, Giazotto, Giordano, Granata, Greverie,
  Grosjean, Guidi, Hamdani, Hebri, Heitmann, Hello, Huet, Kreckelbergh, {La
  Penna}, Laval, Leroy, Letendre, Lopez, Lorenzini, Loriette, Losurdo,
  MacKowski, Majorana, Man, Mantovani, Marchesoni, Marion, Marque, Martelli,
  Masserot, Menzinger, Milano, Minenkov, Moins, Moreau, Morgado, Mosca, Mours,
  Neri, Nocera, Pagliaroli, Palomba, Paoletti, Pardi, Pasqualetti, Passaquieti,
  Passuello, Piergiovanni, Pinard, Poggiani, Punturo, Puppo, Rapagnani,
  Regimbau, Remillieux, Ricci, Ricciardi, Rocchi, Rolland, Romano, Ruggi,
  Russo, Solimeno, Spallicci, Tarallo, Terenzi, Toncelli, Tonelli, Tournefier,
  Travasso, Tremola, Vajente, {Van Den Brand}, {Van Der Putten}, Verkindt,
  Vetrano, Vicer{\'{e}}, Vinet, Vocca, and Yvert]{Acernese2008}
Acernese, F.; Alshourbagy, M.; Amico, P.; Antonucci, F.; Aoudia, S.; Astone,
  P.; Avino, S.; Baggio, L.; Ballardin, G.; Barone, F.;  et~al.
\newblock {Status of Virgo}.
\newblock {\em Classical and Quantum Gravity} {\bf 2008}, {\em 25},~114045.
\newblock
  doi:{\changeurlcolor{black}\href{https://doi.org/10.1088/0264-9381/25/11/114045}{\detokenize{10.1088/0264-9381/25/11/114045}}}.

\bibitem[Grote \em{et~al.}(2005)Grote, Allen, Aufmuth, Aulbert, Babak,
  Balasubramanian, Barr, Berukoff, Bunkowski, Cagnoli, Cantley, Casey,
  Chelkowski, Churches, Cokelaer, Colacino, Crooks, Cutler, Danzmann, Davies,
  Dupuis, Elliffe, Fallnich, Franzen, Freise, Go{\ss}ler, Grant, Grunewald,
  Harms, Heinzel, Heng, Hepstonstall, Heurs, Hewitson, Hild, Hough, Itoh,
  Jones, Huttner, Kawabe, Killow, K\"{o}tter, Krishnan, Leonhardt, L\"{u}ck,
  Machenschalk, Malec, Mercer, Messenger, Mohanty, Mossavi, Mukherjee, Murray,
  Nagano, Newton, Papa, Perreur-Lloyd, Pitkin, Plissi, Quetschke, Re, Reid,
  Ribichini, Robertson, Robertson, Romano, Rowan, R\"{u}diger, Sathyaprakash,
  Schilling, Schnabel, Schutz, Seifert, Sintes, Smith, Sneddon, Strain, Taylor,
  Taylor, Th\"{u}ring, Ungarelli, Vahlbruch, Vecchio, Veitch, Ward, Weiland,
  Welling, Williams, Willke, Winkler, Woan, and Zawischa]{Grote2005}
Grote, H.; Allen, B.; Aufmuth, P.; Aulbert, C.; Babak, S.; Balasubramanian, R.;
  Barr, B.W.; Berukoff, S.; Bunkowski, A.; Cagnoli, G.;  et~al.
\newblock The status of {GEO} 600.
\newblock {\em Classical and Quantum Gravity} {\bf 2005}, {\em 22},~S193--S198.
\newblock
  doi:{\changeurlcolor{black}\href{https://doi.org/10.1088/0264-9381/22/10/009}{\detokenize{10.1088/0264-9381/22/10/009}}}.

\bibitem[and(2010)]{Grote2010}
and, H.G.
\newblock The {GEO} 600 status.
\newblock {\em Classical and Quantum Gravity} {\bf 2010}, {\em 27},~084003.
\newblock
  doi:{\changeurlcolor{black}\href{https://doi.org/10.1088/0264-9381/27/8/084003}{\detokenize{10.1088/0264-9381/27/8/084003}}}.

\bibitem[Ando and Collaboration(2002)]{Ando2002}
Ando, M.; Collaboration, t.T.
\newblock {Current status of TAMA}.
\newblock {\em Classical and Quantum Gravity} {\bf 2002}, {\em 19},~1409.
\newblock
  doi:{\changeurlcolor{black}\href{https://doi.org/10.1088/0264-9381/19/7/324}{\detokenize{10.1088/0264-9381/19/7/324}}}.

\bibitem[Acernese \em{et~al.}(2015)Acernese, Agathos, Agatsuma, Aisa,
  Allemandou, Allocca, Amarni, Astone, Balestri, Ballardin, Barone, Baronick,
  Barsuglia, Basti, Basti, Bauer, Bavigadda, Bejger, Beker, Belczynski,
  Bersanetti, Bertolini, Bitossi, Bizouard, Bloemen, Blom, Boer, Bogaert,
  Bondi, Bondu, Bonelli, Bonnand, Boschi, Bosi, Bouedo, Bradaschia, Branchesi,
  Briant, Brillet, Brisson, Bulik, Bulten, Buskulic, Buy, Cagnoli, Calloni,
  Campeggi, Canuel, Carbognani, Cavalier, Cavalieri, Cella, Cesarini, Mottin,
  Chincarini, Chiummo, Chua, Cleva, Coccia, Cohadon, Colla, Colombini, Conte,
  Coulon, Cuoco, Dalmaz, D'Antonio, Dattilo, Davier, Day, Debreczeni,
  Degallaix, Del{\'{e}}glise, Pozzo, Dereli, Rosa, Fiore, Lieto, Virgilio,
  Doets, Dolique, Drago, Ducrot, Endrczi, Fafone, Farinon, Ferrante, Ferrini,
  Fidecaro, Fiori, Flaminio, Fournier, Franco, Frasca, Frasconi, Gammaitoni,
  Garufi, Gaspard, Gatto, Gemme, Gendre, Genin, Gennai, Ghosh, Giacobone,
  Giazotto, Gouaty, Granata, Greco, Groot, Guidi, Harms, Heidmann, Heitmann,
  Hello, Hemming, Hennes, Hofman, Jaranowski, Jonker, Kasprzack,
  K{\'{e}}f{\'{e}}lian, Kowalska, Kraan, Kr{\'{o}}lak, Kutynia, Lazzaro,
  Leonardi, Leroy, Letendre, Li, Lieunard, Lorenzini, Loriette, Losurdo,
  Magazz{\'{u}}, Majorana, Maksimovic, Malvezzi, Man, Mangano, Mantovani,
  Marchesoni, Marion, Marque, Martelli, Martellini, Masserot, Meacher, Meidam,
  Mezzani, Michel, Milano, Minenkov, Moggi, Mohan, Montani, Morgado, Mours,
  Mul, Nagy, Nardecchia, Naticchioni, Nelemans, Neri, Neri, Nocera, Pacaud,
  Palomba, Paoletti, Paoli, Pasqualetti, Passaquieti, Passuello, Perciballi,
  Petit, Pichot, Piergiovanni, Pillant, Piluso, Pinard, Poggiani, Prijatelj,
  Prodi, Punturo, Puppo, Rabeling, R{\'{a}}cz, Rapagnani, Razzano, Re,
  Regimbau, Ricci, Robinet, Rocchi, Rolland, Romano, Rosi{\'{n}}ska, Ruggi,
  Saracco, Sassolas, Schimmel, Sentenac, Sequino, Shah, Siellez, Straniero,
  Swinkels, Tacca, Tonelli, Travasso, Turconi, Vajente, {Van Bakel}, {Van
  Beuzekom}, {Van Den Brand}, {Van Den Broeck}, {Van Der Sluys}, {Van
  Heijningen}, Vas{\'{u}}th, Vedovato, Veitch, Verkindt, Vetrano, Vicer{\'{e}},
  Vinet, Visser, Vocca, Ward, Was, Wei, Yvert, Zny, and Zendri]{Acernese2015}
Acernese, F.; Agathos, M.; Agatsuma, K.; Aisa, D.; Allemandou, N.; Allocca, A.;
  Amarni, J.; Astone, P.; Balestri, G.; Ballardin, G.;  et~al.
\newblock {Advanced Virgo: A second-generation interferometric gravitational
  wave detector}.
\newblock {\em Classical and Quantum Gravity} {\bf 2015}, {\em 32},~024001,
  \href{https://arxiv.org/abs/1408.3978}{{\normalfont [1408.3978]}}.
\newblock
  doi:{\changeurlcolor{black}\href{https://doi.org/10.1088/0264-9381/32/2/024001}{\detokenize{10.1088/0264-9381/32/2/024001}}}.

\bibitem[Somiya(2012)]{Somiya2012}
Somiya, K.
\newblock Detector configuration of {KAGRA}{\textendash}the Japanese cryogenic
  gravitational-wave detector.
\newblock {\em Classical and Quantum Gravity} {\bf 2012}, {\em 29},~124007.
\newblock
  doi:{\changeurlcolor{black}\href{https://doi.org/10.1088/0264-9381/29/12/124007}{\detokenize{10.1088/0264-9381/29/12/124007}}}.

\bibitem[Aso \em{et~al.}(2013)Aso, Michimura, Somiya, Ando, Miyakawa,
  Sekiguchi, Tatsumi, and Yamamoto]{Aso2013}
Aso, Y.; Michimura, Y.; Somiya, K.; Ando, M.; Miyakawa, O.; Sekiguchi, T.;
  Tatsumi, D.; Yamamoto, H.
\newblock Interferometer design of the {KAGRA} gravitational wave detector.
\newblock {\em Physical Review D} {\bf 2013}, {\em 88}.
\newblock
  doi:{\changeurlcolor{black}\href{https://doi.org/10.1103/physrevd.88.043007}{\detokenize{10.1103/physrevd.88.043007}}}.

\bibitem[Akutsu \em{et~al.}(2021)Akutsu, Ando, Arai, Arai, Araki, Araya,
  Aritomi, Aso, Bae, Bae, Baiotti, Bajpai, Barton, Cannon, Capocasa, Chan,
  Chen, Chen, Chen, Chu, Chu, Eguchi, Enomoto, Flaminio, Fujii, Fukunaga,
  Fukushima, Ge, Hagiwara, Haino, Hasegawa, Hayakawa, Hayama, Himemoto,
  Hiranuma, Hirata, Hirose, Hong, Hsieh, Huang, Huang, Huang, Ikenoue, Imam,
  Inayoshi, Inoue, Ioka, Itoh, Izumi, Jung, Jung, Kajita, Kamiizumi, Kanda,
  Kang, Kawaguchi, Kawai, Kawasaki, Kim, Kim, Kim, Kim, Kimura, Kita, Kitazawa,
  Kojima, Kokeyama, Komori, Kong, Kotake, Kozakai, Kozu, Kumar, Kume, Kuo, Kuo,
  Kuroyanagi, Kusayanagi, Kwak, Lee, Lee, Lee, Leonardi, Lin, Lin, Lin, Liu,
  Luo, Marchio, Michimura, Mio, Miyakawa, Miyamoto, Miyazaki, Miyo, Miyoki,
  Morisaki, Moriwaki, Nagano, Nagano, Nakamura, Nakano, Nakano, Nakashima,
  Narikawa, Negishi, Ni, Nishizawa, Obuchi, Ogaki, Oh, Oh, Ohashi, Ohishi,
  Ohkawa, Okutomi, Oohara, Ooi, Oshino, Pan, Pang, Park, Arellano, Pinto, Sago,
  Saito, Saito, Sakai, Sakai, Sakuno, Sato, Sato, Sawada, Sekiguchi, Sekiguchi,
  Shibagaki, Shimizu, Shimoda, Shimode, Shinkai, Shishido, Shoda, Somiya, Son,
  Sotani, Sugimoto, Suzuki, Suzuki, Tagoshi, Takahashi, Takahashi, Takamori,
  Takano, Takeda, Takeda, Tanaka, Tanaka, Tanaka, Tanaka, Tanaka, Tanioka,
  {Tapia San Martin}, Telada, Tomaru, Tomigami, Tomura, Travasso, Trozzo,
  Tsang, Tsubono, Tsuchida, Tsuzuki, Tuyenbayev, Uchikata, Uchiyama, Ueda,
  Uehara, Ueno, Ueshima, Uraguchi, Ushiba, {Van Putten}, Vocca, Wang, Wu, Wu,
  Wu, Xu, Yamada, Yamamoto, Yamamoto, Yamamoto, Yokogawa, Yokoyama, Yokozawa,
  Yoshioka, Yuzurihara, Zeidler, Zhao, and Zhu]{Akutsu2021}
Akutsu, T.; Ando, M.; Arai, K.; Arai, Y.; Araki, S.; Araya, A.; Aritomi, N.;
  Aso, Y.; Bae, S.; Bae, Y.;  et~al.
\newblock {Overview of KAGRA: Detector design and construction history}.
\newblock {\em Progress of Theoretical and Experimental Physics} {\bf 2021},
  {\em 2021},~49,  \href{https://arxiv.org/abs/2005.05574}{{\normalfont
  [2005.05574]}}.
\newblock
  doi:{\changeurlcolor{black}\href{https://doi.org/10.1093/PTEP/PTAA125}{\detokenize{10.1093/PTEP/PTAA125}}}.

\bibitem[Padma(2019)]{Padma2019}
Padma, T.
\newblock {India's LIGO gravitational-wave observatory gets green light}.
\newblock {\em Nature} {\bf 2019}.
\newblock
  doi:{\changeurlcolor{black}\href{https://doi.org/10.1038/D41586-019-00184-Z}{\detokenize{10.1038/D41586-019-00184-Z}}}.

\bibitem[Adhikari \em{et~al.}(2020)Adhikari, Arai, Brooks, Wipf, Aguiar, Altin,
  Barr, Barsotti, Bassiri, Bell, Billingsley, Birney, Blair, Bonilla, Briggs,
  Brown, Byer, Cao, Constancio, Cooper, Corbitt, Coyne, Cumming, Daw, Derosa,
  Eddolls, Eichholz, Evans, Fejer, Ferreira, Freise, Frolov, Gras, Green,
  Grote, Gustafson, Hall, Hammond, Harms, Harry, Haughian, Heinert, Heintze,
  Hellman, Hennig, Hennig, Hild, Hough, Johnson, Kamai, Kapasi, Komori,
  Koptsov, Korobko, Korth, Kuns, Lantz, Leavey, Magana-Sandoval, Mansell,
  Markosyan, Markowitz, Martin, Martin, Martynov, Mcclelland, Mcghee, Mcrae,
  Mills, Mitrofanov, Molina-Ruiz, Mow-Lowry, Munch, Murray, Ng, Okada, Ottaway,
  Prokhorov, Quetschke, Reid, Reitze, Richardson, Robie, Romero-Shaw, Route,
  Rowan, Schnabel, Schneewind, Seifert, Shaddock, Shapiro, Shoemaker, Silva,
  Slagmolen, Smith, Smith, Steinlechner, Strain, Taira, Tait, Tanner, Tornasi,
  Torrie, {Van Veggel}, Vanheijningen, Veitch, Wade, Wallace, Ward, Weiss,
  Wessels, Willke, Yamamoto, Yap, and Zhao]{Adhikari2020}
Adhikari, R.X.; Arai, K.; Brooks, A.F.; Wipf, C.; Aguiar, O.; Altin, P.; Barr,
  B.; Barsotti, L.; Bassiri, R.; Bell, A.;  et~al.
\newblock {A cryogenic silicon interferometer for gravitational-wave
  detection}.
\newblock {\em Classical and Quantum Gravity} {\bf 2020}, {\em 37},~165003,
  \href{https://arxiv.org/abs/2001.11173}{{\normalfont [2001.11173]}}.
\newblock
  doi:{\changeurlcolor{black}\href{https://doi.org/10.1088/1361-6382/AB9143}{\detokenize{10.1088/1361-6382/AB9143}}}.

\bibitem[Hild \em{et~al.}(2011)Hild, Abernathy, Acernese, Amaro-Seoane,
  Andersson, Arun, Barone, Barr, Barsuglia, Beker, Beveridge, Birindelli, Bose,
  Bosi, Braccini, Bradaschia, Bulik, Calloni, Cella, Mottin, Chelkowski,
  Chincarini, Clark, Coccia, Colacino, Colas, Cumming, Cunningham, Cuoco,
  Danilishin, Danzmann, Salvo, Dent, Rosa, Fiore, Virgilio, Doets, Fafone,
  Falferi, Flaminio, Franc, Frasconi, Freise, Friedrich, Fulda, Gair, Gemme,
  Genin, Gennai, Giazotto, Glampedakis, Gräf, Granata, Grote, Guidi,
  Gurkovsky, Hammond, Hannam, Harms, Heinert, Hendry, Heng, Hennes, Hough,
  Husa, Huttner, Jones, Khalili, Kokeyama, Kokkotas, Krishnan, Li, Lorenzini,
  Lück, Majorana, Mandel, Mandic, Mantovani, Martin, Michel, Minenkov,
  Morgado, Mosca, Mours, Müller{\textendash}Ebhardt, Murray, Nawrodt, Nelson,
  Oshaughnessy, Ott, Palomba, Paoli, Parguez, Pasqualetti, Passaquieti,
  Passuello, Pinard, Plastino, Poggiani, Popolizio, Prato, Punturo, Puppo,
  Rabeling, Rapagnani, Read, Regimbau, Rehbein, Reid, Ricci, Richard, Rocchi,
  Rowan, Rüdiger, Santamar{\'{\i}}a, Sassolas, Sathyaprakash, Schnabel,
  Schwarz, Seidel, Sintes, Somiya, Speirits, Strain, Strigin, Sutton, Tarabrin,
  Thüring, van~den Brand, van Veggel, van~den Broeck, Vecchio, Veitch,
  Vetrano, Vicere, Vyatchanin, Willke, Woan, and
  Yamamoto]{EinsteinTelescope2011}
Hild, S.; Abernathy, M.; Acernese, F.; Amaro-Seoane, P.; Andersson, N.; Arun,
  K.; Barone, F.; Barr, B.; Barsuglia, M.; Beker, M.;  et~al.
\newblock Sensitivity studies for third-generation gravitational wave
  observatories.
\newblock {\em Classical and Quantum Gravity} {\bf 2011}, {\em 28},~094013.
\newblock
  doi:{\changeurlcolor{black}\href{https://doi.org/10.1088/0264-9381/28/9/094013}{\detokenize{10.1088/0264-9381/28/9/094013}}}.

\bibitem[Evans \em{et~al.}(2021)Evans, Adhikari, Afle, Ballmer, Biscoveanu,
  Borhanian, Brown, Chen, Eisenstein, Gruson, Gupta, Hall, Huxford, Kamai,
  Kashyap, Kissel, Kuns, Landry, Lenon, Lovelace, McCuller, Ng, Nitz, Read,
  Sathyaprakash, Shoemaker, Slagmolen, Smith, Srivastava, Sun, Vitale, and
  Weiss]{cehorizonstudy2021}
Evans, M.; Adhikari, R.X.; Afle, C.; Ballmer, S.W.; Biscoveanu, S.; Borhanian,
  S.; Brown, D.A.; Chen, Y.; Eisenstein, R.; Gruson, A.;  et~al.
\newblock A Horizon Study for Cosmic Explorer: Science, Observatories, and
  Community,  2021,  \href{https://arxiv.org/abs/2109.09882}{{\normalfont
  [arXiv:astro-ph.IM/2109.09882]}}.

\bibitem[Baker \em{et~al.}(2019)Baker, Bellovary, Bender, Berti, Caldwell,
  Camp, Conklin, Cornish, Cutler, DeRosa, Eracleous, Ferrara, Francis,
  Hewitson, Holley-Bockelmann, Hornschemeier, Hogan, Kamai, Kelly, Key, Larson,
  Livas, Manthripragada, McKenzie, McWilliams, Mueller, Natarajan, Numata,
  Rioux, Sankar, Schnittman, Shoemaker, Shoemaker, Slutsky, Spero, Stebbins,
  Thorpe, Vallisneri, Ware, Wass, Yu, and Ziemer]{Baker2019}
Baker, J.; Bellovary, J.; Bender, P.L.; Berti, E.; Caldwell, R.; Camp, J.;
  Conklin, J.W.; Cornish, N.; Cutler, C.; DeRosa, R.;  et~al.
\newblock {The Laser Interferometer Space Antenna: Unveiling the Millihertz
  Gravitational Wave Sky} {\bf 2019}.
\newblock  \href{https://arxiv.org/abs/1907.06482}{{\normalfont [1907.06482]}}.
\newblock
  doi:{\changeurlcolor{black}\href{https://doi.org/10.13016/M2WM22-K0UV}{\detokenize{10.13016/M2WM22-K0UV}}}.

\bibitem[Sumner \em{et~al.}(2017)Sumner, Mueller, and Conklin]{Sumner2017}
Sumner, T.J.; Mueller, G.; Conklin, J.W.
\newblock {LISA Pathfinder: First steps to observing gravitational waves from
  space}.
\newblock {\em Journal of Physics: Conference Series} {\bf 2017}, {\em
  840},~012001.
\newblock
  doi:{\changeurlcolor{black}\href{https://doi.org/10.1088/1742-6596/840/1/012001}{\detokenize{10.1088/1742-6596/840/1/012001}}}.

\bibitem[Michelson and Morley(1887)]{Michelson1887}
Michelson, A.A.; Morley, E.W.
\newblock On the Relative Motion of the Earth and the Luminiferous Ether.
\newblock {\em Am. J. Sci.} {\bf 1887}, {\em 34},~333--345.
\newblock
  doi:{\changeurlcolor{black}\href{https://doi.org/10.2475/ajs.s3-34.203.333}{\detokenize{10.2475/ajs.s3-34.203.333}}}.

\bibitem[Willke \em{et~al.}(1999)Willke, Gustafson, Husman, Lawrence, and
  Byer]{Willke1999}
Willke, B.; Gustafson, E.K.; Husman, M.E.; Lawrence, M.J.; Byer, R.L.
\newblock {Dynamic response of a Fabry–Perot interferometer}.
\newblock {\em JOSA B, Vol. 16, Issue 4, pp. 523-532} {\bf 1999}, {\em
  16},~523--532.
\newblock
  doi:{\changeurlcolor{black}\href{https://doi.org/10.1364/JOSAB.16.000523}{\detokenize{10.1364/JOSAB.16.000523}}}.

\bibitem[Rakhmanov \em{et~al.}(2002)Rakhmanov, Savage, Reitze, and
  Tanner]{Rakhmanov2002}
Rakhmanov, M.; Savage, R.; Reitze, D.; Tanner, D.
\newblock Dynamic resonance of light in Fabry-Perot cavities.
\newblock {\em Phys. Lett. A} {\bf 2002}, {\em 305},~239--244.

\bibitem[Bondu and Debieu(2007)]{Bondu2007}
Bondu, F.; Debieu, O.
\newblock {Accurate measurement method of Fabry-Perot cavity parameters via
  optical transfer function}.
\newblock {\em Applied Optics} {\bf 2007}, {\em 46},~2611--2614.
\newblock
  doi:{\changeurlcolor{black}\href{https://doi.org/10.1364/AO.46.002611}{\detokenize{10.1364/AO.46.002611}}}.

\end{thebibliography}

\end{document}